\documentclass[12pt]{article}

\ifx\pdfoutput\undefined
\usepackage[dvips,bookmarks]{hyperref}
\else
\usepackage{hyperref}
\fi
\hypersetup{colorlinks=false,bookmarksopen,bookmarksnumbered,citecolor=blue,
   pdfstartview=FitH}

\usepackage{latexsym}
\usepackage{amssymb,amsfonts,amsmath}
\usepackage{graphicx} 
\usepackage{indentfirst}
\usepackage{bbm}

\parskip=\medskipamount

\newcommand {\cA}{{\cal A}}

\newcommand {\cD}{{\cal D}}
\newcommand {\cE}{{\cal E}}

\newcommand {\cG}{{\cal G}}

\newcommand {\cJ}{{\cal J}}
\newcommand {\cK}{{\cal K}}
\newcommand {\cL}{{\cal L}}
\newcommand {\cM}{{\cal M}}
\newcommand {\cN}{{\cal N}}

\newcommand {\cQ}{{\cal Q}}
\newcommand {\cR}{{\cal R}}

\newcommand {\cW}{{\cal W}}

\newcommand {\cY}{{\cal Y}}

\def\a{\alpha}
\def\b{\beta}
\def\c{\chi}
\def\d{\delta}
\def\e{\epsilon}
\def\f{\phi}
\def\g{\gamma}
\def\G{\Gamma}

\def\j{\psi}

\def\l{\lambda}
\def\m{\mu}
\def\n{\nu}
\def\o{\omega}
\def\p{\pi}
\def\q{\theta}
\def\r{\rho}
\def\s{\sigma}

\def\x{\xi}
\def\z{\zeta}
\def\D{\Delta}
\def\F{\Phi}
\def\J{\Psi}
\def\L{\Lambda}
\def\O{\Omega}

\def\U{\Upsilon}

\def\ri{{\rm i}}
\def\re{{\rm e}}

\def\ra{{\rm a}}

\newcommand{\ad}{{\dot{\alpha}}}                           
\newcommand{\bd}{{\dot{\beta}}}                            
\newcommand{\ve}{\varepsilon}                            
\newcommand{\cDB}{{\bar\cD}}                            

\newcommand{\pa}{\partial}                           
\newcommand{\hf}{\frac12}

\newcommand{\vf}{\varphi}

\newcommand{\be}{\begin{equation}}
\newcommand{\ee}{\end{equation}}
\newcommand{\bea}{\begin{eqnarray}}
\newcommand{\eea}{\end{eqnarray}}
\newcommand{\non}{\nonumber}
\newcommand{\ba}{\begin{array}}
\newcommand{\ea}{\end{array}}

\newcommand{\1}{{\underline{1}}}
\newcommand{\2}{{\underline{2}}}

\def\dt#1{{\buildrel {\hbox{\LARGE .}} \over {#1}}}    

\newcommand{\bm}[1]{\mbox{\boldmath$#1$}}

\def\double #1{#1{\hbox{\kern-2pt $#1$}}}

\newcommand{\gd}{{\dot\g}}

\newcommand{\bsubeq}{\begin{subequations}}
\newcommand{\esubeq}{\end{subequations}}

\newcommand{\eps}{{\epsilon}}

\newcommand{\dalpha}{{\dot{\alpha}}}
\newcommand{\dbeta}{{\dot{\beta}}}

\newcommand{\btheta}{{\bar\theta}}
\newcommand{\N}{{\mathcal N}}
\newcommand{\eol}{\notag \\}
\newcommand{\rd}{\mathrm d}
\newcommand{\BCD}{{\bar\cD}}
\newcommand{\bphi}{{\bar\phi}}

\newcommand{\fint}{{\int \rd^4x\, \rd^4\theta\, E\, }}
\newcommand{\cint}{{\int \rd^4x\, \rd^2\theta\, \cE\, }}
\newcommand{\acint}{{\int \rd^4x\, \rd^2\bar\theta\, \bar \cE\, }}
\newcommand{\compint}{{\int \rd^4x\, e\, }}
\newcommand{\HC}{{\mathrm{c.c.}}}
\newcommand{\veps}{\varepsilon}

\begin{document}

\begin{titlepage}
\begin{flushright}
August 2011\\
\end{flushright}
\vspace{5mm}

\begin{center}
{\Large \bf The structure of $\bm{\cN = 2}$  supersymmetric nonlinear sigma models in $\bm{\text{AdS}_4}$}
\\ 
\end{center}

\begin{center}

{\bf Daniel Butter and  Sergei M. Kuzenko }

\footnotesize{
{\it School of Physics M013, The University of Western Australia\\
35 Stirling Highway, Crawley W.A. 6009, Australia}}  ~\\
\texttt{dbutter,\,kuzenko@cyllene.uwa.edu.au}\\
\vspace{2mm}

\end{center}
\vspace{5mm}

\begin{abstract}
\baselineskip=14pt
We present a detailed study of the most general $\cN=2$ supersymmetric 
sigma models in four-dimensional anti-de Sitter space 
($\rm AdS_4$) formulated in terms of $\cN=1$ chiral superfields.
The target space is demonstrated to be a non-compact hyperk\"ahler manifold
restricted to possess a special Killing vector field which generates an
SO(2) group of rotations on the two-sphere of complex structures and necessarily
leaves one of them invariant.
All hyperk\"ahler cones, that is the target spaces of $\cN=2$ superconformal
sigma models, prove to possess such a vector field that belongs to the
Lie algebra of an isometry group SU(2) acting by rotations on the complex
structures. A unique property of the $\cN=2$ sigma models constructed is that 
the algebra of ${\rm OSp(2|4)}$ transformations closes off the mass shell.
We uncover the underlying $\cN=2$ superfield formulation for the $\cN=2$ sigma
models constructed and compute the associated $\cN=2$ supercurrent.

We give a special analysis of the most general systems 
of self-interacting $\cN=2$ tensor multiplets  in $\rm AdS_4$ and their dual sigma models realized in terms of
$\cN=1$ chiral multiplets. We also briefly discuss the relationship between our results on $\cN=2$ 
supersymmetric sigma models formulated in the $\cN=1$ AdS superspace and 
the off-shell sigma models constructed in the $\cN=2$ AdS superspace in arXiv:0807.3368. 
\end{abstract}
\vspace{0.5cm}

\vfill
\end{titlepage}

\newpage
\renewcommand{\thefootnote}{\arabic{footnote}}
\setcounter{footnote}{0}

\tableofcontents


\numberwithin{equation}{section}



\newpage
\section{Introduction}
In four space-time dimensions, Poincar\'e supersymmetry is intimately connected to
complex geometry. The target space $\cM$ of a rigid supersymmetric $\s$-model is a K\"ahler
manifold in the case of $\cN=1$ supersymmetry \cite{Zumino79} and a hyperk\"ahler space for
$\cN=2$ \cite{A-GF} (see also \cite{CF}). Extending the Poincar\'e supersymmetry to
superconformal  symmetry proves to add the requirement that $\cM$ be endowed
with a homothetic conformal Killing vector which is the gradient of a function
\cite{SezginT} (this is most transparent in three dimensions \cite{BCSS,Kuzenko:2010rp}),
and thus  $\cM$ is globally a cone \cite{GR}. 
The target spaces of rigid superconformal $\s$-models are K\"ahler cones
in the case $\cN=1$, and hyperk\"ahler cones for $\cN=2$ \cite{deWKV0,deWKV,deWRV}.

There exist two standard approaches to describe the most general $\cN=1$ supersymmetric $\s$-models:
(i) in terms of  component physical fields;  or (ii) in terms of $\cN=1$ chiral superfields. 
The latter approach is more efficient, due to its intrinsically geometric form and off-shell supersymmetry.

In the case of $\cN=2$ rigid supersymmetric $\s$-models, it is also natural
to make use of a formulation that permits  some amount of supersymmetry to
be realized manifestly. This requires the use of either $\cN=1$ or $\cN=2$
superspace techniques. In the past, two powerful $\cN=2$ superspace approaches
were developed to construct off-shell $\s$-models, which are: 
(i) the  harmonic superspace \cite{GIKOS,GIOS};
and (ii) the  projective superspace \cite{KLR,LR-projective}. 
One of their conceptual virtues is the possibility to generate supersymmetric
$\s$-model actions (and thus hyperk\"ahler metrics) from Lagrangians of
{\it arbitrary} functional form. Still, many supersymmetry practitioners
consider a formulation in $\cN=1$ superspace as the most transparent and economical
one.

In 1986, Hull et al. \cite{HullKLR} formulated, building on the earlier work
of Lindstr\"om and Ro\v{c}ek  \cite{LR}, the most general  $\cN=2$ rigid
supersymmetric $\s$-models without  superpotentials in terms of $\cN=1$ chiral
superfields.  An extension of \cite{HullKLR}  to include superpotentials was
given in \cite{BX}. Arbitrary $\cN=2$ superconformal $\s$-models were
described in terms of $\cN=1$ chiral superfields in \cite{K-duality}.

Recently, there has been a renewed interest in supersymmetric field
theories in four-dimensional anti-de Sitter space ($\rm AdS_4$) 
\cite{BK2011,Adams:2011vw,Festuccia:2011ws}.
This motivated us in \cite{BK2011-5} to construct, as a nontrivial extension
of \cite{HullKLR}, the most general $\cN=2$ AdS supersymmetric $\s$-models
in terms of covariantly chiral superfields on $\cN=1$ AdS
superspace.\footnote{Historically, the $\cN=1$ AdS superspace,
AdS$^{4|4}:={\rm OSp(1|4) }/  {\rm O(3,1)}$,
was introduced in \cite{Keck,Zumino77}, and the superfield approach to
${\rm OSp(1|4)}$ supersymmetry was developed by Ivanov and Sorin \cite{IS}.}
In the present work we elaborate upon the results announced in \cite{BK2011-5}
by providing technical proofs and detailed explanations. Moreover, we
considerably extend \cite{BK2011-5} by including new results.
In particular, we give a special analysis of the most general systems 
of self-interacting $\cN=2$ tensor multiplets in $\rm AdS_4$ and their
dual $\sigma$-models realized in terms of $\cN=1$ chiral multiplets. We
also develop a manifestly $\cN=2$ superfield formulation corresponding 
to the nonlinear $\s$-model constructed in \cite{BK2011-5}. 

It should be mentioned that off-shell supersymmetric $\s$-models in the $\cN=2$
AdS superspace have already been constructed in \cite{KT-M-ads},
building on the projective superspace formulation for general $\cN=2$
supergravity-matter systems \cite{KLRT-M1,KLRT-M2}. For the series of
$\cN=2$ $\s$-models presented in \cite{KT-M-ads}, one can in principle
derive their reformulation in terms of $\cN=1$ chiral superfields by 
(i) eliminating the (infinitely many) auxiliary superfields; 
and (ii) performing appropriate superspace duality transformations.
These are nontrivial technical problems. Their solution should be similar 
in spirit to the analysis given in \cite{K-duality}, but explicit calculations
remain to be done in the future. Here we only briefly discuss the relationship
between our results on $\cN=2$ supersymmetric $\sigma$-models formulated in
the $\cN=1$ AdS superspace and  the off-shell $\sigma$-models constructed in
the $\cN=2$ AdS superspace \cite{KT-M-ads}.

This paper is organized as follows. In the first half, we present an analysis
of the properties of $\N=2$ tensor multiplet actions when written in terms of $\N=1$
superfields. Section \ref{section2} provides a brief review of the topic in
Minkowski space. Section \ref{section3} deals with superconformal tensor
multiplets, emphasizing their properties in $\N=1$ superspace.
We address the AdS situation in section \ref{section4} and derive an additional
condition on the Lagrangian necessary for $\N=2$ supersymmetry in AdS.
Finally in section \ref{section5} we briefly discuss how the new constraint
required in AdS is naturally understood from a projective superspace setting.

The second half of the paper is devoted to general $\sigma$-models involving
hypermultiplets, which are represented purely in terms of $\N=1$ chiral multiplets.
In section \ref{section6}, we analyze the conditions for $\N=2$ supersymmetry in AdS
and show that the target space must be hyperk\"ahler and possess a special
Killing vector with quite interesting properties. We further analyze the 
geometric implications in section \ref{section7} and briefly discuss how the AdS
condition is naturally fulfilled by superconformal $\sigma$-models in section \ref{section8}.
In section \ref{section9} we discuss a novel superfield formulation of
the hypermultiplet in AdS and in section \ref{section10} we briefly discuss the form
of the supercurrent.

There are four technical appendices. The first deals with $\N=1$ superconformal
Killing vector fields in an AdS background; the second presents the analysis for
$\N=2$. The third appendix provides an alternative technical proof of $\N=2$
supersymmetry for nonlinear $\sigma$-models which is more direct than the
one offered in subsection 6.4. The last addresses a technical issue of the
non-minimal supercurrent in AdS.

\section{Self-interacting $\cN=2$ tensor supermultiplets: Poincar\'e supersymmetry}
\label{section2}
\setcounter{equation}{0}

The $\cN=2$ tensor multiplet \cite{Wess,deWvH,BS,deWvHVP,SSW}
consists of an $\rm SU(2)$ triplet of
scalars $g_{ij}$, a gauge two-form $b_{mn}$, a complex scalar $F$, and a
doublet of Weyl fermions $\chi_{\alpha i}$. This set of
component fields gives an off-shell representation of $\N=2$ supersymmetry.
In $\cN=2$ superspace it is described \cite{BS,SSW}
by an iso-triplet superfield 
$\cG^{ij}$ which has the algebraic properties  
${\bar \cG}_{ij} :=(\cG^{ij})^* =\ve_{ik}\ve_{jl}\cG^{kl}$, and obeys 
the constraints
\begin{align}
D^{(i}_\a \cG^{jk)} =  {\bar D}^{(i}_\ad \cG^{jk)} = 0~.
\end{align}
Upon reduction to $\cN=1$ superspace,  $\cG^{ij}$ decomposes into a real linear superfield
$G$ and a chiral scalar $\vf$.\footnote{The real linear superfield $G$ is used
to describe the $\N=1$ tensor multiplet \cite{Siegel}.}
This is why it is also called the $\cN=2$ linear multiplet 
\cite{BS}.

Models of several $\cN=2$ tensor multiplets can therefore be realized in $\cN=1$
superspace, where the $\N=1$ content involves a set of chiral superfields $\vf^I$,
their conjugates $\bar \vf^I$ and a set of real linear superfields $G^I=\bar G^I$,
obeying the usual constraints
\bea
{\bar D}_\ad \vf^I =0~, \qquad {\bar D}^2G^I = D^2 G^I =0~.
\eea
Here the index $I$ runs over the full set of $n$ tensor multiplets, $I = 1, \cdots, n$.

Our goal in this {\it review} section is to construct the most general nonlinear 
$\sigma$-models\footnote{Recall that the Grassmann measure is defined
${\rm d}^4 \q  := {\rm d}^2 \q \,{\rm d}^2 {\bar \q}$.}
\bea
S=\int {\rm d}^4 x \,{\rm d}^4 \q  \, L (\vf^I, \bar \vf^I , G^I) \label{N=2_tensor_action}
\eea
which are invariant under the second supersymmetry transformations \cite{LR}
\begin{subequations}\label{tensor_mult_second_SUSY}
\bea
\d \vf^I &=& \bar \e \bar D G^I ~,\\
\d G^I &=& - \e D \vf^I - \bar \e \bar D \bar \vf^I~,\\
\d \bar \vf^I &=&  \e  D G^I ~.
\eea
\end{subequations}
This problem was solved by Lindstr\"om and Ro\v{c}ek in 1983 \cite{LR},
but the technical details of the derivation were not included.
Here we give a complete derivation in superspace which is interesting in its
own right, and more importantly can be generalized to the superconformal and
AdS cases. Along the way we will rediscover certain interesting features of
such $\sigma$-models which were noticed in \cite{deWRV} for the case of
superconformal tensor models.

One important feature of the Lagrangian \eqref{N=2_tensor_action} to keep in mind is that it possesses
two classes of trivial transformations
\begin{subequations}\label{eq_TensorTrivialMinkowski}
\begin{align}
L &\rightarrow L + F(\vf) + \bar F(\bar\vf)~, \\
L &\rightarrow L + G^I H_I(\vf) + G^I \bar H_I(\bar\vf)\label{eq_Htransformation}
\end{align}
\end{subequations}
under which the action is invariant. The first is a K\"ahler-like
transformation and is particular to the Minkowski case alone. The second,
which we denote the $H$-transformation, has a much broader applicability,
generalizing not only to AdS but also to arbitrary $\N=1$ supergravity backgrounds.

\subsection{Conditions for $\cN=2$ supersymmetry}
Let us work out what conditions $L$ must obey in order to be $\N=2$
supersymmetric. Since the supersymmetry parameters $\e^\a$ and $\bar \e_\ad$ in 
(\ref{tensor_mult_second_SUSY}) are independent, 
it is sufficient to analyze the condition of invariance under the $\bar \e$-transformation
which is obtained from (\ref{tensor_mult_second_SUSY}) by setting $\e^\a=0$. 
This condition is
\bea
\d_{\bar \e}S &=& \int {\rm d}^4 x \,{\rm d}^4 \q 
\,
\Big\{ \frac{\pa L}{\pa \vf^I} \bar \e \bar D G^I - \frac{\pa L}{\pa G^I} \bar \e \bar D \bar \vf^I \Big\}
=0 ~,
\label{epsilon-var} 
\eea
with ${\bar \e}_\ad$ a constant anti-commuting parameter.

The requirement  (\ref{epsilon-var}) means that the functional
must vanish identically for arbitrary values of the superfields.
Let us vary (\ref{epsilon-var}) with respect to
$\bar \vf^J$. This gives
\bea
 \int 
 {\rm d}^4 x \,{\rm d}^4 \q  \,
  \d \bar \vf^J \Bigg\{ 
\Big( \frac{\pa^2 L}{\pa \bar \vf^J  \pa \vf^I} 
+ \frac{\pa^2 L}{\pa G^J  \pa G^I}\Big)  \bar \e \bar D G^I  
&+&\Big( \frac{\pa^2 L}{\pa   G^J \pa\bar \vf^I  }
- \frac{\pa^2 L}{\pa G^I  \pa \bar \vf^J} \Big)\bar \e \bar D \bar \vf^I \Bigg\} ~.~~~~~~~~
\label{epsilon-var2} 
\eea
${}$For this to vanish, we are forced to require that $L$ obey the Laplace equation \cite{LR}
\bea
\frac{\pa^2 L}{   \pa \vf^I \pa \bar \vf^J } +\frac{\pa^2 L}{\pa G^I  \pa G^J} =0~. 
\label{tensor-Laplace1}
\eea
This equation turns out to imply that the remaining  expression in 
(\ref{epsilon-var2}),
\bea
\int  {\rm d}^4 x \,{\rm d}^4 \q  \, \Big( \frac{\pa^2 L}{\pa   G^J \pa\bar \vf^I  }
- \frac{\pa^2 L}{\pa G^I  \pa \bar \vf^J} \Big) \d \bar \vf^J \bar \e \bar D \bar \vf^I~,
\label{epsilon-var3} 
\eea
is indeed equal to zero. To prove this assertion, we have to make two observations. 
Firstly, the equation (\ref{tensor-Laplace1}) implies that 
\bea
\frac{\pa}{\pa G^K} \Big( \frac{\pa^2 L}{\pa G^J  \pa  \vf^I} - \frac{\pa^2 L}{\pa G^I  \pa  \vf^J} \Big)
= \frac{\pa}{\pa \bar \vf^K} \Big( \frac{\pa^2 L}{\pa G^J  \pa  \vf^I} - \frac{\pa^2 L}{\pa G^I  \pa  \vf^J} \Big)
=0~,
\eea
and hence
 \bea
\frac{\pa^2 L}{\pa G^J  \pa  \vf^I} - \frac{\pa^2 L}{\pa G^I  \pa  \vf^J}  = F_{IJ} (\vf ) = - F_{JI}(\vf)~, 
\label{two-form}
\eea
for some holomorphic two-form $F_{IJ}  (\vf)$ \cite{deWRV}. Secondly, 
making use of the integration rules
\be
\int {\rm d}^4 x \,{\rm d}^4 \q \, U 
= -\frac{1}{4} \int {\rm d}^4 x \,{\rm d}^2 \q \,{\bar D}^2 U
= -\frac{1}{4} \int {\rm d}^4 x \,{\rm d}^2 \bar \q \,{D}^2 U
\label{antichiral_reduction}
\ee
we may show that (\ref{epsilon-var3}) vanishes,
\bea
 \int  {\rm d}^4 x \,{\rm d}^4 \q  \,  
\bar F_{IJ}(\bar \vf)
\d \bar \vf^J \bar \e \bar D \bar \vf^I 
= -\frac{1}{4}  \int  {\rm d}^4 x \, {\rm d}^2 {\bar \q} \,  \bar F_{IJ}(\bar \vf)
\d \bar \vf^J \bar \e_\ad D^2  \bar D^\ad \bar \vf^I \equiv 0~,
\eea
since $ \bar \vf^I $ is antichiral.

It follows from the definition of  the two-form $F_{IJ}  (\vf)$, eq.  (\ref{two-form}),  
that it is closed, 
\bea
\pa_K F_{IJ} +\pa_I F_{JK} +\pa_J F_{KI} =0~.
\eea
On the other hand,  eq. (\ref{two-form}) tells us that  this two-form 
can be written as
\bea
F_{IJ}  = \pa_I \tilde{H}_J  - \pa_J \tilde{H}_I ~, \qquad \tilde{H}_I  := \frac{\pa }{\pa G^I} L(\vf , \bar \vf , G)~.
\eea
Since $F_{IJ}  (\vf)$ does not depend on $\bar \vf$ and $G$, we can 
choose these variables appearing in $\tilde{H}_I (\vf, \bar \vf, G)$ to have any given values, 
say $\bar \vf_0$ and $G_0$. Then the above relation turns into
\bea
F_{IJ} (\vf) = \pa_I {H}_J  (\vf)- \pa_J {H}_I (\vf)~, \qquad {H}_I (\vf) := \frac{\pa }{\pa G^I} L(\vf , \bar \vf_0 , G_0)~.
\eea
Now, we can perform the following transformation of the Lagrangian
\bea
L~ \longrightarrow  ~ L - G^I H_I (\vf) - G^I \bar H_I (\bar \vf )~.
\eea
This transformation does not change the action,  
since it is of the type (\ref{eq_Htransformation}).
Due to (\ref{two-form}),  the transformed Lagrangian obeys the equation
\bea
\frac{\pa^2 L}{\pa G^I  \pa  \vf^J} - \frac{\pa^2 L}{\pa G^J  \pa  \vf^I} =0~.
\label{tensor-Laplace3}
\eea
As a result, we can always choose the Lagrangian to obey (\ref{tensor-Laplace3})
\cite{deWRV}.

With the aid of the equations (\ref{tensor-Laplace1}) and (\ref{tensor-Laplace3}), 
it is not difficult to prove that the variation 
$\d_{\bar \e}S$, eq.  (\ref{epsilon-var}), vanishes identically. 
Making use of the identity (\ref{antichiral_reduction}),
we represent $\d_{\bar \e}S$ as an integral over the chiral subspace:
 \bea
\d_{\bar \e}S &=& -\frac{1}{4} \int {\rm d}^4 x \,{\rm d}^2 \q \,{\bar D}^2
\left\{ 
   \frac{\pa L}{\pa \vf^I} \bar \e \bar D G^I
 -
 \frac{\pa L}{\pa G^I} \bar \e \bar D \bar \vf^I 
 \right\}~.~~~
\eea
Evaluating the integrand and making use of  the equations 
(\ref{tensor-Laplace1}) and (\ref{tensor-Laplace3}),  we indeed observe that $\d_{\bar \e}S =0$.
The calculation is somewhat longer if one does not require eq. (\ref{tensor-Laplace3}) 
to hold, but instead one has to use only the weaker equation (\ref{two-form}). 
Such a calculation also gives  $\d_{\bar \e}S =0$.

\subsection{Projective superspace formulation}
\label{subsection2.2}
In 1984, the most general self-couplings of several $\cN=2$ tensor multiplets were constructed using 
$\cN=2$ superspace techniques \cite{KLR}.
Their manifestly $\cN=2$ supersymmetric action can be rewritten in terms of $\cN=1$ superfields, 
and the result obtained in \cite{KLR} is 
\bea
S= {\rm Re} \oint_\g 
 \frac{\rd\z }{2\pi\ri \z}
\int\rd^4 x\,{\rm d}^4\q \,
\cL\Big(\cG^I {(\z)} , \z \Big)~,
\label{TensorAction}
\eea
for an appropriately chosen closed contour  $\g$ in 
the complex projective space  ${\mathbb C}P^1$ parametrized by 
an inhomogeneous complex variable $\z$.
The dynamical variables in (\ref{TensorAction}) are:
\bea
\cG^I (\z) = \frac{1}{\z}\, \vf^I + G^I - \z \,{\bar \vf}^I~, \qquad {\bar D}_{\dot \a} \vf^I =0~, 
\qquad {\bar D}^2 G^I ={\bar G}^I-G^I=0~.
\eea
The Lagrangian in  (\ref{TensorAction}) is an arbitrary analytic function of its arguments.
Evaluating the contour integral, we find the $\N=1$ action \eqref{N=2_tensor_action}
with Lagrangian
 \bea
L (\vf^I, \bar \vf^I , G^I) = {\rm Re} \oint_\g 
 \frac{\rd\z }{2\pi\ri  \z} \,
\cL\Big(\cG^I {(\z)} , \z \Big)~.
\label{Whittaker}
\eea
Using this representation, it is easy to see that both the equations 
(\ref{tensor-Laplace1}) and (\ref{tensor-Laplace3}) hold \emph{automatically}.\footnote{In
the case of a single  $\cN=2$ tensor multiplet,  eq. (\ref{Whittaker}) coincides with
Whittaker's formula for  harmonic functions in ${\mathbb R}^3$ \cite{Whittaker}.}
In fact, one can also check that this formulation \emph{automatically} selects
out Lagrangians with vanishing $F_{IJ}$ \eqref{two-form}.

\subsection{Dual formulation}
To construct a dual formulation of the theory (\ref{N=2_tensor_action}), we follow \cite{LR}
(see also \cite{Siegel}) and associate with (\ref{N=2_tensor_action})  the first-order action
\bea
S_{\text{first-order}}=\int {\rm d}^4 x \,{\rm d}^4 \q \,\Big\{ L (\vf , \bar \vf , V)
-V^I(\j_I +\bar \j_I ) \Big\}~,
\label{N=2_tensor_action_dual}
\eea
where the real superfields $V^I$ are unconstrained, and the Lagrange multipliers $\j_I$ are chiral, 
$\bar D_\ad \j _I =0$.
This formulation is equivalent to the original theory, since varying $\psi_I$ and its conjugate $\bar \psi_I$
gives the equations $\bar D^2V^I = D^2 V^I =0$, and then 
(\ref{N=2_tensor_action_dual}) turns into (\ref{N=2_tensor_action}).
On the other hand, varying (\ref{N=2_tensor_action_dual}) with respect to $V^I$ gives
the equations
\bea
\frac{\pa }{\pa V^I}  L (\vf , \bar \vf , V)=\j_I +\bar \j_I 
\eea
which can be used to express $V^I$ in terms of the other variables, $V^I= V^I(\vf , \bar \vf , \j + \bar \j) $. 
As a result, we end up with the dual action 
\bea
S_{\text{dual}}=\int {\rm d}^4 x \,{\rm d}^4 \q \,K  (\vf^I , \bar \vf^I , \j_J  + \bar \j_J)~,
\label{tensor-dual}
\eea
where the K\"ahler potential is defined by 
\bea
K  (\vf , \bar \vf , \j + \bar \j) 
:= \Big\{  L (\vf , \bar \vf , V) - V^I(\j_I +\bar \j_I ) \Big\}\Big|_{V= V(\vf , \bar \vf , \j + \bar \j) }~.
\eea
The K\"ahler potential depends on $\j_I $ and $\bar \j_I$ only in the combination 
$\j_I +\bar \j_I$, and therefore the target space has at least $n$ Abelian isometries. 

Assuming that eq. (\ref{tensor-Laplace3}) holds,
the first-order action (\ref{N=2_tensor_action_dual}) 
can be shown to be invariant under the 
second supersymmetry transformations
\begin{subequations}
\bea 
\d \vf^I &=& \hf {\bar D}^2 \Big(\bar \e \bar \q V^I \Big) ~, \\
\d V^I &=& -\e D \vf^I -\bar \e \bar D \bar \vf^I ~, \\
\d \j_I &=& \hf {\bar D}^2 \Big(\bar \e \bar \q \frac{\pa L}{\pa \vf^I} \Big)~. 
\eea
\end{subequations}
For the dual model, eq. (\ref{tensor-dual}), this invariance turns into
\begin{subequations}\label{2.26}
\bea 
\d \vf^I &=& -\hf {\bar D}^2 \Big(\bar \e \bar \q  \frac{\pa K}{\pa \j_I}  \Big) ~,\\
\d \j_I &=&\phantom{-} \hf {\bar D}^2 \Big(\bar \e \bar \q \frac{\pa K}{\pa \vf^I} \Big)~. 
\eea
\end{subequations}

In the general case, when only the equation (\ref{two-form}) holds instead of 
(\ref{tensor-Laplace3}), the supersymmetry transformation (\ref{2.26}) takes the form 
\begin{subequations}\label{2.27}
\bea 
\d \vf^I &=& -\hf {\bar D}^2 \Big(\bar \e \bar \q  \frac{\pa K}{\pa \j_I}  \Big) ~,\\
\d \j_I &=&\phantom{-} \hf {\bar D}^2 \Big(\bar \e \bar \q \frac{\pa K}{\pa \vf^I} 
+\bar \e \bar \q  F_{IJ}  \, \frac{\pa K}{\pa \j_J}
\Big)~. 
\eea
\end{subequations}

Since the $\sigma$-model  (\ref{tensor-dual}) is $\cN=2$ supersymmetric 
and realized in terms of chiral superfields,  
the Lagrangian in (\ref{tensor-dual}) is the K\"ahler potential of a hyperk\"ahler manifold.

\section{$\cN=2$ superconformal  tensor supermultiplets}
\label{section3}

It is of interest to extend the above analysis to the case 
of $\cN=1$ superconformal tensor multiplets. 
The superconformal couplings of $\cN=2$ tensor multiplets were systematically 
discussed in the component approach in \cite{deWRV}.
Within the $\cN=2$ projective superspace approach \cite{KLR,LR-projective}, 
they  were studied in \cite{KLR,K-hyper,KLvU2}.
We are not aware of a detailed discussion 
of the general superconformal $\sigma$-models of $\cN=2$ tensor multiplets
in $\cN=1$ superspace.

It was shown in \cite{K-hyper} that the general $\cN=2$ superconformal 
transformation decomposes, upon reduction to $\cN=1$ superspace, 
into three types of $\cN=1$  transformations:
\begin{enumerate}
\bfseries \item \mdseries
An arbitrary $\cN=1$ superconformal transformation generated by 
\bea
{ \x} = {\overline {\x}} = { \x}^\ra (z) \,\pa_\ra + { \x}^\a (z)\,D_\a
+ {\bar { \x}}_{\dot \a} (z)\, {\bar D}^{\dot \a}
\label{n=1scf1}
\eea  
such that 
\be
[{ \x} \;,\; D_\a ] 
=  -{ \l}_\a{}^\b  D_\b +
\Big(\hf { \s} -  \bar{ { \s}}  \Big) D_\a\quad \Longrightarrow \quad {\bar D}_\ad \s =0~, \quad 
{\bar D}_\ad \l_\a{}^\b =0~.
\label{n=1scf2}
\ee
The lowest component of $\textrm{Re }\sigma$ corresponds to a dilatation,
$\textrm{Im } \sigma$ to a ${\rm U}(1)_R$ rotation, and $\lambda_{\alpha \beta} = \lambda_{\beta\alpha}$
to a Lorentz transformation. (Further details may be found in \cite{BK}.)
This transformation  acts on the $\cN=1$
components of the $\cN=2$ tensor multiplet as
\begin{subequations}\label{N=1superconformal} 
\bea
\d \vf^I &=& -{ \x} \vf^I   -2  { \s} \vf^I~, \\
\d G^I &=& -{ \x} G^I - ({ \s} +\bar{ \s} )G^I~.
\eea
\end{subequations}

\bfseries \item \mdseries
An extended  superconformal transformation generated by  a spinor parameter $\r^\a$
constrained as
\bea
{\bar D}_{\ad} \r^\b =0~, \qquad D^{(\a}\r^{\b )}=0~,
\label{ro}
\eea
and hence
\be
\pa^{{ \ad} (\a} \r^{\b )} = D^2 \r^\b =0~.
\ee
The general solution to (\ref{ro}) is 
 \bea
 \r^\a(x_{(+)}, \q) = \e^\a
+ \l\, \q^\a - {\rm i} \,{\bar \eta}_{\ad}\, x^{{ \ad}\a}_{(+)}~, 
\qquad x^\ra_{(+)} = x^\ra +{\rm i} \q \s^\ra \bar \q~.
\label{ro-exp}
\eea
Here the {\it constant} parameters $\e^\a$, $\l$ and ${\bar \eta}_{\dot \a}$
correspond to  (i) a second Q-supersymmetry 
transformation $(\e^\a$); (ii) an off--diagonal SU(2)-transformation\footnote{In 
the standard $\cN=2$ superspace parametrized by variables
$z^A =(x^\ra, \q^\a_i, {\bar \q}_{\dot \a}^i)$,  this transformation rotates  the Grassmann variable 
$\q^\a_{\1}$ into $\q^\a_{\2}$ and vice versa. }
($\l$); 
and (iii) a second S-supersymmetry transformation (${\bar \eta}_{\dt \a}$). 
The extended superconformal transformation 
acts on  the $\cN=1$ components, $\vf^I$ and $G^I$, of the $\cN=2$ tensor multiplet as
\begin{subequations}\label{extend_superconformal}
\bea
\d \vf^I &=& {\bar \r}_{ \ad} {\bar D}^{ \ad} G^I
+\hf  \big( {\bar D}_{ \ad} {\bar \r}^{ \ad} \big) G^I ~,  \\
\d G^I &=& - D^\a (\r_\a \vf^I ) -{\bar D}_\ad (\bar \r^\ad \bar \vf^I ) ~.
\eea
\end{subequations}

\bfseries \item \mdseries
A shadow chiral rotation generated by a constant parameter $\alpha$.
In $\cN=2$ superspace, this  is a phase transformation
of $\q^\a_{\2}$ only, with $\q^\a_{\1}$ kept unchanged.
Its action on the $\cN=2$ tensor multiplet is
\bea
\d \vf^I &=& {\rm i}  \a\, \vf^I~, \qquad \d G^I =0~.
\label{shadow}
\eea
\end{enumerate}

We have seen that under the conditions (\ref{tensor-Laplace1}) and
(\ref{tensor-Laplace3}),  the action (\ref{N=2_tensor_action}) is $\cN=2$ supersymmetric.
Now we would like to determine conditions for $\cN=2$ superconformal invariance.
The action proves to be invariant under the $\cN=1$ superconformal transformations
(\ref{N=1superconformal}) and the shadow chiral rotations (\ref{shadow}) if 
the Lagrangian obeys the following equations:
\bea
\Big( G^I \frac{\pa }{\pa G^I} + 2\vf^I \frac{\pa }{\pa \vf^I}\Big) L &=& L  - r_I \,G^I~, \qquad 
\bar r_I =r_I ={\rm const}~,
\label{3.9} \\
\vf^I \frac{\pa L}{\pa \vf^I}  &=& \bar \vf^I \frac{\pa L}{\pa \bar \vf^I} ~,
\label{shadow2}
\eea
for some real parameters $r_I$. These are not actually the most general conditions
because one can always modify the Lagrangian by certain trivial transformations
\eqref{eq_TensorTrivialMinkowski} which distort these conditions. However, it is
always possible to make the above choice. In fact, one can even set $r_I$ to zero
by a certain $H$-transformation \eqref{eq_Htransformation}.

As a simple example of this, take the so-called  improved $\cN=2$ tensor multiplet model 
\cite{LR,deWPV}.\footnote{The  improved $\cN=1$ tensor multiplet model was constructed in \cite{deWR}.}
It is described in $\cN=1$ superspace by the Lagrangian \cite{LR}
\bea
 L_{\text{impr}}(G, \vf, \bar \vf) =  \sqrt{ G^2 +4\vf \bar \vf }
    - G \, \ln \Big( G+   \sqrt{ G^2 +4\vf \bar \vf }
    \Big)   ~.
\label{impr-tensor}
\eea
This is the unique $\cN=2$ superconformal theory which can be constructed using a
single $\cN=2$ tensor multiplet. Applying the first-order operator, which appears on the
left of (\ref{3.9}), to $ L_{\text{impr}}$ gives
\bea
\Big( G \frac{\pa }{\pa G} + 2\vf \frac{\pa }{\pa \vf}\Big) L_{\text{impr}}  = 
L_{\text{impr}} -G~.
\label{impr-tensor2}
\eea
However, one may equally well construct the same action using the Lagrangian
\begin{align}
L'_{\text{impr}}(G, \vf, \bar \vf) =  \sqrt{ G^2 +4\vf \bar \vf }
    - G \, \ln \left(\frac{G +   \sqrt{ G^2 +4\vf \bar \vf }}{\sqrt{4 \vf \bar\vf}}\right)
\end{align}
which differs only by a trivial $H$-transformation and indeed obeys
\eqref{3.9} with $r=0$.

Henceforth we will assume that we have modified all superconformal Lagrangians so that
they obey the (weighted) homogeneity condition
\bea
\Big( G^I \frac{\pa }{\pa G^I} + 2\vf^I \frac{\pa }{\pa \vf^I}\Big) L = L  ~.
\label{3.13}
\eea
Taking into account (\ref{shadow2}), this is equivalent to 
\bea
\Big( G^I \frac{\pa }{\pa G^I} + \vf^I \frac{\pa }{\pa \vf^I} + \bar \vf^I \frac{\pa }{\pa \bar \vf^I}
\Big) L = L  ~.
\label{3.14}
\eea
Thus $ L (\vf^I, \bar \vf^I , G^I)$ is a homogeneous function of first degree. 
If we impose also the equation \eqref{tensor-Laplace3}, then we recover the
same conditions imposed in \cite{deWRV}.

It turns out that the above conditions on the Lagrangian guarantee invariance under 
the extended superconformal transformation (\ref{extend_superconformal}). 
To prove this claim, we first point out that it is sufficient to evaluate the corresponding variation of the action 
for $\r_\a=0$ and $\bar \r_\ad \neq 0$, since the parameters $\r_\a$ and $\bar \r_\ad$
are algebraically independent.
Varying the action gives
\bea
\d_{\bar \r} S = \int \frac{\pa L}{\pa \vf^I} \left\{ {\bar \r}_{ \ad} {\bar D}^{ \ad} G^I
+\hf  \big( {\bar D}_{ \ad} {\bar \r}^{ \ad} \big) G^I \right\} 
-  \int \frac{\pa L}{\pa G^I}{\bar D}_\ad (\bar \r^\ad \bar \vf^I ) \equiv I_1 + I_2~,
\label{3.15}
\eea
where we have denoted $\int :=  \int {\rm d}^4 x \,{\rm d}^4 \q$.
The first term may be transformed to take the form:
\bea
I_1&=& -\hf \int \left\{ \frac{\pa^2 L}{\pa \vf^I \pa \bar \vf^J } ({\bar D}_\ad {\bar \vf}^J ) {\bar \r}^\ad G^I
+\frac{\pa^2 L}{\pa \vf^I \pa G^J } ({\bar D}_\ad G^J ) {\bar \r}^\ad G^I
-\frac{\pa L}{\pa \vf^I}  {\bar \r}_{ \ad} {\bar D}^{ \ad} G^I\right\}~.~~~~~
\label{3.16}
\eea
The second term in (\ref{3.15}) may be transformed as
\bea
I_2 = \int  \frac{\pa^2 L}{\pa G^I \pa \bar \vf^J } ({\bar D}_\ad {\bar \vf}^J ) {\bar \r}^\ad \bar \vf^I
+ \int  \frac{\pa^2 L}{\pa G^I \pa G^J } ({\bar D}_\ad G^J ) {\bar \r}^\ad \bar \vf^I~.
\eea
Now, let us make use of the complex conjugate of (\ref{3.13}) to represent
\bea
\frac{\pa L}{\pa \vf^I} = G^J \frac{\pa^2 L}{\pa G^J \pa \vf^I}
 +2 \bar \vf^J \frac{\pa L}{\pa \bar \vf^J \pa \vf^I}~.
 \eea
 Applying this representation to the last term in (\ref{3.16}) gives
 \bea
 I_1+I_2 = -\hf \int  \frac{\pa^2 L}{\pa \vf^I \pa \bar \vf^J } ({\bar D}_\ad {\bar \vf}^J ) {\bar \r}^\ad G^I
 + \int  \frac{\pa^2 L}{\pa G^I \pa \bar \vf^J } ({\bar D}_\ad \bar \vf^J ) {\bar \r}^\ad \bar \vf^I~.
\eea
Here we have used the conditions (\ref{tensor-Laplace1}) and (\ref{tensor-Laplace3}).\footnote{The
$\cN=2$ supersymmetry conditions  (\ref{tensor-Laplace1}) and (\ref{tensor-Laplace3})
are compatible with the superconformal ones, eqs. (\ref{shadow2}) and (\ref{3.14}).}
To show that $I_1+I_2$ is equal to zero, it only remains to make use of the complex conjugate of (\ref{3.13})
to obtain the identity
\bea
 G^J \frac{\pa^2 L}{\pa G^J \pa G^I}
 +2 \bar \vf^J \frac{\pa L}{\pa \bar \vf^J \pa G^I}=0~.
 \eea

The above results have a nice re-formulation in terms of the tensor-multiplet Lagrangian 
(\ref{Whittaker}) which is derived using the $\cN=2$ projective superspace techniques.
The theory is $\cN=2$ superconformal provided $\cL$ has no explicit dependence on $\z$, 
\bea
L (\vf^I, \bar \vf^I , G^I) = {\rm Re} \oint_\g 
 \frac{\rd\z }{2\pi\ri \z } \,
\cL\Big(\cG^I {(\z)}  \Big)~, 
\label{Whittaker2}
\eea
and is homogeneous of degree one, 
\bea
\cG^I \frac{\pa }{\pa \cG^I}  \cL(\cG) = \cL(\cG)~.
\eea
Given this representation, the equations (\ref{3.13}) and (\ref{3.14}) are satisfied identically.
The $\cN=2$ superspace proof of superconformal invariance is given in \cite{K-hyper}.

\section{Self-interacting $\N=2$ tensor supermultiplets: AdS supersymmetry}
\label{section4}

We are now prepared to study self-interactions of several $\cN=2$ tensor multiplets 
in AdS supersymmetry, the symmetry group being $\rm OSp(2|4)$.
In a manifestly $\cN=2$ supersymmetric setting, this problem has been solved in \cite{KT-M-ads}
(the solution will be  discussed in the next section). Here we would like to address the problem
using a formulation in $\cN=1$ AdS superspace 
(all essential information about this superspace is collected in Appendix \ref{Appendix_A}). 
In such an approach, an $\cN=2$ tensor multiplet 
is described by a covariantly chiral superfield $\vf^I$, its conjugate $\bar \vf^I$, and
a real covariantly linear superfield $G^I$.  The constraints are
\bea
{\bar \cD}_\ad \vf^I =0~, \qquad ({\bar \cD}^2 -4\m)G^I = 0~, \qquad \bar G^I= G^I ~, \qquad
I=1,\dots, n~.
\eea
The dynamics is described by an action of the form 
\begin{align}
S = \int \rd^4x\, \rd^4\theta\, E\, 
L(\vf^I, \bar\vf^I, G^I)
\label{4.2}
\end{align}
which is manifestly $\cN=1$  supersymmetric, i.e. invariant under 
arbitrary $\rm OSp(1|4)$ transformations
\bea
\d_\x \vf^I =- \x \vf^I~, \qquad \d_\x G^I =- \x G^I~, \qquad 
\x := \x^{\rm A}\cD_{\rm A} =
\x^{\rm a}\cD_{\rm a} + \x^\a \cD_\a + \bar \x_\ad \bar \cD^\ad 
\label{OSp(1|4)tensorm}
\eea
with the AdS Killing vector field $\x^{\rm A}$ defined by eqs. (\ref{N=1Kv1}) and (\ref{N=1Kv2}).

It should be pointed out that the action (\ref{4.2})
does not change under arbitrary $H$-transformations
of the Lagrangian
\begin{align}
L \to  L + G^I H_I(\vf) + G^I \bar H_I(\bar\vf)~,
\label{H-transfo}
\end{align}
with $H_I(\vf)$ holomorphic functions.

\subsection{The second supersymmetry transformation}
To realize a second supersymmetry transformation, we need a real 
scalar  $\veps$ constrained to obey the conditions \cite{GKS}
\begin{align}
\bar \ve =\ve~, \qquad 
(\bar \cD^2 - 4 \mu) \veps = \bar\cD_\dalpha \cD_\alpha \veps = 0\quad \Longrightarrow \quad
\cD_{\a\ad} \ve = 0~.
\label{epsilon-constraints}
\end{align}
The parameter $\ve$ naturally originates within the $\cN=2$ AdS superspace approach \cite{KT-M-ads}, 
see subsection \ref{appendix_B3} for a review. Along with $\ve$, we will often use the chiral spinor 
\bea
\ve_\a :=\cD_\a \ve~, \qquad {\bar \cD}_\ad \ve_\a=0~.
\eea
The $\q$-dependent parameter $\ve$, due to the constraints (\ref{epsilon-constraints}), 
contains two components:
(i)  a  bosonic parameter $\x$ which is defined by $ \ve |_{\q=0} = \x  |\m|^{-1} $ 
and describes  the O(2) rotations;  
and (ii) a fermionic parameter $\e_\a := \cD_\a \ve|_{\q=0} $ along with its conjugate, 
which generate the second supersymmetry. Schematically, the $\q$-expansion of $\ve$ looks like
\bea
\ve \sim \frac{\x}{|\m|} + \e^\a \q_\a + \bar \e_\ad \bar \q^\ad 
-\x \Big( \frac{\bar \m}{|\m |} \q^2 +  \frac{ \m}{|\m |} \bar \q^2 \Big) ~.
\eea

In accordance with the analysis given in  \cite{KT-M-ads},
for the $\cN=2$ tensor multiplet in AdS the second supersymmetry
transformation is 
\begin{align}
\delta \vf^I = \frac{1}{2} (\bar\cD^2 - 4 \mu) (\ve G^I)~, \qquad
\delta G^I = -\cD^\alpha (\ve_\alpha \vf^I) -\bar \cD_\ad (\bar \ve^\ad \bar \vf^I) ~.
\label{secondSUSY-tensorm}
\end{align}
The transformation laws (\ref{OSp(1|4)tensorm}) and  (\ref{secondSUSY-tensorm})
constitute an off-shell $\rm OSp(2|4)$ supermultiplet.
Our goal is to determine those conditions which  $L$ must obey for the action (\ref{4.2}) to be 
$\cN=2$ supersymmetric.
Varying the action  gives
\begin{align}
\delta_\ve S = 
     \int \rd^4x\, \rd^4\theta\, E\, 
     \Big\{\frac{1}{2} \ve G^I (\bar\cD^2 - 4 \mu) \frac{\partial L}{\partial \vf^I}
     + \ve^\alpha \vf^I \cD_\alpha \frac{\partial L}{\partial G^I} + \HC \Big\}~.
\end{align}
This can be rearranged to
\begin{align}
\delta_\ve S = 
     \int \rd^4x\, \rd^4\theta\, E\, 
     \Big\{\ve^\alpha A_\alpha + 2 \bar\mu \ve G^I \frac{\partial L}{\partial \bar\vf^I}
     - 4 \bar\mu \ve \vf^I \frac{\partial L}{\partial G^I} + \HC \Big\}~,
     \label{4.9}
\end{align}
where
\begin{align}
A_\alpha := \frac{\partial L}{\partial \bar\vf^I} \cD_\alpha G^I
     - \frac{\partial L}{\partial G^I} \cD_\alpha \vf^I~.
\label{4.10}
\end{align}
The combination $\delta_\ve S$ must vanish.

\subsection{Derivation of conditions}
An easy way to derive the conditions that $L$ must obey is to
consider the variation of $\delta_\ve S$ with respect to $\vf^I$, $\delta_\vf \,\delta_\ve S $.
Using the properties of $\ve$, 
in conjunction with the chiral reduction rule 
\bea
 \int \rd^4x\, \rd^4\theta\, E\, U = -\frac{1}{4} \int \rd^4x\, \rd^2\theta\,  \cE\, 
\big(\bar \cD^2 - 4\m\big) U~, 
\eea
it is clear that the variation must
have the form
\begin{align}
\delta_\vf \,\delta_\ve S = 
 \int \rd^4x\, \rd^2\theta\,  \cE\, 
\delta \vf^I\Big\{
     -\frac{1}{4} \ve^\alpha (\bar\cD^2 - 4 \mu) \Gamma_{I\alpha}
     + \bar\ve_\dalpha \bar \Psi^\dalpha_I
     + \ve \Omega_I \Big\}~,
     \label{4.11}
\end{align}
for some fields $\G_{I\a}$, $\bar \J^{\ad}_I$ and $\O_I$.
One can show that
\begin{align}
\Gamma_{I\alpha} =
     \left(\frac{\partial^2 L}{\partial \vf^I \partial \bar\vf^J}
          + \frac{\partial^2 L}{\partial G^I \partial G^J} \right) \cD_\alpha G^J
     + \left(\frac{\partial^2 L}{\partial G^I \partial \vf^J}
          - \frac{\partial^2 L}{\partial \vf^I \partial G^J} \right) \cD_\alpha \vf^J.
\end{align}
This field must be such that
$(\bar\cD^2 - 4 \mu)\G_{I\a}=0$. For this to occur, the Lagrangian must obey the generalized Laplace equations
\begin{align}\label{eq_Laplace}
\frac{\partial^2 L}{\partial \vf^I \partial \bar\vf^J}
     + \frac{\partial^2 L}{\partial G^I \partial G^J} = 0~,
\end{align}
familiar from the Minkowski case, see section \ref{section2}, as well as the condition
\begin{align}
\bar\cD_\dalpha \left(\frac{\partial^2 L}{ \pa \vf^I \partial G^J }
     - \frac{\partial^2 L}{\partial \vf^J \partial G^I}\right) = 0~.
\end{align}
Because $L$ depends on $G$, $\vf$, and $\bar\vf$ only algebraically,
this implies that
\begin{align}
\frac{\partial^2 L}{ \pa \vf^I \partial G^J }
     - \frac{\partial^2 \cL}{\partial \vf^J \partial G^I} = F_{IJ}(\vf)~,
     \label{4.16}
\end{align}
with $F_{IJ}(\vf)$  a closed holomorphic two-form.
This is not actually an independent result; it is implied by \eqref{eq_Laplace}, 
which can be proved as in  section  \ref{section2}. Moreover, 
in complete analogy with the analysis  in section \ref{section2},  the two-form $F_{IJ}(\vf)$  
can be shown to be exact, and therefore it can be set to zero,  $F_{IJ} = 0$,  
by applying an $H$-transformation  (\ref{H-transfo});
but this is not necessary and we will not assume it in what follows.
So far the story is absolutely the same as in the rigid supersymmetric case studied in section 
\ref{section2}. However, now comes a difference.

Assuming that $L$ obeys \eqref{eq_Laplace}, one can then show that
\begin{align}
\bar\Psi_I^\dalpha = \bar\cD^\dalpha G^J \frac{\partial R_I}{\partial \vf_J}
     - \bar\cD^\dalpha \bar\vf^J \frac{\partial R_I}{\partial G_J}~,
\end{align}
where
\begin{align}
\hf R_I := {\rm Re} \left( 
\mu \frac{\partial L}{\partial \vf^I}
     - \mu \frac{\partial^2 L}{\partial \vf^I \partial G^J} G^J
     - 2 \mu \frac{\partial^2 L}{\partial \vf^I \partial \bar\varphi^J} \bar\vf^J \right)
\label{RRi}
\end{align}
is a real quantity. Because $\bar\Psi^\dalpha_I$ must vanish, this implies that
$R_I$ is independent of $\vf$ and $G$. Since  $R_I$ is real, it must also be independent
of $\bar \vf$. This implies that $R_I$ is a constant.
One useful consistency check is to note that $R_I$ is invariant under the
$H$-transformation (\ref{H-transfo}).

The remaining condition, the vanishing of $\Omega_I$ in (\ref{4.11}), can be
shown to give no new results. However, our task is not yet complete.
We need one additional constraint: the constant $R_I$ must actually be zero.
To see why, consider the addition to the Lagrangian of a term
$\bar \mu\, c_I \vf^I +  \mu\, c_I \bar\vf^I$. This obeys all the constraints
we have imposed so far, and shifts $R_I$ by a constant $2 c_I \mu \bar\mu$.
However, this is not actually $\N=2$ supersymmetric, and so arbitrary values of $R_I$
must not be allowed. Such terms have not yet been ruled out by our analysis
since their $\N=2$ variation depends only on $G^I$. We must also analyze the
condition $\delta_G \delta_\ve S = 0$.

Varying $\delta_\ve S$ with respect to $G^I$ leads (after a good deal of algebra) to
\begin{align}
\delta_G \delta_\ve S = -2 \int \rd^4x\, \rd^4\theta\, E\, \delta G^I R_I \ve
\end{align}
after imposing the constraints we have already found.
Because $R_I$ is a constant, the integral involves just $\delta G^I \ve$.
Since $(\bar\cD^2 - 4\mu) \cD_\alpha \ve \neq 0$, this
integral does not vanish unless $R_I$ also vanishes. Our conclusion is that 
\bea
R_I := \frac{1}{2} {\rm Re} \left( 
\mu \frac{\partial L}{\partial \vf^I}
     - \mu \frac{\partial^2 L}{\partial \vf^I \partial G^J} G^J
     - 2 \mu \frac{\partial^2 L}{\partial \vf^I \partial \bar\varphi^J} \bar\vf^J \right)=0~.
\label{R-cond}
\eea
As compared with the situation in the rigid supersymmetric case, 
this is a new condition on the Lagrangian. 

As an example, consider a superconformal tensor multiplet model. In accordance with 
the analysis in section \ref{section3},
the Lagrangian $L(\vf , \bar \vf , G)$ can be taken to obey the equations
\begin{subequations}
\bea
& \Big( G^I \dfrac{\pa }{\pa G^I} + 2{\bar \vf}^I \dfrac{\pa }{\pa {\bar \vf}^I}\Big) L = L~, \\
&\vf^I \dfrac{\pa L}{\pa \vf^I}  = \bar \vf^I \dfrac{\pa L}{\pa \bar \vf^I} ~.
\eea
\end{subequations}
It is obvious that the AdS condition (\ref{R-cond}) is identically satisfied.

\subsection{Proof of invariance}
\label{Tensor_ProofInvariance}
With the conditions derived in the previous subsection, it still remains to be shown that
the variation of $S$ under the second supersymmetry transformation, eq. (\ref{4.9}), vanishes. 
Using the definition (\ref{4.10}) of the quantity $A_\a$, which appears in (\ref{4.9}), 
it can be checked that $A_\alpha$ obeys a particularly useful condition
\begin{align}
\cD_{\beta} A_{\alpha} + \cD_\alpha A_\beta &= F_{IJ} \,\cD_\beta \varphi^J \cD_\alpha \varphi^I
     + \bar F_{IJ} \,\cD_\beta G^J \cD_\alpha G^I~,
\end{align}
with the holomorphic two-form $F_{IJ}(\vf)$ defined by (\ref{4.16}).
As discussed earlier, we may always choose $F_{IJ}$ to vanish,
but we will take it here to be non-vanishing in the interest of full generality.
Since the two-form $F_{IJ}(\vf)$ is exact, we may introduce a holomorphic
one-form $\rho_I$ such that
\begin{align}
\partial_I \rho_J - \partial_J \rho_I = F_{IJ}~.
\end{align}
Then we may introduce the combination
\begin{align}
B_\alpha := A_\alpha - \frac{1}{2} \bar F_{IJ} G^J \cD_\alpha G^I
     + \rho_I \cD_\alpha \varphi^I
\end{align}
which obeys
\begin{align}\label{eq_B3}
\cD_\beta B_\alpha + \cD_\alpha B_\beta = 0~.
\end{align}
The variation of the action can then be written
\begin{align}
\delta_\ve S = \fint
     \Big\{ &\ve^\alpha B_\alpha
     + \frac{1}{2} \veps^\alpha (\cD_\alpha G^I) \bar F_{IJ} G^J 
     - \veps^\alpha (\cD_\alpha \vf^I )\rho_I \non \\
&     + 2 \bar\mu \ve G^I \frac{\partial L}{\partial \bar\vf^I}
     - 4 \bar\mu \ve \vf^I \frac{\partial L}{\partial G^I} + \HC \Big\}~.
\end{align}
The second and third terms which we have added can be shown to vanish
(the second vanishes when we write $\veps^\alpha = \cD^\alpha \veps$ and
integrate this spinor derivative by parts, and the third vanishes under
a chiral projection). We may simplify the first term by noting that the
equation \eqref{eq_B3} is solved by $B_\alpha = \cD_\alpha B$
for some function $B(\vf, \bar\vf, G)$. Inserting this relation into the action
and integrating by parts yields
\begin{align}
\delta_\ve S = -2 \fint
     \veps \,\Omega~, \qquad
\Omega := 2\bar\mu B
     - \bar\mu G^I \frac{\partial L}{\partial \bar\vf^I}
     + 2 \bar\mu \vf^I \frac{\partial L}{\partial G^I} + \HC~.
\end{align}

By construction, the dependence of $B$ on $G^I$ and $\vf^I$ is given by
\begin{align}
\frac{\partial B}{\partial G^I} &= -\frac{1}{2} \bar F_{IJ} G^J + \frac{\partial L}{\partial \bar \vf^I} ~,\eol
\frac{\partial B}{\partial \vf^I} &= \rho_I - \frac{\partial L}{\partial G^I}~.
\end{align}
Its dependence on $\bar\vf^I$ is undetermined. Using the first of these equations,
we may immediately observe that
\begin{align}
\frac{\partial \Omega}{\partial G^I} = R_I = 0~,
\end{align}
where $R_I$ is defined as in (\ref{RRi}).
Therefore the function $\Omega$ can depend only on the variables $\vf$
and $\bar\vf$. Because $\veps$ is a linear superfield, $\Omega$ may freely
be modified by the transformations $\Omega \rightarrow \Omega + \Lambda + \bar\Lambda$,
where $\Lambda = \L(\vf)$ is holomorphic, without affecting the integral.
We may interpret this as a ``K\"ahler transformation''
and so the integral $\delta_\ve S$ can depend only on the ``K\"ahler metric'' constructed
from $\Omega$. However, it is easy to check that
\begin{align}
\frac{\partial^2 \Omega}{\partial \vf^I \partial \bar\vf^J} = -\frac{\partial R_I}{\partial G^J}
     = 0
\end{align}
and so the ``K\"ahler metric'' vanishes. Thus $\delta_\ve S$ must also vanish.

\subsection{Dual formulation}
\label{subsection4.4}
To construct a dual formulation of the theory, we introduce the first-order form of the action
\begin{align}
S = \fint \Big\{ L(\vf,\bar\vf, V) - V^I (\psi_I + \bar \psi_I)\Big\}~,
\end{align}
where $V^I$ is real unconstrained and $\psi_I$ is a covariantly chiral Lagrange multiplier.
We note in passing that the original $H$-invariance, eq. (\ref{H-transfo}), 
manifests here as
\begin{align}
L  \to L + V^I H_I(\vf) + V^I \bar H_I(\bar\vf)~, \qquad
\psi_I \to  \psi_I + H_I~.
\label{H-transfo2}
\end{align}
The variables $V^I$ can be eliminated using their equations of motion
\begin{align}
\frac{\partial L}{\partial V^I} = \psi_I + \bar\psi_I
\end{align}
to express them in terms of the other fields. The resulting Legendre transform of $L$
is given by
\begin{align}
\cK(\vf^I, \bar \vf^I, \j_J +\bar \j_J )= \Big[L(\vf, \bar \vf, V) - V^I (\psi_I + \bar\psi_I)
	\Big]_{V = V(\vf,\bar\vf, \psi+\bar\psi)}
\end{align}
with the properties
\begin{align}
\frac{\partial \cK}{\partial \vf^I} = \frac{\partial L}{\partial \vf^I}~, \qquad
\frac{\partial \cK}{\partial \psi_I} = -V^I~.
\end{align}

The dual theory has an action given purely in terms of chiral and antichiral superfields,
\begin{align}\label{4.40}
S = \fint \cK(\vf^I, \bar \vf^I, \j_J +\bar \j_J )~,
\end{align}
which is invariant under the $\cN=2$ AdS isometry group,  $\rm OSp(2|4)$.
The second supersymmetry transformation acts on the fields as
\begin{align}
\delta \vf^I &= - \frac{1}{2} (\bar\cD^2 - 4 \mu)\left(\ve \frac{\partial \cK}{\partial \psi_I}\right) ~,\eol
\delta \psi_I &=\phantom{-} \frac{1}{2} (\bar\cD^2 - 4 \mu)\left(\ve \frac{\partial \cK}{\partial \varphi^I}
     + \ve F_{IJ} \frac{\partial \cK}{\partial \psi_J} \right)~.
\end{align}
Note the appearance of the closed two-form $F_{IJ}(\vf)$ in the transformation rule of $\psi_I$.

The dual theory (\ref{4.40}) is a special case of the general $\cN=2$ supersymmetric
$\s$-models in AdS which will be studied in section \ref{section7}, so we will
delay a detailed discussion of the geometry of this model until then.
For now, we only  briefly mention the form that the AdS condition \eqref{R-cond}
takes in the dual formulation. We denote $\phi^a := (\vf^I, \psi_I)$
as the complex coordinate of the K\"ahler manifold associated with the
K\"ahler potential $\cK$. The index $a$ runs from $1$ to $2n$.
The AdS condition can be written
\begin{align}
\mu g^{\bar a b} \partial_b \cK + \bar\mu g^{a \bar b} \partial_{\bar b} \cK =
	2 \left(\mu \bar\varphi^I + \bar\mu \varphi^I\right)~,\qquad a = \bar a = I = 1, \cdots, n~.
\end{align}
Here $g^{a \bar b}$ is the inverse of the K\"ahler metric
$g_{a \bar b} = \partial_a \partial_{\bar b} \cK$. The equation in this form
makes sense only in the particular complex coordinates singled out by the
duality transformation.

\subsection{Deriving the AdS condition from a superconformal tensor model}
One last feature of the $\N=1$ formulation of the AdS tensor model case
that we would like to discuss is how to derive it from a superconformal
model. Beginning with a superconformal Lagrangian $L$ obeying the constraints
\begin{subequations}\label{eq_TensorSuperConformal}
\begin{gather}
\Big( G^I \frac{\pa }{\pa G^I} + \vf^I \frac{\pa }{\pa \vf^I} + \bar \vf^I \frac{\pa }{\pa \bar \vf^I}
\Big) L = L \\
\vf^I \frac{\pa L }{\pa \vf^I} = \bar \vf^I \frac{\pa L}{\pa \bar \vf^I} \\
\frac{\partial^2 L}{\partial G^I \partial G^J} + \frac{\partial^2 L}{\partial \vf^I \partial \bar\vf^J} = 0
\end{gather}
\end{subequations}
with the index $I$ running from $0$ to $n$, we single out for
special treatment the $I=0$ tensor multiplet and freeze it at the values
\begin{align}\label{eq_FrozenTensorAdS}
\vf^0 = \ri \mu~, \qquad \bar\vf^0 = -\ri \bar\mu~, \qquad G^0 = 0~.
\end{align}
Starting from a frozen tensor multiplet with $\vf^0 =\text{const}$ and $G^0 =\text{const}$, 
this can always be arranged by applying scale and SU(2) transformations and a shadow chiral rotation.
Our goal is to discover the set of isometries which keep the frozen tensor multiplet
in this form and to determine what conditions the Lagrangian $L$ obeys
in terms of the unfrozen components. From now on, the index $I = 1, \cdots, n$
labels only the dynamical multiplets.

We are interested in those $\N=2$ superconformal transformations which keep the frozen multiplet 
invariant. As discussed earlier (see also Appendix \ref{appendixB}), 
any $\cN=2$ superconformal transformation decomposes 
into three $\cN=1$ transformations, and here we have to analyze only the $\N=1$ superconformal
and the extended superconformal transformations. 
Applying the $\N=1$ superconformal transformation gives
\begin{align}
\delta \vf^0 = -2 \ri \mu \sigma~, \qquad \delta G^0 = 0~.
\end{align}
For consistency with the frozen condition \eqref{eq_FrozenTensorAdS}, we
must restrict $\sigma=0$, and this reduces the   $\N=1$ superconformal Killing vector
to an AdS Killing vector.
The other transformation is the extended superconformal one generated
by a parameter $\rho^\alpha = \cD^\alpha \rho$ for which we find
\begin{align}
\delta \vf^0 = 0~,\qquad \delta G^0 = -\ri \mu \cD^\alpha \rho_\alpha
	+ \ri \bar\mu \bar \cD_\dalpha \bar \rho^\dalpha ~.
\end{align}
Using the constraints \eqref{rho-constraints} and the requirement
$\delta G^0 = 0$, we find $4 \ri \bar \mu \mu (\bar\rho - \rho) = 0$
which implies that $\rho$ must be real. This agrees with the
analysis of Appendix \ref{appendix_B3}, where the second AdS supersymmetry and O(2) rotation
are  generated by a real linear parameter $\veps$ obeying
\begin{align*}
(\bar \cD^2 -4 \m) \ve =(\cD^2 -4\bar \m) \ve =0~, \qquad 
 \cD_\a \bar \cD_\ad \ve =\bar \cD_\ad \cD_\a \ve =0~.
\end{align*}
We conclude that the surviving transformations generate the supergroup $\rm OSp(2|4)$. 

We rewrite the superconformal conditions (singling out the zero components
for special treatment) as
\begin{subequations}
\begin{gather*}
\Big( G^I \frac{\pa }{\pa G^I} + \vf^I \frac{\pa }{\pa \vf^I} + \bar \vf^I \frac{\pa }{\pa \bar \vf^I}
\Big) L = L - \ri \mu \frac{\partial L}{\partial \vf^0} + \ri \bar\mu   \frac{\partial L}{\partial \bar \vf^0}\\
\vf^I \frac{\pa L }{\pa \vf^I} - \bar \vf^I \frac{\pa L}{\pa \bar \vf^I} =
	- \ri \mu \frac{\pa L }{\pa \vf^0} - \ri \bar\mu \frac{\pa L}{\pa \bar \vf^0} \\
\frac{\partial^2 L}{\partial G^I \partial G^J} + \frac{\partial^2 L}{\partial \vf^I \partial \bar\vf^J} = 0
\end{gather*}
\end{subequations}
and make the following observation. Differentiating the first two
equations with respect to $\bar\vf^I$ and rearranging them, we may
derive the formula
\begin{align*}
\frac{\partial^2 L}{\partial \vf^0 \bar \vf^I}
	= \frac{1}{2 \ri \mu} \left(
	\frac{\pa L}{\pa \bar \vf^I}
	- G^J \frac{\pa^2 L}{\pa G^J \pa \bar \vf^I}
	- G^0 \frac{\pa^2 L}{\pa G^0 \pa \bar \vf^I}
	- 2 \vf^J \frac{\pa^2 L}{\pa \vf^J \pa \bar \vf^I}
	\right).
\end{align*}
Inserting this into the relation
\begin{align*}
\frac{\partial^2 L}{\partial \vf^0 \pa \bar \vf^I}
	= \frac{\partial^2 L}{\partial \bar \vf^0 \pa \vf^I}
	\left( =  -\frac{\partial^2 L}{\pa G^0 \pa G^I}= \textrm{real} \right)
\end{align*}
we find
\begin{align}\label{eq_RcondDerived}
\bar \mu \left(
	\frac{\pa L}{\pa \bar \vf^I}
	- G^J \frac{\pa^2 L}{\pa G^J \pa \bar \vf^I}
	- 2 \vf^J \frac{\pa^2 L}{\pa \vf^J \pa \bar \vf^I}\right) + \HC = 0
\end{align}
which is exactly the extra condition \eqref{R-cond} we derived for $\N=2$
tensor models in AdS.

This lends credence to the following hypothesis: \emph{all $\N=2$ tensor models in AdS can be
understood as superconformal tensor models with a single frozen tensor multiplet.}
We will show this more explicitly (and constructively) in the next section.

\section{Manifestly supersymmetric formulation}
\label{section5}

It is of interest to compare the construction of section \ref{section4}  with the manifestly supersymmetric 
description of self-interacting $\cN=2$ tensor multiplets in AdS  developed earlier in \cite{KT-M-ads}.

${}$For general $\cN=2$ supersymmetric theories in AdS, the adequate superspace setting proves
to be the $\cN=2$ AdS projective superspace ${\rm AdS}^{4|8} \times {\mathbb C}P^1$ \cite{KT-M-ads},
which is a natural extension of the flat projective superspace ${\mathbb R}^{4|8} \times {\mathbb C}P^1$
\cite{KLR,LR-projective}. All essential information about the 
$\cN=2$ AdS superspace ${\rm AdS}^{4|8}$ is collected in Appendix \ref{appendixB}.
The complex projective space $ {\mathbb C}P^1 $ is conventionally  parametrized by homogeneous 
coordinates $v^i =(v^{\1}, v^{\2}) \in {\mathbb C}^2 \setminus \{0\}$ 
defined modulo the equivalence relation $v^i \sim c \,v^i$.
Supersymmetric matter in AdS can be described in terms of covariant projective multiplets 
introduced in \cite{KT-M-ads} building on the off-shell formulation for general $\cN=2$
supergravity-matter systems   developed in \cite{KLRT-M1,KLRT-M2}. 
Here we briefly recall the definition (more details can be found in \cite{KT-M-ads}). 

A  projective supermultiplet of weight $n$,
$\cQ^{(n)}({\bm z},v)$,  is defined to be a scalar superfield that
lives on  AdS$^{4|8}$,
is holomorphic with respect to
the isotwistor variables $v^i $ on an open domain of
${\mathbb C}^2 \setminus  \{0\}$,
and is characterized by the following conditions:\\
${}\quad$(1) it obeys the covariant analyticity constraints
\be
\bm\cD^{(1)}_{\a} \cQ^{(n)}  = {\bar {\bm\cD}}^{(1)}_{\ad} \cQ^{(n)}  =0~, \qquad 
\bm\cD^{(1)}_\a := v_i \bm\cD^{i}_\a ~, \quad 
{\bar {\bm\cD}}^{(1)}_\ad := v_i {\bar {\bm\cD}}^{i}_\ad 
~;
\label{ana}
\ee
${}\quad$(2) it is  a homogeneous function of $v^i$
of degree $n$, 
\be
\cQ^{(n)}({\bm z},c\,v)\,=\,c^n\,\cQ^{(n)}({\bm z},v)~, \qquad c\in \mathbb{C}\setminus \{0\}~;
\label{weight}
\ee
${}\quad$(3)  the OSp$(2|4)$ transformation law of  $\cQ^{(n)}$
is as follows:
\bea
\d_\x \cQ^{(n)}
&=& -\Big( {\bm \x}
+2{\bm \ve}  {\bm S}^{ij} J_{ij} \Big)\cQ^{(n)} ~,
\non \\
{\bm S}^{ij} J_{ij}  \cQ^{(n)}&:=& -
 \Big( {\bm S}^{{(2)}} {\bm \pa}^{(-2)}
-n \, {\bm S}^{(0)}\Big) \cQ^{(n)} ~, \qquad
\label{harmult1}
\eea
where 
$$
{\bm \x} := {\bm \x}^{\rm a}{\bm \cD}_{\rm a} + {\bm \x}^\a_i \bm\cD_\a^i + \bar {\bm \x}_\ad^i \bar {\bm\cD}^\ad_i
$$
is  an $\cN=2$ AdS Killing vector field, 
see subsection \ref{appendix_B3}, and the associated scalar parameter 
$\bm \varepsilon$ is given by eq. (\ref{B.333}).
In (\ref{harmult1}) we have introduced 
\bea
{\bm S}^{(2)}:=v_i v_j {\bm S}^{ij}~,\qquad
{\bm S}^{(0)}:= \frac{1}{(v,u)} v_i u_j {\bm S}^{ij}~,
\eea
and also the first-order operator
\bea
{\bm \pa}^{(-2)}= \frac{1}{(v,u)}u^{i} \frac{\pa}{\pa v^{i}}~.
\eea
The transformation law (\ref{harmult1}) involves an additional two-vector,  $u_i$, 
which is only subject 
to the condition $(v,u) := v^{i}u_i \neq 0$, and is otherwise completely arbitrary.
Both  $\cQ^{(n)}$ and $\d_\x  \cQ^{(n)}$ are independent of $u_i$.

In the family of projective multiplets, one can introduce 
a generalized conjugation, $\cQ^{(n)} \to \breve{\cQ}^{(n)}$, defined as
\be
\breve{\cQ}^{(n)} (v):= \bar{\cQ}^{(n)}\big(
\overline{v} \to 
\breve{v}\big)~, 
\qquad \breve{v} = {\rm i}\, \s_2\, v~, 
\ee
with $\bar{\cQ}^{(n)}(\overline{v}) $ the complex conjugate of $\cQ^{(n)}(v)$.\footnote{In what follows, 
we do not indicate explicitly the $\bm z$-dependence of projective supermultiplets.}
It is easy to check that $\breve{\cQ}^{ (n) } (v)$ is a projective multiplet of weight $n$.
One can also see that
$\breve{\breve{\cQ}}{}^{(n)}=(-1)^n \cQ^{(n)}$,
and therefore real supermultiplets can be consistently defined when 
$n$ is even.
The $\breve{\cQ}^{(n)}$ is called the smile-conjugate\footnote{The   smile-conjugation is the real structure
pioneered by Rosly \cite{Rosly} and re-discovered some time later in \cite{GIKOS,KLR,HitchinKLR}.} 
of ${\cQ}^{(n)}$.

Let us also recall that there is a regular procedure to construct 
$\cN=2$ supersymmetric field theories in AdS  \cite{KT-M-ads}.
The supersymmetric action principle is 
\bea
S&=& \oint  \frac{ v_i \rd v^i}{2\p}  \int \rd^4 x \,{\rm d}^8\q \,
{\bm E}\, \frac{\cL^{(2)}}{({\bm S}^{(2)})^2 }~, \qquad {\bm E}^{-1} = {\rm Ber} ({\bm E}_{\cA}{}^\cM)
\label{InvarAc-AdS}
\eea
where the Lagrangian $\cL^{(2)}(v)$ is a real weight-two projective supermultiplet
 constructed in terms of the dynamical projective supermultiplets. 

In this section, we restrict our attention to the  $\cN=2$ supersymmetric models 
of $n$ interacting tensor multiplets,
$\cG^{I\,(2)} (v)$, with $I=1,\dots, n$. Each $\cN=2$ tensor multiplet is a real weight-two 
projective multiplet of the following functional form:
\bea
\cG^{I\, (2)} ( v) = \cG^{I\, ij}  v_i v_j~, \qquad \overline{\cG^{I\, ij} } = \cG^I_{ij} =\ve_{ik}\ve_{jl} \cG^{I\, kl}~.
\eea
A general self-coupling of $\cN=2$ tensor multiplets in AdS
is generated by a Lagrangian of the following  type:
\bea
\cL^{(2)}_{\text{tensor}} = \cL ( \cG^{I\, (2)}, {\bm S}^{(2)})~, \qquad 
\Big( \cG^{I\,(2)} \frac{\pa}{\pa \cG^{I\, (2)}} + {\bm S}^{(2)} \frac{\pa}{\pa {\bm S}^{(2)}} \Big) \cL
= \cL~.
\label{5.9}
\eea 
Here $\cL$ is an analytic  homogeneous function of its arguments of degree one.
In the case of superconformal tensor multiplets, the Lagrangian is independent of $ {\bm S}^{(2) }$,
\bea
 \frac{\pa}{\pa {\bm S}^{(2)}}  \cL=0~.
\eea 
The Lagrangian (\ref{5.9}) is obtained from that describing 
the most general self-coupling of $n+1$ superconformal 
tensor multiplets by freezing one of these multiplets to coincide with ${\bm S}^{(2)}$.

The action (\ref{InvarAc-AdS}) is constructed as an integral over the  $\cN=2$ AdS superspace.
It can be reduced to $\cN=1$ AdS superspace, AdS$^{4|4}$, 
according to the scheme worked out in \cite{KT-M-ads}.
This involves two stages. First of all, assuming (without loss of generality) that the closed integration contour in 
(\ref{InvarAc-AdS}) lies outside of the north pole of ${\mathbb C}P^1$, $v^i \propto (0,1)$, 
all projective multiplets should be expressed in terms of the inhomogeneous complex coordinate
$\z \in \mathbb C$ for ${\mathbb C}P^1$ which
is defined as\footnote{In this chart, we can choose $u_i =(1,0)$.}
\begin{align}
v^{i} =v^{\1}(1,\z)~. \label{north}
\end{align}
In particular, associated with the Lagrangian  $\cL^{(2)}(v)$ is  
the superfield $\cL(\z)$ defined as
\bea
\cL^{(2)}(v):=\ri v^{\1}v^{\2}\cL(\z)=\ri (v^{\1})^2\z\,\cL(\z)~.
\eea
Similarly, associated with ${\bm S}^{(2)}(v) $ is the superfield $ {\bm S}(\z)$ defined as
(see eq. (\ref{S11S22})) 
\bea
{\bm S}^{(2)}(v):=\ri (v^{\1})^2\z\,{\bm S}(\z)~,
\qquad {\bm S}(\z)=\ri \Big( {\bar \mu} \,\z+\mu\,\frac{1}{ \z}\Big)~.
\label{S(zeta)}
\eea
The components of $\bm S^{ij}$, involving the parameters $\mu$ and $\bar\mu$, are defined according to
\eqref{S11S22} and correspond to the constant torsion of AdS$^{4|4}$.
Secondly, the $\cN=2$ superspace integral in (\ref{InvarAc-AdS})  should be reduced to that over
AdS$^{4|4}$ by  making use of the analyticity conditions (\ref{ana}).
Let ${\cL}(\z)|$ denote the $\cN=1$ projection of the Lagrangian  $\cL(\z)$, 
see subsection \ref{appendixB2}.
Then,  the  $\cN=2$ supersymmetric action
(\ref{InvarAc-AdS}) 
can be shown to be equivalent to the following functional  in AdS$^{4|4}$:
\bea
S = \oint \frac{\rd \z}{2\pi \rm i\z}
\int \rd^4 x \, {\rm d}^4\q \,{E}\,{\cL}(\z)|~.
\label{InvarAc5-N=2-N=1}
\eea
In what follows, we do not indicate the symbol of $\cN=1$ projection. 

Given a projective supermultiplet $\cQ^{(n)} (v)$, it can equivalently be described in terms 
of a properly defined superfield $\cQ (\z) \propto \cQ^{(n)} (v)$ such that 
the smile-conjugation $\cQ^{(n)} \to \breve{\cQ}^{(n)}$ operates as follows:
\bea
\cQ (\z) =
\sum \cQ_k  \z^k \quad \longrightarrow \quad
\breve{\cQ} (\z ) 
=\sum (-1)^k {\bar \cQ}_{-k}  { \z^k}~.
\eea
If ${\cQ} (\z ) $ is a real supermultiplet, $\breve{\cQ} (\z ) = {\cQ} (\z ) $, then the corresponding 
component superfields $\cQ_k$ obey the reality conditions $\bar \cQ_k = (-1)^k \cQ_{-k}$.
The Lagrangian $\cL (\z)$ in (\ref{InvarAc5-N=2-N=1}) is real, $\breve{\cL} (\z ) = {\cL} (\z ) $, which implies
the action (\ref{InvarAc5-N=2-N=1}) is real.

In the case of $\cN=2$  tensor multiplets, we represent
\bea
\cG^{I \,(2)} (v):=\ri (v^{\1})^2\z\, \cG^I(\z)~, \qquad
\cG^I (\z) = \frac{1}{\z}\, \vf^I + G^I - \z \,{\bar \vf}^I~, \quad \breve{\cG}^I (\z ) = {\cG}^I (\z )~.~~~
\eea
The analyticity conditions (\ref{ana}) on $\cG^{I\,(2)}(v)$ can be shown to imply that 
the $\cN=1$ scalar superfields $\vf^I$ and $G^I$ obey the constraints 
\bea
{\bar \cD}_{\dot \a} \vf^I =0~, 
\quad ({\bar \cD}^2 -4\m) G^I =0~, \quad
{\bar G}^I=G^I~.~~~
\eea
Reformulated in $\cN=1$ AdS superspace, the
tensor multiplet model generated by (\ref{5.9}) becomes
\bea
S_{\text{tensor}} =  \oint \frac{\rd \z}{2\pi \rm i\z}
\int \rd^4 x \, {\rm d}^4\q \,{E}\,{\cL} \Big(\cG^I (\z),  {\bm S} (\z)\Big)~.
\eea
Upon evaluation of the contour integral, this action takes the form  (\ref{4.2}) 
with
\bea \label{eq_Lcontour}
L(\vf^I, \bar\vf^I, G^I) =   \oint \frac{\rd \z}{2\pi \rm i\z}
{\cL} \Big( \cG^I(\z),  \ri {\bar \mu} \,\z+\ri \mu\,{ \z}^{-1} \Big)~.
\eea
In contrast to the rigid supersymmetric case, see subsection \ref{subsection2.2},
the integrand on the right is not allowed to depend on $\z$ in an arbitrary way, 
but only through the real combination $\ri ({\bar \mu} \,\z+\mu\,{ \z}^{-1} )$.

One way of understanding this $\zeta$ dependence is to investigate how the
condition \eqref{R-cond} is satisfied by the integral \eqref{eq_Lcontour}. We begin
by observing that the quantity $R_I$ \eqref{RRi} can be written
\begin{align}
R_I = \oint \frac{\rd\zeta}{2\pi \ri \zeta} \left(
	\frac{\mu}{\zeta} \frac{\partial \cL}{\partial \cG^I}
	- \frac{\mu}{\zeta} \frac{\partial^2 \cL}{\partial \cG^I \partial \cG^J} G^J
	+ 2\mu \frac{\partial^2 \cL}{\partial \cG^I \partial \cG^J} \bar\vf^J
	\right)
	+ \textrm{c.c.}
\end{align}
The first term can be rewritten (neglecting a total contour derivative) as
\begin{align}
R_I = \oint \frac{\rd\zeta}{2\pi \ri \zeta} \left(
	\mu \frac{\rd}{\rd\zeta} \frac{\partial \cL}{\partial \cG^I}
	- \frac{\mu}{\zeta} \frac{\partial^2 \cL}{\partial \cG^I \partial \cG^J} G^J
	+ 2\mu \frac{\partial^2 \cL}{\partial \cG^I \partial \cG^J} \bar\varphi^J
	\right)
	+ \textrm{c.c.}
\end{align}
Now if we make use of the fact that the $\zeta$ dependence of $\cL$
occurs implicitly in the superfields $\cG^I$ and explicitly in the
combination ${\bm S}(\zeta)$ \eqref{S(zeta)}, this integral may be further
simplified to
\begin{align}
R_I = \oint \frac{\rd\zeta}{2\pi \ri \zeta} \left(
	\mu \frac{\rd {\bm S}}{\rd\zeta} \frac{\partial^2 \cL}{\partial \cG^I \partial {\bm S}}
	- \frac{\mu}{\zeta} \frac{\partial^2 \cL}{\partial \cG^I \partial \cG^J} \cG^J
	\right)
	+ \textrm{c.c.}
\end{align}
Because the $\N=2$ Lagrangian $\cL$ is homogeneous of degree one in terms of
$\cG^I$ and ${\bm S}$ \eqref{5.9}, we have
\begin{align}
R_I &= \oint \frac{\rd\zeta}{2\pi \ri \zeta} \left(
	\mu \frac{\rd {\bm S}}{\rd\zeta} \frac{\partial^2 \cL}{\partial \cG^I \partial {\bm S}}
	+ \frac{\mu}{\zeta} \frac{\partial^2 \cL}{\partial \cG^I \partial {\bm S}} {\bm S}
	\right)
	+ \textrm{c.c.} \eol
	&= \oint \frac{\rd\zeta}{2\pi \ri \zeta} \left(
	2\ri \mu \bar \mu \frac{\partial^2 \cL}{\partial \cG^I \partial {\bm S}}
	\right)
	+ \textrm{c.c.} = 0
\end{align}
after making use of the explicit form \eqref{S(zeta)} of ${\bm S}(\zeta)$.

\section{$\cN=2$ supersymmetric $\sigma$-models in AdS}\label{section6}

$\cN=2$ supersymmetric $\s$-models in AdS can be formulated using off-shell hypermultiplets
in $\cN=2$ AdS superspace \cite{KT-M-ads}. The virtue of the approach pursued in \cite{KT-M-ads}
is that the superfield Lagrangian can have an arbitrary functional form. 
The technical disadvantage of this approach is that, upon reduction to $\cN=1$ AdS superspace
(elaborated in \cite{KT-M-ads}), the action functional involves not only physical superfields
(chiral and complex linear ones) but also an infinite set of auxiliary
fields which have to be eliminated using their nonlinear algebraic equations of motion.
Here we will pursue a complementary approach, in the spirit of \cite{HullKLR}, 
by formulating the action in terms of the physical $\cN=1$ chiral superfields only 
and then determining conditions on the Lagrangian for the theory to possess 
$\cN=2$ supersymmetry. The main results of this work were previously reported
in a brief letter \cite{Butter:2011zt}. We refer the reader there for details
of the component form of the action and the supersymmetry transformation rules.

\subsection{$\cN=1$ supersymmetric $\sigma$-models in AdS}

The most general  nonlinear $\s$-model in $\N=1$ AdS superspace is given by
\begin{align}
S = \int \rd^4x\, \rd^4\theta \, E\, \cK (\f^a, \bar \f^{\bar b})~.
\label{sigma-action}
\end{align}
The dynamical variables $\f^a$ are covariantly chiral superfields, ${\bar \cD}_\ad \f^a =0$,
and at the same time local complex coordinates of a complex  manifold $\cM$.
Unlike in the Minkowski case, the action does not possess K\"ahler
invariance since 
\begin{align}\label{2.6}
\int \rd^4x\, \rd^4\theta \, E\, F (\f^a)= \int \rd^4x\, \rd^2\theta \, \cE\, \mu F (\f^a)  \neq 0~,
\end{align}
with $\cE$ the chiral density. Nevertheless, 
K\"ahler invariance naturally emerges if we represent the Lagrangian as
\bea\label{eq_KW}
\cK (\f , \bar \f ) = K(\f , \bar \f) + \frac{1}{\m} W(\f) +  \frac{1}{\bar \m} \bar W( \bar \f) ~, 
\eea
for some K\"ahler potential $K$ and superpotential $W$. 
Under a K\"ahler transformation, these transform as
\begin{align}\label{eq_AdSKahler}
K \rightarrow K + F + \bar F, \qquad
W \rightarrow W - \mu F~.
\end{align}
The K\"ahler metric defined by 
\begin{align} \label{Kahler_metric}
g_{a \bar b} := \partial_a \partial_{\bar b} \cK = \partial_a \partial_{\bar b} K
\end{align}
is obviously invariant under (\ref{eq_AdSKahler}).

The nonlinear $\sigma$-model (\ref{sigma-action}) is manifestly invariant under 
arbitrary $\cN=1$ AdS isometry transformations 
\bea
\d_\x \f^a =- \x \f^a~, \qquad \x := \x^{\rm A}\cD_{\rm A} =
\x^{\rm a}\cD_{\rm a} + \x^\a \cD_\a + \bar \x_\ad \bar \cD^\ad 
\label{N=1AdStransfo6.6}
\eea
with the AdS Killing vector field $\x^{\rm A}$ defined by eqs. (\ref{N=1Kv1}) and (\ref{N=1Kv2}).

Because of (\ref{2.6}), the Lagrangian $\cK$ in  (\ref{sigma-action}) should be  a globally defined function 
on the K\"ahler target space $\cM$. This immediately implies that the K\"ahler two-form, 
 $ \O=2\ri \,g_{a \bar b} \, \rd \f^a \wedge \rd \bar \f^{\bar b}$,  associated with 
(\ref{Kahler_metric}), is exact and hence  $\cM$ is necessarily non-compact. 
We see that the $\s$-model couplings in AdS are more restrictive than in the Minkowski case.
The same conclusion follows from our earlier analysis of AdS supercurrent multiplets \cite{BK2011}.
In \cite{BK2011} we demonstrated that $\cN=1$ AdS supersymmetry allows the existence 
of just one minimal ($12+12)$ supercurrent, unlike the case of Poincar\'e supersymmetry
which admits three ($12+12)$ supercurrents. The corresponding AdS supercurrent 
is associated with the old minimal supergravity   and coincides with  the AdS extension of the Ferrara-Zumino 
multiplet \cite{FZ}. An immediate application of this result is that all supersymmetric $\s$-models in AdS
must possess a well-defined  Ferrara-Zumino  multiplet. The same conclusion also follows
from the exactness of $\O$ and earlier results of Komargodski and Seiberg \cite{KS}
who demonstrated that all {\it rigid} supersymmetric $\s$-models with an exact K\"ahler two-form 
possess a well-defined  Ferrara-Zumino  multiplet. 
The exactness of $\O$ for the general $\cN=1$ $\s$-models in AdS 
was independently  observed in 
\cite{Adams:2011vw} and \cite{Festuccia:2011ws} which appeared shortly after \cite{BK2011}.

We should discuss briefly how the structure \eqref{sigma-action}
emerges within a supergravity description  (see also \cite{Adams:2011vw}). 
Recall that nonlinear $\sigma$-models
may be coupled to supergravity via
\begin{align}
S = -\frac{3}{\kappa^2} \int \rd^4x\, \rd^4\theta \, E\, {\rm e}^{-\kappa^2 K / 3}
     + \int \rd^4x\, \rd^2\theta \,  \cE\, W_{\rm sugra}
     + \int \rd^4x\, \rd^2\btheta \, \bar\cE\, \bar W_{\rm sugra}
\end{align}
where the K\"ahler potential $K$ and the superpotential $W_{\rm sugra}$
transform under K\"ahler transformations as
\begin{align}\label{eq_SGKahler}
K \rightarrow K + F + \bar F, \quad
W_{\rm sugra} \rightarrow {\rm e}^{-\kappa^2 F} W_{\rm sugra}~.
\end{align}
The parameter $\kappa$ corresponds to the inverse Planck mass
which we will take to be vanishingly small to freeze out the gravitational
dynamics. The cleanest way to derive an AdS model from this supergravity
model is to assume $W_{\rm sugra}$ is dominated by a cosmological
term with a (relatively) small correction associated with the AdS superpotential,
\begin{align*}
W_{\rm sugra} = \frac{\mu}{\kappa^2} + W + \mathcal O(\kappa^2)~.
\end{align*}
The precise choice of the $\mathcal O(\kappa^2)$ corrections is
irrelevant once the small $\kappa$ limit is taken, but the
\emph{cleanest} choice is to choose
\begin{align}
W_{\rm sugra} = \frac{\mu}{\kappa^2} \exp\left(\frac{\kappa^2}{\mu} W \right)~.
\end{align}
For this choice, the AdS K\"ahler transformation \eqref{eq_AdSKahler}
matches the supergravity K\"ahler transformation \eqref{eq_SGKahler}.
The terms which diverge in a small $\kappa$ limit correspond to pure supergravity
with a cosmological constant and the supergravity equations of motion may be
solved to yield an AdS solution, freezing the supergravity structure.
The terms which remain as $\kappa$ tends to zero can be shown to
take the form \eqref{sigma-action} with $\cK$ given by \eqref{eq_KW}.

\subsection{The second supersymmetry transformation}
Next we look for those restrictions on the target space geometry 
which guarantee that the action (\ref{sigma-action}) is $\cN=2$ supersymmetric.
We make the following ansatz for the action of a second supersymmetry on the chiral
superfield $\phi^a$:
\begin{align}
\delta_\ve \phi^a = \frac{1}{2} (\bar\cD^2 - 4 \mu) (\veps \bar \Omega^a)~,
\label{N=2SUSYtr}
\end{align}
where $\bar\Omega^a$ is a function of $\phi$ and $\bar\phi$ which has to be determined.

The transformation law (\ref{N=2SUSYtr}) is a generalization 
of that derived in \cite{KT-M-ads} 
in the case of  a free off-shell $\cN=2$ hypermultiplet $\f^a = (\F, \J)$ described by the action
\bea
S= \int \rd^4 x \, {\rm d}^4\q \, E\Big(
\bar{{ \F}}{ \F}
+\bar{{ \Psi}}{ \Psi}
+\ri\frac{m}{\mu } 
{ \Psi}{ \F} 
-\ri\frac{m}{\bar \mu } 
{\bar { \Psi}} {\bar { \F}}
\Big) ~,~~~
\label{S-CC-massive}
\eea
with  $m$ a mass parameter.\footnote{The choice $m=0$ corresponds to the superconformal massless hypermultiplet.}
This action is invariant under the second supersymmetry transformation\footnote{In conjunction with (\ref{N=1AdStransfo6.6}), the transformation law (\ref{FS1}) defines an off-shell 
(Fayet-Sohnius-type) hypermultiplet in AdS.} 
\bea
\d_\ve \F =  \frac{1}{2} (\bar\cD^2 - 4 \mu) (\veps \bar \J)~, \qquad
\d_\ve \J =  - \frac{1}{2} (\bar\cD^2 - 4 \mu) (\veps \bar \F)~.
\label{FS1}
\eea
The ansatz (\ref{N=2SUSYtr}) also has a correct  super-Poincar\'e limit \cite{HullKLR}
(see also \cite{BX}).

On the mass shell, the right-hand side of (\ref{N=2SUSYtr})
should transform as a vector field of type $(1,0)$ under
reparametrizations of the target space.
Due to the constraints (\ref{epsilon-constraints}), the transformation $\delta \phi^a$ may be
rewritten
\begin{align}
\delta_\ve \phi^a = \bar\veps_\dalpha \bar\cD^\dalpha \bar\Omega^a
     + \frac{1}{2} \veps \,\bar\cD^2 \bar\Omega^a~.
     \label{N=2SUSYtr2}
\end{align}
This makes clear that $\bar\Omega^a$ is defined only up to a holomorphic vector,
\begin{align}
\bar\Omega^a \rightarrow \bar\Omega^a + H^a(\phi)~.
\label{3.5}
\end{align}

\subsection{Deriving the conditions of invariance}
Let us derive the conditions on $\cK$ and $\bar\Omega^a$ so that the variation
of the action
\begin{align}\label{eq_varyS}
\delta_{\ve} S = \int \rd^4x\, \rd^4\theta\, E\, \Big\{
     \frac{1}{2} \cK_{b} (\bar \cD^2 - 4 \mu) (\veps \bar\Omega^b)
     + \frac{1}{2} \cK_{\bar b} (\cD^2 - 4 \bar \mu) (\veps \Omega^{\bar b})
     \Big\}
\end{align}
is zero. An easy way to find a set of \emph{necessary} conditions is to require
not $\delta S$ itself to vanish but rather its variation under arbitrary deformations
of a chiral field $\phi^a$:
\begin{align}\label{eq_vary2S}
\delta_\phi \delta_\ve S &= -\frac{1}{8} \int \rd^4x\, \rd^2\theta\, \cE\, \delta \phi^a
     (\bar \cD^2 - 4 \mu) \Big\{
      \cK_{ab} (\bar \cD^2 - 4 \mu) (\veps \bar\Omega^b)
     +  \veps \bar\Omega^b{}_{,a} (\bar \cD^2 - 4 \mu) \cK_{b} 
     \eol & \quad
     +  g_{a \bar b} (\cD^2 - 4 \bar \mu) (\veps \Omega^{\bar b})
     +  \veps \Omega^{\bar b}{}_{,a} (\cD^2 - 4 \bar \mu) \cK_{\bar b} 
     \Big\}~.
\end{align}
Certainly if \eqref{eq_varyS} vanishes, then so should this quantity.
It turns out that ensuring the vanishing of \eqref{eq_vary2S} gives
us all the constraints required to show that \eqref{eq_varyS} vanishes.
Let us work these out now.

We begin by looking for all terms which will yield $\veps_\alpha$ under
the chiral projection. There is only one, and it is found in the third term
of \eqref{eq_vary2S}
\begin{align}
-\frac{1}{8} (\bar \cD^2 - 4 \mu) \Big\{g_{a \bar b} (\cD^2 - 4 \bar \mu) (\veps \Omega^{\bar b})\Big\}
	&= -\frac{1}{4} \veps^\alpha (\bar \cD^2 - 4 \mu) \Big\{g_{a\bar b} \,\Omega^{\bar b}{}_{,c} \,
	\cD_\alpha \phi^c\Big\} + \cdots
\end{align}
where we have made use of the chirality of $\veps_\alpha$ and neglected in
the ellipsis all terms involving $\veps$ and $\bar\veps_\dalpha$.
The term in braces must vanish; this implies that the quantity
\begin{align}\label{eq_HKcond1}
\omega_{a b} := g_{a\bar c} \,\Omega^{\bar c}{}_{,b} 
\end{align}
must be holomorphic, 
\bea
 \omega_{ab} = \omega_{ab}(\phi) \quad \Longleftrightarrow \quad 
 \nabla_{\bar c} \o_{ab}=0~.
 \eea

Next we consider all the other contributions from the third and
fourth terms of \eqref{eq_vary2S}. These amount to
\begin{align}
&\phantom{=} -\frac{1}{8} (\bar \cD^2 - 4 \mu) \Big\{\veps\, g_{a \bar b} \cD^2 \Omega^{\bar b}
     + \veps \cD^2 \cK_{\bar b} \Omega^{\bar b}{}_{,a}
     - 4 \veps \bar\mu \cK_{\bar b} \Omega^{\bar b}{}_{,a}\Big\} \eol
&= -\frac{1}{8} (\bar \cD^2 - 4 \mu) \Big\{
	\veps \cD^2 \cK_{\bar b} \,g^{\bar b c} (\omega_{ac} + \omega_{ca})
     + \veps \cD^\alpha \phi^c \cD_\alpha\phi^d \, \nabla_d \omega_{ac}\Big\}~.
\end{align}
When chirally projected, these are the only terms which will
give contributions proportional to $\cD^{\dalpha \alpha} \cD_{\alpha \dalpha} \phi^c$ and
$\cD_{\alpha \dalpha} \phi^c \cD^{\dalpha \alpha} \phi^d$ respectively, so
we must require both of these to vanish. This leads to the conditions
\begin{align}\label{eq_HKcond2}
\omega_{ab} = -\omega_{ba}, \qquad \nabla_c \omega_{ab} = 0~,
\end{align}
which tells us that $\omega_{ab}$ is indeed a covariantly constant holomorphic
two-form. These conditions completely eliminate the third and fourth
terms of \eqref{eq_vary2S}.

${}$Finally, we must ensure cancellation of the first and second terms of
\eqref{eq_vary2S}. Making use of the identity
\begin{align}
-\frac{1}{4} (\bar \cD^2 - 4 \mu) \bar\cD_\dalpha \bar\psi^\dalpha = 0
\end{align}
for arbitrary $\bar\psi^\dalpha$, we may rearrange the first and
second terms to
\begin{align}
\frac{1}{8} (\bar \cD^2 - 4 \mu)\Big\{
	\bar\veps_\dalpha \bar\cD^\dalpha \bphi^{\bar b} \,\nabla_a \bar\Omega_{\bar b}
     + 4 \veps\,\mu \,\partial_a (\bar\Omega^b \cK_b)
     + 4 \veps \,\bar\mu \,\cK_{\bar b} \,g^{\bar b b} \omega_{ba}\Big\}
\end{align}
where we have defined $\bar\Omega_{\bar b} := g_{\bar b c} \bar\Omega^c$
and made use of the anti-holomorphy of $\omega_{\bar a \bar b}$
to eliminate an extraneous term. We must apply the chiral projection
operator and check that the coefficients of $\bar\veps_\dalpha$ and
$\veps$ vanish separately.

At this point it is useful to observe that so far we have established the same
set of constraints \eqref{eq_HKcond1} and \eqref{eq_HKcond2} as imposed in the
globally supersymmetric case \cite{HullKLR}. It follows that only terms which explicitly
depend on $\mu$ (or $\bar\mu$) after the chiral projection
will need to be checked. Taking the chiral projection and selecting out just
the terms involving $\mu$ (or $\bar\mu$) and $\bar\veps_\dalpha$, we find
\begin{align}
\Big\{
     \mu \partial_a (\cK_b g^{b \bar b} \bar \omega_{\bar b \bar c})
     + \bar \mu \partial_{\bar c} (\cK_{\bar b} g^{\bar b b} \omega_{b a})
     \Big\}\, \bar\veps_\dalpha \bar\cD^\dalpha \bphi^{\bar c} ~.
\end{align}
The term in braces must vanish. Defining
\begin{align}
V^a :=  \frac{\mu}{2|\mu|} \, \omega^{ab} \cK_b~, \qquad
V^{\bar a} := \frac{\bar  \mu}{2|\mu|} \,\bar \omega^{\bar a \bar b} \cK_{\bar b}
\label{6.25}
\end{align}
we find the cancellation condition amounts to
\begin{align}
\nabla_a V_{\bar b} + \nabla_{\bar b} V_a = 0~.
\end{align}
In addition, we observe that by construction
\begin{align}
\nabla_a V_b = - \frac{\bar \mu}{2|\mu|} \,\omega_{ab}~,
\end{align}
which leads to
\begin{align}
\nabla_a V_{b} + \nabla_{b} V_a = 0~.
\end{align}

There still remain several terms in $\delta_\phi \delta_\ve S $ 
which we have not yet analyzed. However,  their total contribution can be shown
to vanish by using the conditions we have already established, and thus
$\delta_\phi \delta_\ve S =0$. These conditions are: 

(i)  the existence of a
covariantly constant two-form 
\begin{gather}\label{eq_omegaCond}
\omega_{a b}:= g_{a\bar b} \,\Omega^{\bar b}{}_{,b}=-\o_{ba}~,\qquad
\nabla_{c} \,\omega_{ab} =\nabla_{\bar c} \,\omega_{ab} = 0
\quad \Longrightarrow \quad \o_{ab}= \o_{ab}(\f)~;
\end{gather}

(ii) the existence of a certain Killing vector field obeying
\begin{gather}
V^a :=  \frac{\mu}{2|\mu|} \,\omega^{ab} \cK_b, \qquad
\nabla_a V_{\bar b} + \nabla_{\bar b} V_a = 0~, \qquad
\nabla_a V_b = -\frac{\bar \m}{2|\mu|} \,\omega_{ab} =- \nabla_b V_a~. 
\label{eq_KillingCond}
\end{gather}
The first condition occurs both in the Minkowski and AdS cases. The second condition 
is characteristic of AdS supersymmetry only.

It should be remarked that, modulo transformations (\ref{3.5}),  we can choose
\bea
\bar \O^a (\f , \bar \f ) = \o^{ a b} (\f) \, \cK_b (\f , \bar \f)~,
\label{6.32}
\eea
similarly to the super-Poincar\'e case \cite{HullKLR}.  The specific feature of the AdS case
is that $ \cK_b $ is a (globally-defined) one-form, and thus $\bar \O^a $ is necessarily a vector field.
Comparing the expression for $\bar \O^a$ with (\ref{6.25}) shows  that $ V^a =\m \bar \O^a / 2|\mu|$.
The choice (\ref{6.32}) will be assumed in what follows.

There is an important piece of information encoded in the first relation in (\ref{eq_omegaCond})
which leads to 
\bea
2\o_{ab}(\f) = \nabla_b \O_a- \nabla_a \O_b = \pa_b \O_a(\f, \bar \f) - \pa_a \O_b (\f, \bar \f) ~.
\label{6.33}
\eea
(It should be recalled that we have made the choice $\O_a =g_{a\bar b} \O^{\bar b}$.)
The left-hand side of (\ref{6.33}) does not depend on $\bar \f$, which makes it possible to evaluate
the right-hand side locally
by giving these variables  any given values $\bar \f_0$, that is 
\bea
\o_{ab} (\f) = \pa_a \r_b(\f) - \pa_b \r_a(\f) ~, \qquad 
\r_a(\f) := -\hf \O_a(\f,\bar \f_0) ~.
\label{6.33rho}
\eea
This gives an explicit local expression for $\r_a$.\footnote{A
similar trick was used by Bagger and Xiong in \cite{BX}
in a slightly different context.}

\subsection{Proof of invariance}\label{HK_indirect}
Having derived the conditions \eqref{eq_omegaCond} and \eqref{eq_KillingCond}, we must
still show that the action is indeed invariant. We begin by noting that the variation
of the action $\delta_\ve S$ may be written
\begin{align}
\delta_\ve S = -\frac{1}{2} \int \rd^4x\, \rd^4\theta\, E\, \Big(
     \veps^\alpha A_\alpha
     + \bar\veps_\dalpha \bar A^\dalpha
     \Big)~, \qquad
A_\alpha := \cD_\alpha \phi^{b} \omega_{b c} \,g^{c \bar c} \cK_{\bar c}~.
\end{align}
${}$From this point we may proceed in a manner quite analogous to the $\N=2$ tensor
model considered in subsection \ref{Tensor_ProofInvariance}.

We first observe that the quantity $A_\alpha$ obeys the condition
\begin{align}
\cD_\beta A_{\alpha} + \cD_\alpha A_{\beta}
     = -2 \cD_{\beta} \phi^{b} \cD_{\alpha} \phi^{c}\,\omega_{b c}~.
\end{align}
The form $\o_{ab} (\f)$ is 
locally  exact and given by (\ref{6.33rho}).
This representation is crucial for the proof below.  
We include a more direct proof of invariance, which does not make use of (\ref{6.33rho}), 
 in Appendix \ref{HK_direct}.

Because $\rho_a$ is holomorphic, it is possible to add trivial terms to the
integrand,
\begin{align}\label{eq_Smodified}
\delta_\ve S = -\frac{1}{2} \int \rd^4x\, \rd^4\theta\, E\, \Big\{
     \veps^\alpha (A_\alpha + 2 \cD_\alpha \phi^{b} \rho_{b})
     + \bar\veps_\dalpha (\bar A^\dalpha + 2 \bar\cD^\dalpha \bphi^{\bar b} \bar\rho_{\bar b})
     \Big\}~.
\end{align}
These additional terms contribute nothing since the chiral projection of
$\veps^\alpha \cD_\alpha \phi^b\, \rho_b$ vanishes using the chirality of $\veps^\alpha$
and $\rho_b$. Defining
\begin{align}\label{eq_B1}
B_\alpha := A_\alpha + 2 \cD_\alpha \phi^{b} \rho_{b} 
=\cD_\alpha \phi^{b} \Big(\omega_{b c} \,g^{c \bar c} \cK_{\bar c} +2\r_b \Big)~,
\end{align}
we observe that $\cD_\beta B_\alpha + \cD_\alpha B_\beta = 0$.
This equation is solved by
\begin{align}\label{eq_B2}
B_\alpha = \cD_\alpha B
\end{align}
for some function  $B(\f, \bar \f)$ on the target space.
Not much may be said about $B$ in the general case except that
\begin{align}
\partial_{b} B = \omega_{b c} g^{c \bar c} \cK_{\bar c} + 2 \rho_b
\end{align}
which follows from \eqref{eq_B1} and \eqref{eq_B2}.
The formula \eqref{eq_Smodified} may be then integrated by parts to yield
\begin{align}\label{eq_S_IBPd}
\delta_\ve S = 2 \int \rd^4x\, \rd^4\theta\, E\, \veps \left(\bar\mu B + \mu \bar B\right)~.
\end{align}
Our task is essentially complete. Since $\veps$ is a real linear superfield, the
quantity resembles the integrand
$\int \rd^4x\, \rd^4\theta\, E\, L K$
which is well-known to depend only on the value of the K\"ahler
metric constructed from $K$ when $L$ is a linear superfield. In
\eqref{eq_S_IBPd}, $\veps$ is indeed a linear superfield while
$\bar \mu B + \mu \bar B$ plays the role of a ``K\"ahler potential'' in this analogy.
Its corresponding K\"ahler metric,
\begin{align}
\partial_a \partial_{\bar b} \left(\bar\mu B + \mu \bar B \right)
     &= \bar \mu \partial_{\bar b} (\omega_{a c} g^{c \bar c} \cK_{\bar c} + 2 \rho_{a})
     + \mu \partial_a (\omega_{\bar b \bar c} g^{\bar c c} \cK_c + 2 \bar\rho_{\bar b}) \eol
     &= \bar \mu \partial_{\bar b} (\omega_{a c} g^{c \bar c} \cK_{\bar c})
     + \mu \partial_a (\omega_{\bar b \bar c} g^{\bar c c} \cK_c)~,
\end{align}
vanishes due to Killing vector condition \eqref{eq_KillingCond}.
This implies that \eqref{eq_S_IBPd} is indeed equal to zero.

\subsection{Closure of the supersymmetry algebra}
Let us calculate the commutator of two second supersymmetry transformations (\ref{N=2SUSYtr}).
This calculation is rather short and the result is 
\bea
[ \delta_{\ve_2} , \delta_{\ve_1}  ]\phi^a =  \omega^{ac} \omega_{cb} 
\Big( -\hf \tilde{\x}^{\a\ad} \cD_{\a \ad} + \tilde{\x}^\a \cD_\a\Big) \f^b ~,
\eea
where 
\bea
\tilde{\x}^{\a\ad} &:=& 4\ri \big(\ve^\a_1 \bar \ve^\ad_2 - \ve^\a_2 \bar \ve^\ad_1 \big)~, \qquad 
\tilde{\x}^\a:= 2\m \big( \ve^\a_2 \ve_1 - \ve^\a_1 \ve_2\big)
\eea
are the components of the first-order operator
$\x_{[\ve_2, \ve_1]} = -\hf \tilde{\x}^{\a\ad} \cD_{\a\ad} + \tilde{\x}^\a \cD_\a + \bar{\tilde{\x}}_\ad \bar \cD^\ad $
which proves to be an AdS Killing vector field, see eqs. (\ref{N=1Kv1}) and (\ref{N=1Kv2}).
If we impose
\begin{align}
\omega^{ac} \omega_{cb} = -\delta^a{}_b~, 
\label{3.18}
\end{align}
then the above result turns into
\bea
[ \delta_{\ve_2} , \delta_{\ve_1}  ]\phi^a = - \x_{[\ve_2, \ve_1]}  \f^a ~.
\label{3.19}
\eea

We see from (\ref{3.19}) that the commutator $[ \delta_{\ve_2} , \delta_{\ve_1}  ]\phi^a $
closes off the mass shell. 
This is similar to the supersymmetry structure  
within the Bagger-Xiong formulation \cite{BX} for $\cN=2$ rigid supersymmetric $\sigma$-models.
However, in the case of flat superspace, the commutator of the first and the second supersymmetries 
closes {\it only} on-shell \cite{BX}.  What about the AdS case? Computing 
the commutator of the $\cN=1$ AdS transformation and the second supersymmetry 
transformation gives
\bea
[\d_\x , \d_\ve ] \f^a 
= - \frac{1}{2} (\bar\cD^2 - 4 \mu) 
\Big ((\x  \veps )\bar \Omega^a \Big )~.
\eea 
Since $\x$ is an $\cN=1$ Killing vector field, the parameter $\ve'= \x \ve$ obeys the constraints 
(\ref{epsilon-constraints}) and hence generates a second supersymmetry transformation. 
We observe that commuting 
the $\cN=1$ AdS transformation and the second supersymmetry gives
a second supersymmetry transformation, 
\bea
[\d_\x , \d_\ve ] \f^a 
= - \d_{\x\ve} \f^a ~.
\eea
As a result, the algebra of ${\rm OSp(2|4)}$ transformations is closed 
off the mass shell!\footnote{It should be mentioned that the linearized action for {\it all} massless supermultiplets
of arbitrary superspin in $\cN=1$ AdS superspace \cite{GKS} is also invariant under 
$\cN=2$  supersymmetry transformations which close off-shell. }

Let us return to the equation (\ref{3.18}). Its implications are the same 
as in the super-Poincar\'e case \cite{HullKLR}.
In addition to the canonical complex structure
\begin{align}\label{complex_structure1} 
J_3 = \left(\begin{array}{cc}
\ri \,\delta^a{}_b & 0 \\
0 & -\ri \,\delta^{\bar a}{}_{\bar b}
\end{array}\right),
\end{align}
we may construct two more using $\omega^a{}_{\bar b}$
\begin{align}
\label{complex_structure2}
J_1 = \left(\begin{array}{cc}
0 & \omega^a{}_{\bar b} \\
\omega^{\bar a}{}_b & 0
\end{array}\right), \qquad
J_2 = \left(\begin{array}{cc}
0 & \ri\, \omega^a{}_{\bar b} \\
-\ri\, \omega^{\bar a}{}_b & 0
\end{array}\right)
\end{align}
such that $\cM$ is K\"ahler with respect to each of them.
The operators $J_A = (J_1, J_2, J_3)$ obey the quaternionic algebra
\begin{align}
J_A J_B = -\delta_{A B} \mathbb{I} + \eps_{ABC} J_C~.
\label{6.70}
\end{align}
Thus, $\cM$ is a hyperk\"ahler manifold. 
In accordance with the discussion in subsection 6.1, this manifold is non-compact. 
The above analysis also shows that $\cM$ must possess a special Killing vector.

Using (\ref{3.18}), it is easy to establish the equivalence 
\bea
(\bar \cD^2 - 4 \mu) \cK_a =0 \quad \Longleftrightarrow \quad (\bar \cD^2 - 4 \mu) (\o^{ab}\cK_b)=0~.
\eea
This results implies that the following rigid symmetry of the $\cN=2$ $\sigma$-model 
\bea
\d \f^a = \z (\bar \cD^2 - 4 \mu) (\o^{ab}\cK_b) ~, \qquad \z \in {\mathbb C} 
\label{6.72}
\eea
is trivial.

It is well-known that when $\N=2$ $\sigma$-models are coupled to supergravity,
their target spaces must be quaternionic K\"ahler manifolds \cite{BW}.
Unlike the hyperk\"ahler spaces which are Ricci-flat, their quaternionic K\"ahler 
cousins are Einstein spaces with a non-zero constant scalar curvature
(see, e.g., \cite{Besse} for a review).
Since AdS is a curved geometry, one may wonder whether the target spaces
of $\cN=2$ $\sigma$-models in AdS should also be   quaternionic K\"ahler.
Yet we have shown here that within AdS, the geometry is hyperk\"ahler
just as in Minkowski space. The reason is simple. As shown in
\cite{BW}, the  scalar curvature in the target space of locally supersymmetric $\sigma$-models 
must be nonzero and proportional to $\kappa^2$,
\begin{align}
R = -8 \kappa^2 (n^2 + 2n)~,
\end{align}
where the real dimension of the target space is $4n$.
But AdS (or Minkowski) space can be interpreted
as the $\kappa^2 \rightarrow 0$ limit of supergravity with (or without) a cosmological
constant $\mu$. In such a limit, we find indeed that the quaternionic
K\"ahler requirement from supergravity reduces to a hyperk\"ahler requirement.

\section{Geometric aspects of $\N=2$ $\sigma$-models in AdS}\label{section7}

In this section we would like to take a closer look at the geometric properties of the
Killing vector field (\ref{6.25}) which is characteristic of the target space of any $\cN=2$ 
supersymmetric $\s$-model in AdS. For that it is useful to  
recall the key facts about (tri-)holomorphic (Killing) vector fields.

\subsection{Holomorphic (Killing) vector fields}
Consider a K\"ahler manifold
$(\cM,g_{\m\n}, J^\m{}_\n)$, with $g_{\m\n}$ the K\"ahler metric and
$J^\m{}_\n$ the complex structure, which obeys $\nabla_\l J^\m{}_\n =0$.
A vector field ${ V}=V^\mu \pa_\m$ is said to be 
holomorphic with respect to  $J$ if
\begin{align}
\cL_V J = -J^\rho{}_\nu \nabla_\rho V^\mu + J^\mu{}_\rho \nabla_\nu V^\rho = 0~.
\label{7.1}
\end{align}
If in addition $ V$ is  a Killing vector field, 
\be
\nabla_\m V_\n + \nabla_\n V_\m=0 ~,
\label{7.2}
\ee
then $ V$  proves to be a Hamiltonian vector field with respect to
the symplectic two-form $\cJ= J_{\m\n}\rd \f^\m \wedge \rd \f^\n \equiv g_{\m\l}J^\l{}_\n \rd \f^\m \wedge \rd \f^\n$,
that is 
\begin{align}\label{eq_KillingHolo}
\rd (i_{ V} \cJ )=0 \quad \Longleftrightarrow \quad \cL_{ V} \cJ=0~. 
\end{align}
Any two of the conditions (\ref{7.1}), (\ref{7.2}) and (\ref{eq_KillingHolo}) imply the third one 
\cite{HitchinKLR}.

We choose local complex coordinates,
 $\f^\m =(\f^a , \bar \f^{\bar a})$, such that the complex structure becomes diagonal,
 and the K\"ahler metric takes the form ${\rm d}s^2 = 2g_{a \bar b } \,{\rm d} \f^a {\rm d} {\bar \f}^{\bar b}$.
Then, the holomorphy condition (\ref{7.1}) imposed on our vector field 
\bea
 V =V^a \frac{\pa}{\pa \f^a} + {\bar V}^{\bar a}  \frac{\pa}{\pa {\bar \f}^{\bar a}} 
\label{7.vf}
\eea
amounts to the requirement that $V^a$ is independent of $\bar \f$, $V^a = V^a(\f)$.
If $ V$ is also a Killing vector, then the condition  (\ref{eq_KillingHolo}) 
is equivalent to 
\bea\label{eq_KillingPotential}
V_a :=g_{a\bar b} {\bar V}^{\bar b} = \ri \, \pa_a \U ~, \qquad \bar \U = \U
\eea
with $\U (\f, \bar \f)$ a Killing potential \cite{BWitten}.

Consider now  a hyperk\"ahler manifold $(\cM, g_{\m\n}, J_A{}^\m{}_\n)$,
where $J_A{}^\m{}_\n$ are the three integrable complex structures 
obeying the quaternion algebra (\ref{6.70}).  On such a manifold we can define a tri-holomorphic vector field
$ V$ that  is holomorphic with respect to all the complex structures.
Let us  choose local complex coordinates,
 $\f^\m =(\f^a , \bar \f^{\bar a})$, such that the complex structure $J_3$ becomes diagonal,
eq. (\ref{complex_structure1}), and the other complex structures can be brought to  the form  
(\ref{complex_structure2}). As before, holomorphy with respect to $J_3$ means that the component  
$V^a$ of the vector field (\ref{7.vf}) is holomorphic, $V^a = V^a(\f)$. 
Holomorphy with respect to $J_1$ amounts to the conditions
\begin{subequations}
\begin{align}\label{eq_J1eq1}
0 &= \omega^c{}_{\bar b} \nabla_c V^a - \omega^a{}_{\bar c} \nabla_{\bar b} V^{\bar c} \\ \label{eq_J1eq2}
0 &= \omega^{\bar c}{}_{b} \nabla_{\bar c} V^a - \omega^a{}_{\bar c} \nabla_{b} V^{\bar c}
\end{align}
\end{subequations}
along with their complex conjugates. Holomorphy with respect to $J_2$ amounts to
\begin{subequations}
\begin{align}\label{eq_J2eq1}
0 &= \omega^c{}_{\bar b} \nabla_c V^a - \omega^a{}_{\bar c} \nabla_{\bar b} V^{\bar c} \\ \label{eq_J2eq2}
0 &= \omega^{\bar c}{}_{ b} \nabla_{\bar c} V^a + \omega^a{}_{\bar c} \nabla_{b} V^{\bar c}~.
\end{align}
\end{subequations}
Note that \eqref{eq_J1eq1} and \eqref{eq_J2eq1} are identical while
\eqref{eq_J1eq2} and \eqref{eq_J2eq2} differ in the relative sign of the two terms.
If $V$ is holomorphic with respect to $J_3$, then the conditions  \eqref{eq_J1eq2} and \eqref{eq_J2eq2} 
are satisfied.
It is easy to check that if a vector is holomorphic with respect to any two of these
complex structures, it is automatically holomorphic with respect to the third.

\subsection{Superpotential in $\cN=2$ rigid supersymmetric theories}
Tri-holomorphic Killing vector fields naturally occur in $\cN=2$ rigid supersymmetric $\sigma$-models
with non-vanishing superpotentials. Such a model is described by  the action 
\bea
S=\int {\rm d}^4 x \,{\rm d}^4 \q  \, K(\f , \bar \f)
+ \int {\rm d}^4 x \,{\rm d}^2 \q  \, W(\f) + \int {\rm d}^4 x \,{\rm d}^2 \bar \q  \, \bar W(\bar \f) ~.
\eea
As is well-known (see \cite{BX} and references therein), $\cN=2$ supersymmetry requires 
 that (i) $K(\f , \bar \f)$ is the K\"ahler potential of a hyperk\"ahler manifold $\cM$; (ii) 
$W(\f)$ must be such  that
\begin{align}\label{eq_VMink}
V^a = \omega^{ab} W_b~,\qquad V^{\bar a} = \bar\omega^{\bar a \bar b} \bar W_{\bar b}~
\end{align}
is a tri-holomorphic Killing vector field on $\cM$.
Holomorphy  with respect to $J_3$ means that $V^a= V^a(\f)$,
while holomorphy with respect to $J_1$ (and $J_2$) amount to
\begin{align}
\omega_{\bar a}{}^c \nabla_b W_c + \omega_{b}{}^{\bar c} \nabla_{\bar a} \bar W_{\bar c} = 0~.
\end{align}
The Killing equations
\begin{align}
\nabla_a V_b + \nabla_b V_a = \nabla_a V_{\bar b} + \nabla_{\bar b} V_a = 0
\end{align}
are satisfied as a consequence.

\subsection{Geometry of $\N=2$ AdS $\sigma$-models}\label{geometry_AdSsigma}
The geometric structure we have uncovered in AdS is quite interesting.
First of all, we have found that AdS supersymmetry demands the existence of 
a vector field $V^\m=(V^a , V^{\bar a})$ of the form
\begin{align}\label{eq_VAds}
V^a = \frac{\mu}{2|\mu|} \omega^{ab} \cK_b~,\qquad
V^{\bar a} = \frac{\bar\mu}{2|\mu|} \bar\omega^{\bar{ab}} \cK_{\bar b}~,
\end{align}
which obeys the Killing equations
\begin{align}\label{eq_VKilling}
\nabla_a V_b + \nabla_b V_a = \nabla_a V_{\bar b} + \nabla_{\bar b} V_a = 0~.
\end{align}
It is clearly not holomorphic with respect to $J_3$. In fact, it is easy
to show that $V$ \emph{rotates} the complex structures:
\begin{subequations}\label{eq_CSrotate}
\begin{align}
\cL_V J_1 &= \frac{\textrm{Im}\,\mu}{|\mu|} \, J_3 = J_3 \,\sin\theta \\
\cL_V J_2  &= -\frac{\textrm{Re}\,\mu}{|\mu|} J_3 = -J_3 \,\cos\theta \\
\cL_V J_3  &= \frac{\textrm{Re}\,\mu}{|\mu|}J_2 - \frac{\textrm{Im}\,\mu}{|\mu|} J_1
	= J_2 \,\cos\theta  - J_1\, \sin\theta 
\end{align}
\end{subequations}
where $\theta := \arg\mu$. There is a preferred complex structure
\begin{align}\label{eq_JAdS}
J_{\rm AdS} := \frac{\textrm{Re } \mu}{|\mu|} J_1 + \frac{\textrm{Im } \mu}{|\mu|} J_2
	= \frac{1}{|\mu|} \left(\begin{array}{cc}
	0 & \mu\, \omega^a{}_{\bar b} \\
	\bar\mu\, \omega^{\bar a}{}_b & 0
	\end{array}\right)
\end{align}
(normalized as usual so that $J^2 = -\mathbbm 1$) with respect to which
$V^\mu$ is \emph{holomorphic},
\begin{align}\label{eq_JAdSInv}
\cL_V J_{\rm AdS} = 0~.
\end{align}

It turns out that the conditions \eqref{eq_CSrotate}, which imply
\eqref{eq_JAdSInv}, \emph{also} imply \eqref{eq_VAds} in an elementary way.
As a consequence of \eqref{eq_JAdSInv}, one can always introduce some
function $\cK$ so that
\begin{align}
V^\mu = \frac{1}{2} J_{\rm AdS}{}^\mu{}_\nu \nabla^\nu \cK~.
\end{align}
This function $\cK$ is (up to a numerical factor) the real \emph{Killing potential} for $V^\mu$ as
defined in \eqref{eq_KillingPotential}, if we work in the basis where
$J_{\rm AdS}$ is diagonal. Moreover, using eqs. \eqref{eq_CSrotate},
it is a simple exercise to show that
\begin{align}
g_{\mu \nu} = \frac{1}{2} (\delta_\mu{}^\rho \delta_\nu{}^\sigma + J_3{}_\mu{}^\rho J_3{}_\nu{}^\sigma) \nabla_\rho \nabla_\sigma \cK
\end{align}
or equivalently (in complex coordinates where $J_3$ is diagonalized)
\begin{align}
g_{a b} = 0~,\qquad g_{a \bar b} = \partial_a \partial_{\bar b} \cK~.
\end{align}
In other words, the function $\cK$ is not only the Killing potential
with respect to $J_{\rm AdS}$, but \emph{also} the \emph{K\"ahler potential}
with respect to $J_3$. In fact, it is the K\"ahler potential
with respect to \emph{any} complex structure orthogonal to $J_{\rm AdS}$.
Thus the specification of a Killing vector \eqref{eq_VAds} in terms of
the K\"ahler potential $\cK$ is \emph{completely equivalent} to the
geometric requirement that the hyperk\"ahler manifold permit a Killing
vector which rotates the complex structures as \eqref{eq_CSrotate}.\footnote{Hyperk\"ahler 
manifolds with such properties were discussed by Hitchin et al. \cite{HitchinKLR}.}
In accordance with our earlier discussion, the function $\cK$ must be globally defined, 
and therefore the K\"ahler two-form associated with \emph{any} complex structure orthogonal to $J_{\rm AdS}$
must be exact.

It is quite remarkable that many of the features described above have also been
noticed recently in the context of supersymmetric nonlinear $\sigma$-models in
AdS$_5$ \cite{BX2011, BaggerLi}. As argued in \cite{BX2011}, 
the ${\rm AdS}_5$ supersymmetry requires the $\sigma$-model target space to be hyperk\"ahler and
possess a holomorphic Killing vector field (i.e. holomorphic with respect to $J_3$).\footnotetext{The 
Killing vector turns out to
be holomorphic due to a certain imbedding of the hypermultiplets into 4D $\N=1$
chiral superfields.}
It was noted in \cite{BaggerLi} that in fact, the holomorphic Killing vector field acts
as a \emph{rotation} on the complex structures. 

It is also worth mentioning that there is an interesting formal similarity
between \eqref{eq_VMink} and \eqref{eq_VAds}.
Recall that the general AdS Lagrangian $\cK$ may be interpreted as arising
from a real K\"ahler potential $K$ and a holomorphic superpotential $W$
via
\begin{align}
\cK = K + \frac{W}{\mu} + \frac{\bar W}{\bar \mu}
\end{align}
where $K$ and $W$ transform under K\"ahler transformations as
\begin{align}
K \rightarrow K + F + \bar F~,\qquad W \rightarrow W - \mu F~.
\end{align}
Because $W$ transforms nonlinearly, the K\"ahler-covariant derivative of
$W$ is naturally defined as
\begin{align}
\nabla_a W := \partial_a W + \mu \,\partial_a K.
\end{align}
However, one easily sees that
\begin{align}
\nabla_a W = \mu \,\partial_a \cK
\end{align}
and so $V^a$ can be equally as well written
\begin{align}
V^a = \frac{1}{2|\mu|} \omega^{ab} \nabla_b W~.
\end{align}
This formally resembles \eqref{eq_VMink} (up to the factor of $1 / 2 |\mu|$).
In AdS, the obstruction to tri-holomorphy arises since $[\nabla_{\bar a}, \nabla_b]$
involves a curvature associated with K\"ahler transformations. Thus,
\begin{align}
\nabla_{\bar a} \nabla_b W = [\nabla_{\bar a}, \nabla_b] W = \mu g_{\bar a b} \neq 0
\end{align}
implies that $V^a$ cannot be tri-holomorphic. In fact, it acts as a rotation
on the complex structures \eqref{eq_CSrotate}.

\subsection{Retrofitting hyperk\"ahler metrics and nonlinear $\sigma$-models}
We have identified the key criterion for whether a hyperk\"ahler metric may be used
as the target space for the a nonlinear $\sigma$-model in AdS: it must possess a
Killing vector which rotates the complex structures. Moreover, we have even
found the Lagrangian $\cK$: it is the Killing potential with respect to the
invariant complex structure. Given a hyperk\"ahler manifold with the requisite
geometric property, we may directly construct (at least in principle) the correct
Lagrangian $\cK$.

To demonstrate this, we take the simplest example possible: a four-dimensional
hyperk\"ahler manifold with vanishing curvature. This space is easily described by the K\"ahler
potential $K = x \bar x + y \bar y$ for complex coordinates $x$ and $y$. This
manifold is naturally equipped with a canonical holomorphic two-form
\begin{align}\label{eq_canonicalCS}
\omega_{ab} = \ri \sigma_2 = 
\begin{pmatrix}
0 & 1\\
-1 & 0 
\end{pmatrix}~.
\end{align}

This manifold possesses a U(2) group of holomorphic isometries associated
with the Killing vectors
\begin{align}
V_0^a = (\ri x, \ri y)~, \qquad
V_1^a = (\ri y, \ri x)~, \qquad
V_2^a = (y, -x), \qquad
V_3^a = (\ri x, -\ri y)~.
\end{align}
The Killing vectors $V_I = \{V_1, V_2, V_3\}$ obey an SU(2) algebra and
are tri-holomorphic. The Killing vector $V_0$ is holomorphic with respect to
$J_3$ alone and acts as a rotation in the plane of $J_1$ and $J_2$,
\begin{align}
\cL_{V_0} J_1 = -2 J_2 ~,\qquad
\cL_{V_0} J_2 = +2 J_1 ~.
\end{align}

Now let us introduce a new Killing vector $V_{\rm AdS}$ given by
\begin{align}
V_{\rm AdS} = \frac{1}{2} V_0 + \frac{1}{2} c_I V_I
\end{align}
where $c_I$ are arbitrary real constants. Let us denote $J_{\rm AdS} = J_3$ in this basis.
Because $V_I$ are tri-holomorphic, this Killing vector rotates the complex structures
around the axis selected out by $J_{\rm AdS}$. Since $V_{\rm AdS}$ is
a holomorphic Killing vector in this basis, we can easily construct its Killing potential, using
\begin{align}
V_{\rm AdS}^\mu = \frac{1}{2} J_{\rm AdS}{}^\mu{}_\nu \nabla^\nu \cK~.
\end{align}
The result is
\begin{align}
\cK = x \bar x + y \bar y + c_1 (x \bar y + y \bar x) + \ri c_2 (x \bar y - y \bar x)
	+ c_3 (x \bar x - y \bar y)~.
\end{align}
In the basis where $J_{\rm AdS}$ is given by \eqref{eq_JAdS}
\begin{align}\label{eq_JAdS2}
J_{\rm AdS} = \frac{1}{|\mu|} \begin{pmatrix}
0 & \ri \mu \sigma_2 \\
\ri \bar\mu \sigma_2 & 0
\end{pmatrix}~,
\end{align}
the function $\cK$ will coincide with an acceptable AdS Lagrangian.
It is easy to construct this new coordinate basis. Let $\phi$ and $\psi$ be
new complex coordinates given by
\begin{subequations}
\begin{align}
x &= \frac{1}{\sqrt {2 |\mu|}} \left(\bar \mu^{1/2} \phi - \ri \mu^{1/2} \bar\psi \right)~, \\
y &= \frac{1}{\sqrt {2 |\mu|}} \left(\bar \mu^{1/2} \psi + \ri \mu^{1/2} \bar\phi \right)~.
\end{align}
\end{subequations}
It is easy to see that in the new complex coordinates, the metric remains K\"ahler
and of canonical form. It again possesses a holomorphic two-form given by
\eqref{eq_canonicalCS}.
We easily find now that $\cK$ is given by
\begin{align}
\cK = \phi \bar\phi + \psi \bar\psi
	+ \frac{1}{|\mu|}\Big(\bar \mu \phi^2 (c_2 -\ri c_1)
	+ \bar \mu \psi^2 (c_2 + \ri c_1)
	+ \ri c_3 \bar \mu \phi \psi
	+ \HC \Big)~.
\end{align}
For the choice $c_1 = c_2 = 0$ and $c_3 = m / |\mu|$, this function $\cK$
matches the free off-shell $\cN=2$ hypermultiplet model given
in eq. \eqref{S-CC-massive}. For other values of $c_I$, we have found
a family of allowed AdS Lagrangians with a flat hyperk\"ahler metric.
It is also hardly a coincidence that the terms proportional to $c_I$ are, in this
final formulation, the real part of a holomorphic function. This is simply
a consequence of the Killing vectors $V_I$ being tri-holomorphic.

Although this was a fairly simple example, the technique appears to be quite general.
As a \emph{nontrivial} example, let us consider the target space
hyperk\"ahler metric associated with the Eguchi-Hanson gravitational instanton
in four dimensions. It possesses a K\"ahler potential (see e.g. \cite{HullKLR})
\begin{align}\label{eq_EgHK}
K = \sqrt{a^4  + \rho^4} - a^2 \log \left(\frac{a^2 + \sqrt{a^4 + \rho^4}}{\rho^2}\right)~,\qquad
\rho^2 := x \bar x + y \bar y~,
\end{align}
in terms of two complex coordinates $x$ and $y$.
The dimensionful parameter $a$ represents the size of the
gravitational instanton in the target space. For
$\rho \gg a$, the K\"ahler potential reduces to
the free hypermultiplet.

In these coordinates, the K\"ahler metric and its inverse are given by
\begin{subequations}
\begin{align}
g_{a \bar b} &= \frac{1}{\rho^4 \sqrt{a^4 + \rho^4}}
\begin{pmatrix}
\rho^6 + a^4 y \bar y & - a^4 y \bar x \\
-a^4 x \bar y & \rho^6 + a^4 x \bar x
\end{pmatrix}~, \\
g^{\bar a b} &= \frac{1}{\rho^4 \sqrt{a^4 + \rho^4}}
\begin{pmatrix}
\rho^6 + a^4 x \bar x & a^4 \bar x y \\
a^4 \bar y x & \rho^6 + a^4 y \bar y
\end{pmatrix}~.
\end{align}
\end{subequations}
The holomorphic two-form $\omega_{ab}$ again has the canonical form
\begin{align}
\omega_{ab} = \omega_{\bar a \bar b} = \omega^{a b} = \omega^{\bar a \bar b} = \ri \sigma_2~.
\end{align}
As before, there is a family of holomorphic isometries involving
the vectors $V_0$ and $V_I$. $V_I$ are tri-holomorphic while $V_0$
rotates the complex structures. Taking the same combination as before for
$V_{\rm AdS}$, we are led to construct the Killing potential $\cK$ as
\begin{align}\label{eq_EgHKAdS}
\cK = \frac{\sqrt{a^4+\rho^4}}{\rho^2} \Big(\rho^2 + c_1 (x \bar y + y \bar x) + \ri c_2 (x \bar y - y \bar x)
	+ c_3 (x \bar x - y \bar y)\Big)~.
\end{align}
Evidently this scalar function, when rewritten in the new coordinates
which transform $J_{\rm AdS}$ to \eqref{eq_JAdS2}, is an AdS Lagrangian
with an Eguchi-Hanson target space;
furthermore, the terms proportional to $c_I$ are the real part of a holomorphic function
in the new complex coordinates. What makes this procedure nontrivial is
\emph{finding} these complex coordinates. One set was given, e.g., in 
\cite{Bonneau:1993wz}.

An important feature worthy of note is the behavior of these two K\"ahler
potentials as $\rho$ tends to zero. The original potential \eqref{eq_EgHK}
is clearly singular at $\rho=0$. The coordinates $x$ and $y$ cover only
a single patch of the full target space manifold and a K\"ahler transformation
is necessary to connect the different patches. However, the function $\cK$,
eq. \eqref{eq_EgHKAdS},
in order to be a physical AdS Lagrangian, must be globally defined. Indeed
we see that it possesses a well-defined $\rho \rightarrow 0$ limit.

\subsection{Hyperk\"ahler structure of dual tensor models}
We are now prepared to resume the study of the dual tensor model in AdS 
derived in subsection \ref{subsection4.4}.
The original tensor model Lagrangian $L (\vf, \bar \vf, G)$ obeyed the conditions
\begin{align}
\frac{\pa^2 L}{   \pa \vf^I \pa \bar \vf^J } +\frac{\pa^2 L}{\pa G^I  \pa G^J} &= 0 ~,\\
\frac{\pa^2 L}{\pa  \vf^I \pa G^J} - \frac{\pa^2 L}{\pa  \vf^J \pa G^I}  &= F_{IJ} (\vf)~,
\end{align}
where $F_{IJ}$ is an exact  holomorphic two-form. As discussed already, the second
condition follows from the first and one can always make a trivial transformation
on the Lagrangian to set $F_{IJ} = 0$. However, we will keep a nonzero value
for full generality.

It is a straightforward task to calculate the K\"ahler metric in the dual geometry.
The result is \cite{HitchinKLR}
\begin{subequations}
\begin{align}
\frac{\partial^2 \cK}{\partial \varphi^I \partial \bar \varphi^J} &=
	\frac{\partial^2 L}{\partial \varphi^I \partial \bar \varphi^J} +
	\frac{\partial^2 L}{\partial \varphi^I \partial G^K}
		\left(\frac{\partial^2 L}{\partial \varphi^L \partial \bar \varphi^K}\right)^{-1}
		\frac{\partial^2 L}{\partial G^L \partial \bar\varphi^J} ~,\\
\frac{\partial^2 \cK}{\partial \varphi^I \partial \bar \psi_J} &=
	- \frac{\partial^2 L}{\partial \varphi^I \partial G^K}
		\left(\frac{\partial^2 L}{\partial \varphi^J \partial \bar \varphi^K}\right)^{-1} ~,\\
\frac{\partial^2 \cK}{\partial \psi_I \partial \bar \varphi^J} &=
	- \left(\frac{\partial^2 L}{\partial \varphi^K \partial \bar \varphi^I}\right)^{-1} 
		\frac{\partial^2 L}{\partial G^K \partial \varphi^J } ~,\\
\frac{\partial^2 \cK}{\partial \psi_I \partial \bar \psi_J} &=
	\left(\frac{\partial^2 L}{\partial \bar\varphi^J \partial \varphi^I}\right)^{-1}~.
\end{align}
\end{subequations}
We may cast this in matrix form by introducing
\begin{subequations}
\begin{align}
(\mathbf M)_{IJ} &:= \frac{\partial^2 L}{\partial \varphi^I \partial \bar \varphi^J}
	= - \frac{\partial^2 L}{\partial G^I \partial \bar G^J} ~,\\
(\mathbf Q)_{IJ} &:= \frac{\partial^2 L}{\partial \varphi^I \partial G^J} ~, \\
(\mathbf Q^\dag)_{IJ} &= \frac{\partial^2 L}{\partial G^I \partial \bar\varphi^J}~.
\end{align}
\end{subequations}
Note that $\mathbf M$ is symmetric and real.
The K\"ahler metric may be written in matrix form as
\begin{align}
g_{a \bar b} := \left(\begin{array}{cc}
\mathbf{M + Q M^{-1} Q^\dag} \,\,& -\mathbf{Q M^{-1}}\\
-\mathbf{M^{-1} Q^\dag} & \mathbf{M^{-1}} 
\end{array}
\right)~.
\end{align}
In this matrix form, the Monge-Amper\`e equation $\det g_{a \bar b} = 1$,
see e.g. \cite{HitchinKLR,Besse}, is clearly obeyed. The inverse
K\"ahler metric also takes a very simple form
\begin{align}
g^{\bar a b} = \left(
\begin{array}{cc}
\mathbf{M^{-1}} & \mathbf{M^{-1} Q} \\
\mathbf{Q^\dag M^{-1}}\,\, & \mathbf{M + Q^\dag M^{-1} Q}
\end{array}
\right)~.
\end{align}

The complex structure in the dual model is easily calculated. In similar
notation, it is given by
\begin{align}
\omega^{a b} = \left(
\begin{array}{rr}
\mathbf 0 & -\mathbbm 1 \\
\mathbbm 1 & \mathbf F
\end{array}
\right)~,\qquad
\omega_{a b} = \left(
\begin{array}{rr}
-{\mathbf F} & -\mathbbm 1 \\
{\mathbbm  1} & \mathbf 0
\end{array}
\right)
\end{align}
where $(\mathbf F)_{IJ} := F_{IJ}(\varphi)$. It is straightforward to check that the
complex structure is covariantly constant. Because $F_{IJ}$ may always be
taken to zero by shifting the original Lagrangian by a certain $H$-transformation,
one can always choose the dual complex structures to be of a simple canonical form.

We already know that any $\N=2$ nonlinear $\sigma$-model in AdS must possess a special
Killing vector $V^\mu$ which acts as a rotation on the complex structures.
It is enlightening to see how this comes about for nonlinear $\sigma$-models
which are dual to $\N=2$ tensor multiplet models. We easily calculate
$V^\mu = (V^a, V^{\bar a})$ where
\begin{align}
V^a = \begin{cases}
-\dfrac{\mu}{2 |\mu|} \dfrac{\partial \cK}{\partial \psi_I}~, & a = I \\
\phantom{+}\dfrac{\mu}{2 |\mu|} \left(\dfrac{\partial \cK}{\partial \varphi^I} + F_{IJ} \dfrac{\partial \cK}{\partial \psi_J}\right)~,
	& a = I + n~.
\end{cases}
\end{align}
Using the solution $G^I = G^I(\vf, \bar\vf, \psi+\bar\psi)$ from the duality
transformation, the first term above can be rewritten as
\begin{align}
V^a = \frac{\mu}{2|\mu|} G^I~,\qquad a = I
\end{align}
and so the Killing vector acts on the coordinates $\vf$ as
$V \vf^I = \mu G^I / 2 |\mu|$. If we calculate how the Killing vector
acts on the function $G^I(\vf, \bar\vf, \psi+\bar\psi)$, we find
\begin{align}
V^\mu \partial_\mu G^I = - \frac{1}{|\mu|}(\bar\mu \vf^I + \mu \bar\vf^I)~.
\end{align}
If we interpret this as a transformation of the original tensor variables
$(\vf^I, \bar\vf^I,G^I)$, this is simply an SO(2) transformation rotating
the $\N=2$ tensor multiplet! Unsurprisingly, the specific SO(2) subgroup of SU(2)
appearing here is that which preserves the form of $\bm S^{ij}$ in the underlying
projective superspace description of section \ref{section5}.

Using the complex structure and K\"ahler metric, we may calculate
$V_\mu = (V_a, V_{\bar a})$. One finds
\begin{align}
2|\m| V_{\bar a} = \begin{cases}
\mu \dfrac{\pa^2 \cL}{\pa \vf^I \pa \bar\vf^J} G^J
	+ \mu \dfrac{\pa^2 \cL}{\pa \bar \vf^I G^J} 
	\left(
	\left(\dfrac{\partial^2 L}{\partial \varphi^J \partial \bar \varphi^I}\right)^{-1} C_{J}
	+ 2 \bar\vf^I 
	\right)~, & a = I \\
\mu \left(\dfrac{\partial^2 L}{\partial \varphi^J \partial \bar \varphi^I}\right)^{-1} C_{J}
	+ 2 \mu \bar\vf^I~, & a = I + n
\end{cases}
\end{align}
where
\begin{align}
C_J := \frac{\partial L}{\partial \vf^I}
     - \frac{\partial^2 L}{\partial \vf^I \partial G^J} G^J
     - 2 \frac{\partial^2 L}{\partial \vf^I \partial \bar\vf^J} \bar\vf^J~.
\end{align}
Recall that the additional AdS condition for tensor models \eqref{R-cond} amounts to
\begin{align}
R_J = \mu C_J + \bar\mu \bar C_J = 0
\end{align}
and so, for example,
\begin{align}
V_{a} + V_{\bar a} = \frac{1}{|\m|} \left(\mu \bar\varphi^I + \bar\mu \varphi^I\right) ~,\qquad a = I+n~.
\end{align}
In order for $V^\mu$ to be a Killing vector, it must obey a number of conditions.
The only nontrivial one is $\partial_a V_{\bar b} + \partial_{\bar b} V_a = 0$.
This is straightforward (but tedious) to check.

\section{$\cN=2$ superconformal $\sigma$-models}\label{section8}
Both Minkowski and AdS $\cN=2$ superspaces have the same superconformal group 
$\rm SU(2,2|2)$. Thus all $\cN=2$ rigid superconformal $\sigma$-models should be invariant under
the $\cN=2$ AdS supergroup $\rm OSp(2|4)$. Here we elaborate on this point.

\subsection{Hyperk\"ahler cones}

Target spaces for $\cN=2$ superconformal $\s$-models are hyperk\"ahler cones
(see \cite{deWKV0,deWKV,deWRV} and references therein). By definition, 
a hyperk\"ahler cone is a hyperk\"ahler manifold
possessing an infinitesimal dilatation or, equivalently,
a homothetic conformal Killing vector field which is the gradient of a function.
Let us recall the salient facts about  
homothetic conformal Killing vector fields (see \cite{GR,deWRV} for more details).

Consider a Riemannian manifold $(\cM, g_{\m\n})$ admitting an infinitesimal dilatation.
The  required homothetic conformal Killing vector field ${\bm \c}=\c^\m\pa_\m$
is defined to obey the equation 
\bea
\nabla_\n \c^\m = \d_\n{}^\m \quad \Longleftrightarrow \quad 
\nabla_\n \c_\m = g_{\n\m}~,
\eea
and hence 
\bea
\c_\m =\hf \nabla_\m \c^2~, \qquad \c^2 = g_{\m\n} \c^\m \c^\n~.
\eea
The manifold $\cM$ is globally a (Riemannian) cone
\cite{GR}.

We next specify the Riemannian manifold under consideration to be a K\"ahler space, 
$(\cM,g_{\m\n}, J^\m{}_\n)$, with $J^\m{}_\n$ the complex structure.
 We choose local complex coordinates,
 $\f^\m =(\f^a , \bar \f^{\bar a})$, such that the complex structure becomes diagonal,
 and the K\"ahler metric takes the form ${\rm d}s^2 = 2g_{a \bar b } \,{\rm d} \f^a {\rm d} {\bar \f}^{\bar b}$.
The required homothetic conformal Killing vector field 
\bea
{\bm \c} = \c^\m \frac{\pa}{\pa \f^\m}
= \c^a \frac{\pa}{\pa \f^a} + {\bar \c}^{\bar a}  \frac{\pa}{\pa {\bar \f}^{\bar a}} 
\eea
is defined to obey the constraint
\bea
\nabla_\n \c^\m = \d_\n{}^\m \quad \Longleftrightarrow \quad 
\nabla_b \c^a = \d_b{}^a~, \qquad 
\nabla_{\bar b} \c^a = \pa_{\bar b} \c^a = 0~.
\label{hcKv}
\eea
In particular,  the vector field $\bm \c $ is holomorphic, $\c^a =\c^a(\f)$. Its properties include:
\begin{subequations}\label{hcKv-pot}
\bea
\c_\m := g_{\m \n} \c^\n = \pa_\m \cK
& \quad \Longleftrightarrow \quad &
 \c_a := {g}_{a \bar b} \,{\bar \c}^{\bar b} = \pa_a {\cK}~, \\
\hf g_{\m \n} \c^\m \c^\n = \cK  
& \quad \Longleftrightarrow \quad & 
{ g}_{a \bar b} \, \c^a {\bar \c}^{\bar b} ={ \cK}~. \label{8.3a} 
\eea
\end{subequations}
It follows from the above properties that 
$g_{a\bar b }=\pa_a\pa_{\bar b}\cK$, and thus $\cK$ is the K\"ahler potential. 
These relations imply that 
\be
\c^a (\f) \,\pa_a \cK (\f, \bar \f)  = \cK (\f , \bar \f) ~.
\label{idei}
\ee
The K\"ahler potential $\cK$ is a globally defined scalar function over $\cM$, 
in accordance with (\ref{8.3a}).\footnote{Although we call $\cK$  the K\"ahler potential, 
it should be kept in mind that 
 there is no K\"ahler invariance, for $\cK$ is uniquely determined, eq.  (\ref{8.3a}).}
This implies that the K\"ahler two-form, 
 $ \O=2\ri \,g_{a \bar b} \, \rd \f^a \wedge \rd \bar \f^{\bar b}$,  associated with 
the K\"ahler metric ${g}_{a \bar b} = \pa_a \pa_{\bar b}\cK$ 
is exact, and hence  $\cM$ is necessarily non-compact. This manifold is  called a K\"ahlerian cone
\cite{GR}.  It should be remarked that associated with the conformal Killing vector $\bm \c$ is 
 a U(1)  Killing vector 
\bea
X = {\rm i} \,\c^a (\f) \frac{\pa}{\pa \f^a} -{\rm i} \, {\bar \c}^{\bar a} (\bar \f)  \frac{\pa}{\pa {\bar \f}^{\bar a}} ~,
\label{U(1)Killing}
\eea
which leaves the K\"ahler potential invariant,
\bea
X \cK =0~.
\eea

A hyperk\"ahler cone is simply a hyperk\"ahler manifold $(\cM, g_{\m\n}, J_A{}^\m{}_\n)$
 admitting an infinitesimal  dilatation. Here $J_A{}^\m{}_\n$ are the three integrable complex structures 
 obeying the quaternion algebra (\ref{6.70}).
Associated with the conformal Killing vector field
$\bm \c$ are three Killing vectors $X_A{}^\mu := J_A{}^\mu{}_\nu \chi^\nu$,
which leave the K\"ahler potential invariant, $X_A{}^\mu \pa_\m \cK=0$.
These obey the $\rm{SU}(2)$ algebra
\begin{align}
[X_A, X_B] = -2 \eps_{ABC} X_C~.
\end{align}

\subsection{The AdS condition}
Given a hyperk\"ahler cone $\cM$,  our goal is to make use of the above properties of $\bm \c$ to show that
\bea
V^\m = (V^a, V^{\bar a})
:= \Big( \frac{\mu}{2|\mu|} \, \omega^{ab} \cK_b\, , \,\frac{\bar\mu}{2|\mu|}  \, \omega^{\bar a \bar b} \cK_{\bar b}\Big)
=   \Big(\frac{\mu}{2|\mu|} \, \omega^{ab} \c_b\, , \,\frac{\bar \mu}{2|\mu|}  \, \omega^{\bar a \bar b} \c_{\bar b}\Big)
\label{eq_Vconf}
\eea
is a Killing vector field,
for any non-zero complex parameter $\m$. By representing $ V_a = \bar \m\,  \o_{ab}\c^b / 2|\mu|$ and 
using the facts that $\o_{ab}$ and $\c^b$ are holomorphic, the conditions \eqref{eq_VKilling} follow.
It is also straightforward to check \eqref{eq_CSrotate}.

It is instructive to give a slightly different proof that (\ref{eq_Vconf}) is a Killing vector
and which shows that $V$ belongs to the Lie algebra of the group SU(2) isometrically acting on
the hyperk\"ahler cone.
As shown e.g. in \cite{deWRV,GR}, associated with the complex structures $(J_A)^\m{}_\n$, eqs. 
(\ref{complex_structure1}) and (\ref{complex_structure2}),  are   
the three Killing vectors $X_A^\m := (J_A)^\m{}_\n \c^\n$ which span  the Lie algebra of SU(2).
In particular, we have that $X_1^\mu = (\o^{ab}\cK_b\, , \, \omega^{\bar a \bar b} \cK_{\bar b})$
and $X_2^\mu = (\ri \, \o^{ab}\cK_b\, , -\ri\,  \omega^{\bar a \bar b} \cK_{\bar b})$
are Killing vectors. Moreover, it is a simple exercise to check that
\begin{align}
\cL_{X_A} J_B = - [J_A, J_B] = -2 \eps_{ABC} J_C~.
\end{align}
In the superconformal case there is a unique scalar function $\cK$ which serves as the
\emph{Killing potential} for each SU(2) isometry and the \emph{K\"ahler potential} for each
complex structure.

The Killing vector (\ref{eq_Vconf}) particular to the AdS case is simply a real
combination of  $X_1 = (J_1)^\m{}_\n \c^\n$ and  $X_2 = (J_2)^\m{}_\n \c^\n$, 
and thus $V^\m$ belongs to the Lie algebra of SU(2) and acts as a rotation on
the complex structures.

\subsection{Superconformal invariance}
Let $\cK (\f^a , \bar \f^{\bar b})$ be the K\"ahler potential of a hyperk\"ahler cone. 
We demonstrate here that the $\s$-model
\begin{align}
S = \int \rd^4x\, \rd^4\theta \, E\, \cK (\f , \bar \f )
\label{8.6}
\end{align}
is $\cN=2$ superconformal. As shown in Appendix \ref{appendixB}, an $\cN=2$ 
superconformal transformation is described in $\cN=2$ AdS superspace in terms of an
$\cN=2$  superconformal Killing vector. Upon reduction to $\cN=1$ AdS superspace, such
a transformation turns into three different ones: (i) an $\cN=1$ superconformal
transformation; (ii) an extended superconformal transformation; and (iii) a shadow
chiral rotation.  See subsection \ref{appendixB2} for more details.

The action (\ref{8.6}) is invariant under the $\cN=1$ superconformal transformation 
\bea
\d_{ \x}  \f^a &=& -{ \x} \f^a  -{ \s}\, \c^a (\f)~, 
\eea
with $\x=  { \x}^{\rm a} \cD_{\rm a} + { \x}^\a {\cD}_\a + \bar{\x}_\ad \bar{\cD}^\ad$ an arbitrary 
$\cN=1$ superconformal Killing vector, and the covariantly chiral parameter $\s$ defined by (\ref{A.15}).
This invariance follows from the homogeneity condition (\ref{idei}).
The action  (\ref{8.6}) is also invariant under the shadow chiral rotation of $\f^a$: 
\bea
\d \f^a = \frac{\ri \a}{2} \, \c^a (\f) ~, \qquad \bar \a = \a~,
\label{scr}
\eea
as a consequence of the identity (\ref{idei}).
It should be remarked that the shadow chiral rotation is generated 
by the Killing vector (\ref{U(1)Killing}).

${}$Finally, we define the extended superconformal transformation  of 
the chiral fields:
\bea 
\d_{\r, \bar \r} \f^a &=&\hf
({\bar \cD}^2 -4\m) \Big( \bar{ \r} \,  { \o}^{ab} \c_b \Big)
\label{hkc-esc}
\eea
where $\bar\rho$ is an $\cN=1$ superfield obeying certain constraints (see
eq. \eqref{rho-deff} for its definition and \eqref{rho-constraints}
and \eqref{rho-constraints2} for its constraints). In the flat superspace limit,
this correctly reduces to the transformation given in \cite{K-duality}.
The invariance of (\ref{8.6}) under this transformation can be proved in complete analogy with 
the rigid supersymmetric case \cite{K-duality}.
It is sufficient to evaluate the variation 
$\d_{\bar { \r} }S$ which corresponds to the choice $ \r =0$ and $\bar { \r} \neq 0$, 
since the parameters $\r$ and $\bar \r$ are independent.
The variation of the action is 
\bea
\d_{ {\bar \r}} S&=& -\hf    \int \rd^4 x\,{\rm d}^4\q \, E
\big({\bar \cD}_{\dot \a} \c_a \big)
\big(  {\bar \cD}^{\dot \a} {\bar \r} \big) \, \o^{ab}\c_b
= -\hf    \int \rd^4 x\,{\rm d}^4\q \, E \,{\bar \r}_{\dot \a} \big({\bar \cD}^{\dot \a} {\bar \f}^{\bar c} \big)
g_{\bar c\, a}  \, \o^{ab}\c_b \non \\
&=&  -\hf    \int \rd^4 x\,{\rm d}^4\q
\, E\,{\bar \r}_{\dot \a} \big({\bar \cD}^{\dot \a} {\bar \f}^{\bar a} \big)
{\bar \o}_{\bar a \bar b} \,{\bar \c}^{\bar b}~.
\eea
Since the tensor fields ${\bar \o}_{\bar a \bar b} $ and ${\bar \c}^{\bar b}$ are anti-holomorphic, 
and the parameter $ {\bar \r}_{\dot \a} $ is antichiral, 
the combination $ {\bar \r}_{\dot \a} {\bar \o}_{\bar a \bar b} \,{\bar \c}^{\bar b}$
appearing in the integrand is antichiral. 
Therefore, antichirally projecting the variation  gives
\bea
\d_{\bar { \r}} S&=& \frac{1}{8}  \int \rd^4 x\,{\rm d}^2{\bar \q}\, \bar\cE\,
 {\bar \r}_{\dot \a} {\bar \o}_{\bar a \bar b} \,{\bar \c}^{\bar b}\,
 (\cD^2 -4\bar \m){\bar \cD}^{\dot \a} {\bar \f}^{\bar a} 
=0~, 
\eea
for $(\cD^2 -4\bar \m) {\bar \cD}_{\dot \a} {\bar \F} $ is identically zero for any covariantly 
antichiral superfield $\bar \F$.

In the case of $\cN=1$ supersymmetry, more general superconformal $\s$-models exist 
than those described by  the action (\ref{8.6}) in which 
the Lagrangian is subject to  the homogeneity condition (\ref{idei}).
In fact, the most general $\cN=1$ superconformal $\s$-model is given by 
\bea
S &=& \int \rd^4x\, \rd^4\theta \, E\, {\mathbb K} (\f , \bar \f )~,
\eea
where the Lagrangian obeys a generalized homogeneity condition 
\begin{subequations}
\bea
\c^a (\f) \,\pa_a {\mathbb K} (\f, \bar \f)  
&=& {\mathbb K} (\f , \bar \f) + \frac{2}{\m} \cW(\f) - \frac{1}{\bar \m} \bar \cW (\bar \f) ~,
\label{8.18a}
\eea
for some homogeneous holomorphic function $\cW(\f)$ of degree three,
\bea
\c^a (\f) \,\pa_a \cW (\f) = 3 \cW(\f)~.
\eea
\end{subequations}
The general solution of eq. (\ref{8.18a}) is 
\bea
{\mathbb K} (\f, \bar \f) = \cK (\f, \bar \f) + \frac{1}{\m} \cW(\f) + \frac{1}{\bar \m} \bar \cW (\bar \f) ~,
\eea
with $ \cK (\f, \bar \f) $ obeying the homogeneity condition (\ref{idei}).
The above model in AdS can easily be related to the most general $\cN=1$ superconformal
$\s$-model in Minkowski space
\bea
S=\int {\rm d}^4 x \,{\rm d}^4 \q  \, \cK(\f , \bar \f)
+ \int {\rm d}^4 x \,{\rm d}^2 \q  \, \cW(\f) + \int {\rm d}^4 x \,{\rm d}^2 \bar \q  \, \bar \cW(\bar \f) ~.
\eea

In the case of $\cN=2$ superconformal symmetry, the $\s$-model action must also be invariant 
under the shadow chiral rotation (\ref{scr}) and 
the extended superconformal transformation (\ref{hkc-esc}). These symmetries 
prove to require the  superpotential to vanish,
\bea
\cW(\f) =0~.
\eea

\subsection{Analysis of the commutation relations}
Let us calculate the commutator of two extended superconformal transformations.
We find
\begin{align}\label{eq_SCcommutator}
[\delta_2, \delta_1] \phi^a &=
	\frac{1}{2} \xi^{\dalpha \alpha} \cD_{\alpha \dalpha} \phi^a
	- \xi^\alpha \cD_\alpha \phi^a
	- \frac{1}{4} (\bar \cD^2 - 4 \mu)\Big((\bar \rho_{1} \rho_{2} - \bar\rho_2 \rho_1) g^{a \bar b} \cD^2 \chi_{\bar b}\Big)
\end{align}
where
\begin{subequations}
\begin{align}
\xi_{\alpha \dalpha} &:= - 4\ri (\cD_\alpha \rho_{2} \bar\cD_\dalpha \bar\rho_{1} - 
	\cD_\alpha \rho_{1} \bar\cD_\dalpha \bar\rho_{2})~, \\
\xi_\alpha &:= \frac{\ri}{8} \bar\cD^\dalpha \xi_{\alpha \dalpha}
	= 2 \mu (\bar\rho_1 \cD_\alpha \rho_2 - \bar\rho_2 \cD_\alpha \rho_1)~.
\end{align}
\end{subequations}
The third term in \eqref{eq_SCcommutator} can be rearranged into a piece which
involves the equation of motion and a remainder. The result is
\begin{align}
[\delta_2, \delta_1] \phi^a &=
	\frac{1}{2} \xi^{\dalpha \alpha} \cD_{\alpha \dalpha} \phi^a
	- \xi^\alpha \cD_\alpha \phi^a
	- \sigma \chi^a
	+ \frac{\ri}{2} \alpha \chi^a
	\eol & \qquad
	- \frac{1}{4} (\bar \cD^2 - 4 \mu)\Big((\bar \rho_{1} \rho_{2} - \bar\rho_2 \rho_1) g^{a \bar b}
	(\cD^2 - 4 \bar\mu) \chi_{\bar b}\Big)
\label{8.25}
\end{align}
where $\sigma$ is given in terms of $\xi_{\alpha \dalpha}$ as in 
\eqref{A.15} and $\alpha$ is given by
\begin{align}
\alpha = 2\ri (\sigma -  \bar\sigma ) 
	- 8\ri \mu \bar \mu (\bar \rho_1 \rho_2 - \bar\rho_2 \rho_1)~.
\end{align}
The first line in (\ref{8.25}) is clearly an $\N=1$ superconformal transformation
combined with a shadow chiral rotation with real parameter
$\alpha$.\footnote{The parameter $\alpha$ must
be constant, and this can be shown to be the case using the formulae given in
Appendix B.} The second line vanishes on-shell and ensures on-shell closure
of the algebra. We see the algebra is open in the superconformal case but becomes 
closed under the $\rm OSp(2|4)$ transformations for which $\bar \r = \r$.

\section{$\N=2$ superfield formulation}\label{section9}
One important feature of the $\N=1$ AdS construction we have presented
is that the algebra closes off-shell. For this reason, there ought to
exist  a formulation in terms of $\N=2$ superfields. In this section,
we present just such a formulation.  As a brief warm-up, we describe 
an AdS generalization of the Fayet-Sohnius hypermultiplet \cite{Fayet,Sohnius}.

\subsection{Warm-up: Fayet-Sohnius hypermultiplet in $\rm{AdS}_4$}
Recall the algebra of covariant derivatives in AdS:
\begin{subequations}
\begin{gather}
\{  {\bm \cD}_\alpha^i, {\bm \cD}_\beta^j  \} = 4 {\bm S}^{ij} M_{\alpha \beta}
     + 2 \ve_{\alpha \beta} \ve^{ij} {\bm S}^{kl} J_{kl}~, \qquad
\{ {\bm \cD}_\alpha^i, \bar {\bm \cD}_{\dalpha j }  \} = -2\ri  \delta^i_j {\bm \cD}_{\alpha \dalpha}~, \\
[  {\bm \cD}_{ \alpha \dalpha },  {\bm \cD}_\beta^i] = - \ri \ve_{\alpha \beta} {\bm S}^{ij} \bar{\bm \cD}_{\dalpha j}~, 
\qquad [{\bm \cD}_{\rm a}, {\bm \cD}_{\rm b}]   = -{\bm S}^2 M_{\rm ab}~.
\end{gather}
\end{subequations}
A key feature of this algebra is that only an ${\rm SO}(2)_R \cong {\rm U}(1)_R$
subgroup of ${\rm SU}(2)_R$ generated by $J := {\bm S}^{kl} J_{kl}$
is respected. The generator $J$ acts as
\begin{align}\label{eq_SK0}
[J, {\bm\cD}_\alpha^i] = {\bm S}^i{}_j {\bm\cD}_\alpha^j~,\qquad
[J,\bar{\bm\cD}^\dalpha_i] = - {\bm S}_i{}^j \bar{\bm\cD}^\dalpha_j~,\qquad
[J, {\bm S}^{ij}] = 0~.
\end{align}
The constant isotriplet $\bm S^{ij}$ is chosen to obey \eqref{S12} and \eqref{S11S22}.

We introduce the Fayet-Sohnius hypermultiplet $q^i$, which is defined to obey
the constraints
\begin{align}\label{eq_FSconstraint}
{\bm\cD}_\alpha^{(i} q^{j)} = \bar {\bm\cD}_\dalpha^{(i} q^{j)}  = 0~.
\end{align}
However, the action of the generator $J$ on $q^i$ is \emph{not} fixed in advance.
It must be determined by the constraints \eqref{eq_FSconstraint}. Making use
of the algebra of covariant derivatives, one can show that
\begin{align}
J = \frac{1}{4} \{\bar{\bm\cD}_{\dalpha\1},\bar{\bm\cD}^\dalpha_{\2}\}
\end{align}
and so we may easily deduce
\begin{align}\label{eq_SK2}
J q_\1 = -\frac{1}{4} (\bar {\bm\cD}_\1)^2 q_\2~,\qquad
J \bar q_\1 = -\frac{1}{4} (\bar {\bm\cD}_\1)^2 \bar q_\2~.
\end{align}
Similarly,
\begin{align}
J = \frac{1}{4} \{{\bm\cD}^{\alpha\1},{\bm\cD}_\alpha^{\2}\}
\end{align}
which leads to
\begin{align}\label{eq_SK4}
J q_\2 = \frac{1}{4} ({\bm\cD}^\1)^2 q_\1~,\qquad
J \bar q_\2 = \frac{1}{4} ({\bm\cD}^\1)^2 \bar q_\1~.
\end{align}

We note that the superfield $q_\1$ is $\N=1$ chiral
\begin{align}
\bar \cD_{\dalpha \1} q_\1 = 0~.
\end{align}
However, the right-hand side of $J q_\1$ in \eqref{eq_SK2} is \emph{not} chiral.
We may rewrite it then in the form
\begin{align}\label{eq_SK6}
J q_\1 + \mu q_\2 = -\frac{1}{4} \Big[(\bar {\bm \cD}_\1)^2 - 4 \mu\Big] q_\2~.
\end{align}
Let us denote by $\mathbb J$ the ${\rm U}(1)$ operator transforming $q_i$ as
an isospinor,
\begin{align}\label{eq_FSdefJ2}
\mathbb J \,q_i = -{\bm S}_i{}^j q_j = -{\bm S}^j{}_i q_j~.
\end{align}
Then
\begin{align}\label{eq_SK7}
\mathbb J \,q_\1 = - {\bm S}_\1{}^\2 q_\2 = {\bm S}^{\2\2} q_\2 = -\mu \,q_\2~.
\end{align}
This allows \eqref{eq_SK6} to be rewritten as
\begin{align}
\Delta q_\1 = -\frac{1}{4} \Big[(\bar {\bm \cD}_\1)^2 - 4 \mu\Big] q_\2~,\qquad
\Delta := J - \mathbb J~.
\end{align}
The operator $\Delta$ commutes with the covariant derivatives, as
\eqref{eq_SK0} gives
\begin{align}\label{eq_SK8}
[\Delta, {\bm\cD}_\alpha^i] = [\Delta, \bar{\bm\cD}_{\dalpha i}] = 0~,
\end{align}
and therefore it can be interpreted as an intrinsic central charge.

Note also that the first expression in \eqref{eq_SK4} can be rewritten
\begin{align}\label{eq_SK9}
\Delta q_\2 = \frac{1}{4} \Big[({\bm \cD}^\1)^2 - 4 \bar\mu\Big] q_\1~.
\end{align}
Combined with \eqref{eq_SK7} and \eqref{eq_SK8}, this yields
\begin{align}
(\Delta^2 + \Box_{\rm c}) q_\1 = 0
\end{align}
where
\begin{align}
\Box_{\rm c} := \frac{1}{16} \Big[(\bar {\bm \cD}_\1)^2 - 4 \mu\Big] \Big[({\bm \cD}^\1)^2 - 4 \bar\mu\Big]
\end{align}
is the covariantly chiral d'Alembertian. 
In the case of massless Fayet-Sohnius hypermultiplet, the equation of motion 
is $\D=0$. In the massive case, it takes the form $\D =\ri m =\text{const}$, with $m$ a real mass parameter. 

There are several very important lessons we can take away from
this discussion. First we impose the constraints \eqref{eq_FSconstraint}
and then \emph{derive} the action of the ${\rm SO}(2)$ generator
on the hypermultiplet as a consequence. However, it is still
possible to separate the ${\rm SO}(2)$ generator into a
``natural'' generator, whose action is specified, along with a separate
piece $\Delta$ which commutes with everything else,
\begin{align}
{\bm S}^{jk} J_{jk} = \mathbb J
+ \Delta~.
\end{align}
In addition, the operator  $\Delta$, at least in this example, is constant
on-shell.

\subsection{$\N=2$ hypermultiplets as deformed Fayet-Sohnius}
We turn now to our real task: constructing an $\N=2$ superfield
formulation for the off-shell structure we have constructed.
Recall that we have an $\N=1$ superfield $\phi^a$ transforming as
\begin{align}\label{eq_dphi}
\delta \phi^a &= \frac{1}{2} (\bar \cD^2 - 4 \mu) (\veps \bar\Omega^a) 
	= \omega^a{}_{\bar b} \bar\veps_\dalpha \bar\cD^\dalpha \bar\phi^{\bar b}
	+ \frac{1}{2} \veps \bar \cD_\dalpha \left(\omega^a{}_{\bar b} \bar\cD^\dalpha \bar\phi^{\bar b}\right)
\end{align}
under the second supersymmetry and ${\rm SO}(2)$ transformation.
On-shell, this takes the simpler form
\begin{align}
\delta \phi^a
	&= \omega^a{}_{\bar b} \bar\veps_\dalpha \bar\cD^\dalpha \bar\phi^{\bar b}
	+ 4 |\mu| \veps V^a
\end{align}
where $V^a = \mu \omega^{ab} \cK_b / 2|\mu|$ is a Killing vector.

We would like to interpret $\phi^a$ as the $\N=1$ projection
of some $\N=2$ superfield $\Phi^a$. In doing so, we should
identify $\delta \phi^a$ as the lowest component of
a corresponding $\N=2$ transformation $\delta \Phi^a$
\begin{align}
\delta \Phi^a &=
	-{\bm \xi}^{\alpha}_\2 {\bm \cD}_\alpha^\2 \Phi^a
	- \bar{\bm \xi}_{\dalpha}^\2 \bar{\bm \cD}^\dalpha_\2 \Phi^a
	- 2 {\bm \veps} {\bm S}^{jk} {J}_{jk} \Phi^a~.
\end{align}
Because of the chirality of $\phi^a$ and the absence of
$\veps^\alpha = {\bm\xi}_\2^\alpha\vert$ in \eqref{eq_dphi},
we are led to conclude
\begin{align}\label{eq_dFSconstraint1}
\bar{\bm\cD}_{\dalpha \1} \Phi^a = {\bm\cD}_\alpha^\2 \Phi^a = 0~.
\end{align}
Similarly, because of the form of the $\bar\veps_\dalpha = {\bm\xi}^\2_\dalpha\vert$
term, we must also choose
\begin{align}\label{eq_dFSconstraint2}
\bar{\bm\cD}_{\dalpha \2} \Phi^a = -\omega^a{}_{\bar b} \bar{\bm \cD}_{\dalpha \1} \bar \Phi^{\bar b}~.
\end{align}
It is now a simple task to use these constraints to work out the
action of the ${\rm SO}(2)$ generator. We find
\begin{align}
{\bm S}^{jk} {J}_{jk} \Phi^a = -\frac{1}{4} \bar {\bm\cD}_{\dalpha \1}
	\left(\omega^a{}_{\bar b} \bar {\bm\cD}_{\1}^\dalpha \bar\Phi^{\bar b}\right)
	= -\frac{1}{4} (\bar {\bm\cD}_{\1})^2 (\omega^{ab} \cK_b)~.
\end{align}

As with the Fayet-Sohnius case, it is useful to split the ${\rm SO}(2)$ generator
into two pieces, one of which acts upon $\Phi^a$ in a geometric way and
a remainder which preserves $\N=1$ chirality and commutes with the
covariant derivative. Taking as before ${\bm S}^{jk} {J}_{jk} = \mathbb J + \Delta$
we \emph{define}
\begin{align}\label{eq_JPhi}
\mathbb J \Phi^a := -\mu \omega^{ab} \cK_b = -2 |\mu| V^a
\end{align}
in analogy with \eqref{eq_FSdefJ2}. Moreover, one can easily check that
the function $\cK$ is invariant
\begin{align}
\mathbb J \cK = -\mu \omega^{ab} \cK_a \cK_b + \textrm{c.c.} = 0
\end{align}
under the action of $\mathbb J$. This is exactly as one would
except from an ${\rm SO}(2)$ generator. From the definition
\eqref{eq_JPhi} it follows that the residual
piece $\Delta$ given by
\begin{align}\label{eq_DeltaPhi}
\Delta \Phi^a = -\frac{1}{4} \Big[(\bar {\bm\cD}_{\1})^2 - 4 \mu\Big] (\omega^{ab} \cK_b)
\end{align}
is chiral, which is consistent with the requirement \eqref{eq_SK8}.
Note also that this quantity vanishes on-shell,
\begin{align}
\Delta \Phi^a = -\frac{1}{4} \omega^{ab} \,\Big[(\bar {\bm\cD}_{\1})^2 - 4 \mu\Big] \cK_b
	= 0 \qquad \textrm{(on-shell)}~,
\end{align}
due to the chirality of $\omega^{ab}$.\footnote{It should be remarked that 
eq. (\ref{6.72}) defines a trivial symmetry transformation involving the
operator $\Delta$.}

We may provide some additional justification for the choice \eqref{eq_JPhi}
by considering a superconformal model, where this choice is quite natural.
For the constraints \eqref{eq_dFSconstraint1} and \eqref{eq_dFSconstraint2}
to be consistent with the superconformal algebra, the dilatation generator
$\mathbb D$, chiral U(1)$_R$ generator $\bm J$,
and SU(2)$_R$ generators $J_{ij}$ must act on $\Phi^a$ as
\begin{subequations}
\begin{gather}
\mathbb D \Phi^a = \chi^a ~,\qquad \bm J \Phi^a= 0~ \\
J_{\1\2} \Phi^a = \frac{1}{2} \chi^a~,\qquad
J_{\2\2} \Phi^a = \omega^{ab} \chi_b~,\qquad
J_{\1\1} \Phi^a = 0~.
\end{gather}
\end{subequations}
The three generators $J_{ij}$ may be naturally associated with the three
Killing vectors $X_A=(J_A)^\mu{}_\nu \chi^\nu$, where
$\chi^\mu = (g^{a \bar b} \cK_{\bar b}, g^{\bar a b} \cK_b)$.
We necessarily find that $\bm S^{ij} J_{ij} \Phi^a = -\mu \omega^{ab} \chi_b$
in accordance with \eqref{eq_JPhi}. Note that for this choice of $J_{ij}$
action, $\Phi^a$ must be on-shell since the operator $\Delta$ must be
chosen to vanish.

Note the superconformal case is special since we have a triplet of Killing
vectors to match the triplet of ${\rm SU}(2)_R$ transformations;
in the non-superconformal case we have only a single Killing vector,
corresponding to the single ${\rm SO}(2)_R$ transformation available.

\subsection{An ${\rm SU}(2)_R$ covariant geometric reformulation}
Before moving on, we would like to discuss the reformulation of the $\N=2$ superfield
and constraints we have imposed in a way which is ${\rm SU}(2)_R$ covariant
\`a la Sierra and Townsend \cite{Sierra-Townsend} (see also \cite{BW}).

We begin by taking $\Phi^\mu$ to be a coordinate of a $4n$-dimensional hyperk\"ahler manifold
with structure group ${\rm Sp}(1) \times {\rm Sp}(n)$. We use the index
$i=\1, \2$ as the ${\rm Sp}(1) \cong {\rm SU}(2)$
index and $a = 1, \cdots, 2n$ as the ${\rm Sp}(n)$ index.
Following \cite{Sierra-Townsend} we introduce a vielbein $f_{\mu}{}^{a i}$ and
its inverse $f_{a i}{}^\mu$ to convert between world-index vectors and tangent-space vectors.
They obey the usual conditions
\begin{align}
f_{\mu}{}^{a i} f_{a i}{}^\nu = \delta_\mu{}^\nu~,\qquad
f_{a i}{}^\mu f_{\mu}{}^{b j} = \delta_a{}^b \delta_i{}^j~.
\end{align}
In terms of these one can construct the metric $g_{\mu \nu}$ via
\begin{align}
g_{\mu \nu} := f_\mu{}^{ai} f_{\nu}{}^{bj} \,\eps_{ij} \,\omega_{ab}
\end{align}
where $\eps_{ij}$ and $\omega_{ab}$ are the tangent space 
${\rm Sp}(1)$ and ${\rm Sp}(n)$ metrics, respectively.
In addition one can construct the covariantly constant complex structures
\begin{align}
(J_A)^\mu{}_\nu := -\ri f_{a i}{}^\mu\, (\sigma_A)^i{}_j \,f_\nu{}^{a j}
\end{align}
which obey the quaternionic algebra
\begin{align}
(J_A)^\mu{}_\nu (J_B)^\nu{}_\rho = - \delta_{AB} \delta^\mu{}_\rho + \eps_{ABC} (J_C)^\mu{}_\rho~.
\end{align}
Finally, the $\N=2$ superfield $\Phi^\mu$ is assumed to obey the constraint
\begin{align}
f_\mu{}^{a (i} {\bm\cD}_\beta{}^{j)} \Phi^\mu = f_\mu{}^{a (i} \bar{\bm\cD}_\dbeta{}^{j)} \Phi^\mu = 0~.
\end{align}
This implies the relation
\begin{align}
\frac{1}{\sqrt 2} {\bm\cD}_{\beta j} \Phi^\mu = f_{a j}{}^\mu \chi^a_\beta~, \qquad
\frac{1}{\sqrt 2} \bar{\bm\cD}_{\dbeta j} \Phi^\mu = f_{a j}{}^\mu \bar\chi^a_\dbeta
\end{align}
for Weyl fermion superfields $\chi^a_\beta$ and $\bar\chi^a_\dbeta$, both carrying
${\rm Sp}(n)$ indices.

For the situation we considered in the previous subsection, there is a natural
identification between the ${\rm Sp}(n)$ index $a$ and half of the world indices
$\mu = (a, \bar a)$. The coordinate is $\Phi^\mu = (\Phi^a, \bar\Phi^{\bar a})$
and the corresponding vielbein $f_{\mu}{}^{b j}$ is given by
\begin{gather}
f_{a}{}^{b \1} = 0~, \qquad
f_{a}{}^{b \2} = \delta_a{}^b~, \qquad
f_{\bar a}{}^{b \1} = \ri \,\omega_{\bar a}{}^b ~, \qquad
f_{\bar a}{}^{b \2} = 0~.
\end{gather}
The metric is easily calculable and has the usual form. Similarly, the complex
structures $J_1$, $J_2$, and $J_3$ are given by eqs.
 (\ref{complex_structure1}) and 
(\ref{complex_structure2}).

\section{Supercurrents of the $\N=2$ supersymmetric $\sigma$-model in AdS}\label{section10}
We turn next to a brief discussion of the supercurrent of this
$\N=2$ nonlinear $\sigma$-model. In order to have a self-contained
presentation, we discuss in the first subsection the purely $\N=1$ supercurrent 
of the nonlinear $\sigma$-model. Then drawing upon our previous
work on $\N=2$ supercurrents, we construct in the second subsection
the $\N=2$ supercurrent associated with the model.

\subsection{$\N=1$ supercurrent in AdS}\label{section10.1}
Recall that the most general nonlinear $\sigma$-model action involving only
chiral superfields in AdS can be written
\begin{align}\label{eq_Sads}
S &= \fint K + \cint W + \acint \bar W \eol
	&= \fint \cK~,\qquad \cK = K + \frac{W}{\mu} + \frac{\bar W}{\bar \mu}~.
\end{align}
We would like to discuss the supercurrent associated with this model.

As we showed in \cite{BK-dual}, the most general $\cN=1$ supercurrent multiplet
in AdS is characterized by the conservation equation
\begin{align}\label{eq_AdSgenSC}
\BCD^\dalpha J_{\alpha \dalpha} = \cD_\alpha X -\frac{1}{4} \BCD^2 \zeta_\alpha~,
\end{align}
where $J_{\a\ad} $ is the supercurrent, and $X$ and $\z_\a$ the trace multiplets constrained by 
\bea
\bar \cD_\ad X =0~, \qquad \cD_{(\a} \z_{\b)}=0~.
\eea
The case $\z_\a =0$ corresponds to the Ferrara-Zumino multiplet which is associated with the old minimal 
formulation of AdS supergravity.   On the other hand, the supercurrent with $X=0$ corresponds to 
the non-minimal formulation of AdS supergravity  \cite{BK-dual}.\footnote{As
mentioned in \cite{BK-dual}, there is a certain freedom in choosing how
to define the non-minimal supercurrent, leading to a one-parameter family of
supercurrents. We show in Appendix \ref{appendixD} that these alternative
supercurrents can easily be represented in the form \eqref{eq_AdSgenSC}.}
The specific feature of the AdS supersymmetry is that the trace multiplets 
are defined modulo a gauge transformation of the form
\begin{align}\label{eq_genAdSgauge}
X \rightarrow X + \mu \Lambda~,\qquad
\zeta_\alpha \rightarrow \zeta_\alpha + \cD_\alpha \Lambda~
\end{align}
for chiral $\Lambda$, $\bar \cD_\ad \L =0$. This gauge symmetry allows one to set $X=0$
and so the supercurrent \eqref{eq_AdSgenSC} is completely \emph{equivalent} to the
non-minimal AdS supercurrent.

The general supercurrent \eqref{eq_AdSgenSC} can be modified by an
improvement transformation\footnote{If the gauge $X=0$ is chosen, it is
straightforward to modify the below improvement transformation to maintain
this gauge.}
\begin{subequations}
\bea
J_{\a\ad } &~ \to ~ & J_{\a\ad} +\cD_\a \bar \cD_\ad \bar \U - \bar \cD_\ad \cD_\a \U~, \\
X &~ \to ~ & X+\hf (\bar \cD^2 -4\m) \bar \U~, \\
\z_{\a} &~ \to ~ & \z_\a - 2 \cD_\a( \bar \U + 2 \U)~,
\eea
\end{subequations}
with $\U$ a well-defined complex scalar operator. The important feature of AdS superspace is that 
$\z_\a$ can always be represented as a gradient, 
\bea
\z_\a = \cD_\a \z~, 
\eea
for some globally defined scalar operator $\z$. As a result, the above improvement transformation allows 
us to choose 
\bea
\z_\a =0~.
\label{10.6}
\eea
In other words, the AdS supercurrent can be improved to a Ferrara-Zumino one.
If the condition (\ref{10.6}) holds, then the above improvement transformation reduces to  
\begin{subequations}\label{eq_FZimprovement}
\bea
J_{\a\ad } &~ \to ~ & J_{\a\ad} +2\ri \, \cD_{\a \ad} (\J -\bar \J)~, \\
X &~ \to ~ & X+ 4\m \J +\hf (\bar \cD^2 -4\m) \bar \J~, 
\eea
\end{subequations}
for chiral $\J$, $\bar \cD_\ad \J=0$.

For the model \eqref{eq_Sads} under consideration, the general AdS
supercurrent \eqref{eq_AdSgenSC} is
\begin{align}
J_{\alpha \dalpha} = - \frac{1}{2} K_{a \bar b} \cD_\alpha \phi^a \bar\cD_\dalpha \bar\phi^{\bar b}~,\qquad
\zeta_\alpha = \cD_\alpha K~,\qquad
X = -W~.
\end{align}
The gauge invariance \eqref{eq_genAdSgauge} coincides with the K\"ahler transformation
in AdS, which allows the supercurrent to be recast purely in the non-minimal form
and in terms of the function $\cK$ alone:
\begin{align}
J_{\alpha \dalpha} = - \frac{1}{2} \cK_{a \bar b} \cD_\alpha \phi^a \bar\cD_\dalpha \bar\phi^{\bar b}~,\qquad
\zeta_\alpha = \cD_\alpha \cK~.
\end{align}

For the same model, the Ferrara-Zumino supercurrent is given by
\begin{subequations}
\begin{align}
J_{\alpha \dalpha}
	&= - \frac{1}{6} K_{a \bar b} \cD_\alpha \phi^a \bar\cD_\dalpha \bar\phi^{\bar b}
	+ \frac{\ri}{3} \left(K_a \cD_{\alpha \dalpha} \phi^a - K_{\bar a} \cD_{\alpha \dalpha} \bar\phi^{\bar a}\right) \\
X &= \frac{1}{12} (\bar\cD^2 - 4 \mu) K - W~.
\end{align}
\end{subequations}
The K\"ahler transformation corresponds to a Ferrara-Zumino improvement
transformation \eqref{eq_FZimprovement} for the choice $\Psi = F / 6$, with $F$
a holomorphic function of the target space coordinates $F = F(\phi)$.
If we choose $F = W / \mu$, we find exactly
\begin{subequations}\label{eq_FZcK}
\begin{align}
J_{\alpha \dalpha}' &=
	- \frac{1}{6} \cK_{a \bar b} \cD_\alpha \phi^a \bar\cD_\dalpha \bar\phi^{\bar b}
	+ \frac{\ri}{3} \left(\cK_a \cD_{\alpha \dalpha} \phi^a - \cK_{\bar a} \cD_{\alpha \dalpha} \bar\phi^{\bar a}\right) \\
X' &= \frac{1}{12} (\bar\cD^2 - 4 \mu) \cK
\end{align}
\end{subequations}
with the supercurrent determined entirely by the function $\cK$ alone.

\subsection{$\N=2$ supercurrent in AdS}
Let us now specialize to an $\N=1$ Lagrangian $\cK$ which possesses
$\N=2$ supersymmetry in AdS. We would like to construct its supercurrent.
Recall in \cite{BK2011} we showed that the natural supercurrent arising in $\N=2$
supersymmetric theories AdS takes the form\footnote{In the rigid supersymmetric limit,
the supercurrent (\ref{10.8}) reduces to that constructed in \cite{Butter:2010sc}.}
\begin{align}
\frac{1}{4} (\bar {\bm\cD}_{ij} + 4 \bm{S}_{ij}) {\bm\cJ} = w {\bm T}_{ij} - g_{ij} {\bm \cY}
\label{10.8}
\end{align}
where ${\bm \cJ}$ is a real superfield corresponding to the $\N=2$
supercurrent while ${\bm T}_{ij}$ and ${\bm \cY}$ correspond to contributions
to the $\N=2$ trace multiplet. They obey
\begin{align}
	{\bm \cD}_\alpha^{(k} {\bm T}^{ij)} &= \bar{\bm \cD}_\dalpha^{(k} {\bm T}^{ij)} = 0 ~,
	 \qquad 
	({\bm T}_{ij})^* = {\bm T}^{ij}
	\\
{\bm \cD}^\dalpha_i {\bm \cY} &= 0~, \qquad 
	\frac{1}{4} ({\bm\cD}^{ij} + 4 {\bm S}^{ij}) {\bm\cY} =
	\frac{1}{4} \bar({\bm\cD}^{ij} + 4 {\bm S}^{ij}) \bar {\bm\cY}~.
\end{align}
The first condition says that ${\bm T}_{ij}$ is an $\N=2$ linear multiplet;
the second condition says that ${\bm \cY}$ is an $\N=2$ reduced chiral multiplet.
The constant parameters $w$ and $g_{ij} = \overline{g^{ij}}$ in (\ref{10.8}) obey
\begin{align*}
w \bar w = \sqrt{\frac{1}{2} g^{ij} g_{ij}} = 1~, \qquad g_{ij} \propto \bm S_{ij}~.
\end{align*}
These parameters  can always be chosen so that
the supercurrent equation takes the simpler form
\begin{align}
\frac{1}{4} (\bar {\bm\cD}_{ij} + 4 \bm{S}_{ij}) {\bm\cJ} = {\bm T}_{ij} - \frac{{\bm S}_{ij}}{\bm S} {\bm \cY}~.
\end{align}

One can derive the $\N=1$ supercurrent from the $\N=2$ supercurrent. Within AdS, this
requires the choice ${\bm S}_{\1\2} = 0$. Then the $\N=1$ supercurrent 
takes the Ferrara-Zumino form with
\begin{align}
J_{\alpha \dalpha} = \frac{1}{4} [{\bm\cD}_\alpha^\2, \bar {\bm\cD}_{\dalpha \2}] {\bm \cJ} \vert
	- \frac{1}{12} [\cD_\alpha, \bar \cD_\dalpha] \bm \cJ\vert
\end{align}
where $\vert$ denotes projection to $\N=1$. The corresponding trace
multiplet turns out to be
\begin{align}
X = \frac{1}{3} \bm T_{\1\1}\vert - \frac{2}{3} \frac{\mu}{|\mu|} \bm\cY\vert~.
\end{align}

The $\N=2$ analogue of the improvement transformation \eqref{eq_FZimprovement} is
\begin{subequations}
\begin{align}
\bm \cJ &\rightarrow \bm \cJ + \bm \cR + \bar {\bm \cR} \\
\bm T_{ij} &\rightarrow \bm T_{ij} + \frac{1}{4} (\bar{\bm \cD}_{ij} + 4 \bm S_{ij}) \bar{\bm \cR} \\
\bm \cY &\rightarrow \bm \cY - \bm S \bm \cR
\end{align}
\end{subequations}
where $\bm\cR$ is a reduced chiral superfield. This transformation
allows us to eliminate $\bm \cY$. It is an easy exercise to check that
this leads to the $\N=1$ improvement transformation \eqref{eq_FZimprovement}
with $F = \bm \cR\vert$.

Now we would like to postulate the form of the $\N=2$ supercurrent
for the nonlinear $\sigma$-model in AdS. Because there is a single
function $\cK$ that parametrizes the $\N=1$ action, we will
make a guess that
\begin{align}
\bm \cJ = - \frac{1}{2} \cK
\end{align}
and examine whether this is reasonable. 
First by explicit calculation,
one can check that on-shell
\begin{align}
\frac{1}{4} (\bar{\bm\cD}_{ij} + 4 \bm S_{ij}) \bm \cJ = - \frac{\bm S_{ij}}{\bm S} \bm\cY
\end{align}
where
\begin{align}
\bm \cY = -\frac{|\mu|}{2} \cK_a g^{a \bar b} \cK_{\bar b} + \frac{|\mu|}{2} \cK
	- \frac{\bar\mu}{8 |\mu|} \nabla_{\bar a} \cK_{\bar b}\,
		\bar{\bm\cD}_{\dalpha \1} \bar\Phi^{\bar a} \bar{\bm\cD}^\dalpha_\1 \bar\Phi^{\bar b}
\end{align}
is a reduced chiral superfield on-shell.
We may provide further justification of this $\N=2$ supercurrent
by considering its $\N=1$ reduction. The result is exactly
\eqref{eq_FZcK}.

\section{Concluding remarks}
We have covered a great deal of ground, so let us briefly recap the
main results of this work.
Our focus in sections \ref{section2} through \ref{section4} was $\N=2$ tensor multiplet models.
The key result there was that within an AdS background the $\N=1$
Lagrangian $L(\vf^I, \bar \vf^I, G^I)$ must obey not only the usual Laplace
equation 
\begin{align}
\frac{\pa^2 L}{   \pa \vf^I \pa \bar \vf^J } +\frac{\pa^2 L}{\pa G^I  \pa G^J} =0
\end{align}
but also an additional constraint
\begin{align}
{\rm Re} \left( 
\mu \frac{\partial L}{\partial \vf^I}
     - \mu \frac{\partial^2 L}{\partial \vf^I \partial G^J} G^J
     - 2 \mu \frac{\partial^2 L}{\partial \vf^I \partial \bar\vf^J} \bar\vf^J \right) = 0
\end{align}
arising from the requirement that the models respect the SO(2) invariance
of AdS. It was shown long ago \cite{KLR} how the first condition finds its solution
most naturally expressed in the language of $\N=2$ projective superspace.
We have briefly discussed in section \ref{section5} how the second constraint also
emerges naturally in the same setting and moreover may easily be understood by 
requiring that the AdS Lagrangian arise from a superconformal tensor model where
one of the tensor multiplets has been ``frozen.'' Such an analysis was anticipated
in \cite{KT-M-ads}.

These tensor multiplet models are dual to a subclass of nonlinear $\sigma$-models --
those with $n$ Abelian isometries. In sections \ref{section6} and \ref{section7} we discussed the properties
of the most general nonlinear $\sigma$-models and uncovered a number of fascinating
features, which were previously reported in \cite{Butter:2011zt}.

First, the supersymmetry algebra closes off-shell.
This is a new type of structure that has no analogue in Minkowski space. 
Indeed, in order to have off-shell supersymmetry for general $\cN=2$ nonlinear $\sigma$-models 
in Minkowski space, one has to use the harmonic \cite{GIKOS,GIOS}
or the projective \cite{KLR,LR-projective} superspace approaches
in which the off-shell hypermultiplet realizations involve  an infinite number of auxiliary fields.
In our construction, the hypermultiplet is described using a minimal realization of two ordinary $\cN=1$ chiral superfields with $8+8$ degrees of freedom.
Off-shell supersymmetry is also characteristic of the gauge models for massless higher spin 
$\cN=2$ supermultiplets in AdS constructed in \cite{GKS} using $\cN=1$ superfields. 
Since those theories are linearized, one may argue that their  off-shell supersymmetry is not really impressive.
However, now we have demonstrated that the formulation of the most general nonlinear $\cN=2$ supersymmetric 
$\sigma$-models in terms of $\cN=1$ chiral superfields is also off-shell. 
This gives us some evidence to believe that, say, general $\cN=2$ super Yang-Mills theories in AdS 
possess an off-shell formulation in which the hypermultiplet is realized in terms of two chiral superfields.
If this conjecture is correct, there may be nontrivial implications for quantum effective actions.

The second intriguing feature is that the target space geometry must be hyperk\"ahler
with a special Killing vector $V^\mu$ which rotates the complex structures
\begin{align}
\cL_V J_1 = J_3 \,\sin\theta~, \qquad
\cL_V J_2  = -J_3 \,\cos\theta~, \qquad
\cL_V J_3  = J_2 \,\cos\theta  - J_1\, \sin\theta
\end{align}
where $\theta := \arg\mu$. It necessarily leaves invariant one linear combination
of complex structures, which we denote $J_{\rm AdS}$. There is an underlying physical
reason for this: $\N=2$ AdS supersymmetry requires an SO(2) isometry on the target space
as well as on the covariant derivatives. Most importantly, in the coordinates where $J_{\rm AdS}$
in diagonal, $V^\mu$ is holomorphic and is associated with a Killing potential $\cK$ via
\begin{align}
V^\mu = \frac{1}{2} J_{\rm AdS}{}^\mu{}_\nu \nabla^\nu \cK~.
\end{align}
In the usual coordinates where $J_3$ is diagonal, $\cK$ is the K\"ahler
potential and the AdS Lagrangian for the nonlinear $\sigma$-model.
This discussion shows which hyperk\"ahler manifolds can be used as
target spaces of $\N=2$ nonlinear $\sigma$-models in AdS and gives
the procedure for constructing the $\sigma$-model action.

In section \ref{section8}, we discussed 
$\cN=2$  superconformal $\sigma$-models for which the target spaces are 
hyperk\"ahler cones.
In this class, the additional SO(2) isometry required by AdS is naturally realized
within a larger SU(2) isometry group required by the $\N=2$ superconformal algebra.
Unlike the situation in AdS, however,  the full $\N=2$ superconformal algebra
closes only \emph{on-shell}. This demonstrates the particular importance
of AdS for off-shell closure.

In section \ref{section9}, we described a simple $\N=2$ superfield formulation within AdS
which reproduces all of the features of the $\N=1$ model and explains their
off-shell closure. By allowing the behavior of the SO(2) generator to
be dictated by algebraic consistency rather than target space geometry,
it can be made to perform the same function in AdS that a
central charge does in Minkowski. This formulation also makes clear why
the $\N=2$ superconformal algebra does not close. Finally, in section \ref{section10}, we
discussed briefly the $\N=1$ and $\N=2$ supercurrents of nonlinear $\sigma$-models
in AdS. Such supercurrents have been of interest recently \cite{KS-FI,
DT, K-FI, KS, K-var, Butter-FI, ZH, K-Noether, BK_supercurrent, DS,
Butter:2011zt}.

There remain a number of open questions. The most apparent
lies in the contrast between our understanding of tensor models and that
of general $\sigma$-models. For the former, the projective superspace
approach elegantly solves the problem with a minimum of technical
difficulty. In particular, it clearly explains how the geometry of
tensor models in AdS can be understood as a superconformal tensor
model with an additional ``frozen'' tensor multiplet. For general
$\sigma$-models, we are led to ask two corresponding questions.
First, can we think of general $\sigma$-models in AdS as arising from a
frozen superconformal model? Second, can we use the projective superspace
formulation of  \cite{KT-M-ads} involving polar multiplets to gain insight about the 
geometric features of general $\sigma$-models? In practice, this is a
daunting task requiring the elimination of an infinite number of auxiliary
fields. In Minkowski space, this problem was solved for a large class of
$\cN=2$ supersymmetric $\s$-models on cotangent bundles of Hermitian symmetric spaces 
\cite{GK1,GK2,AN,AKL1,AKL2,KN}. We expect these results can be generalized to the AdS case.

Another open question is how the selection of the SO(2) isometry
reflects the choice of which supersymmetry is made manifest.
Recently, work on nonlinear $\sigma$-models in ${\rm AdS}_5$ with eight
supercharges \cite{BX2011, BaggerLi} has uncovered a similar
SO(2) isometry in a different context. In these works, which are inspired by
brane world scenarios, the AdS$_5$
geometry is foliated with flat four dimensional hypersurfaces perpendicular
to the fifth dimension with metric
\begin{align}\label{eq_AdS5Metric}
\rd s^2 = \re^{-2\lambda z} \eta_{\rm mn} \rd x^{\rm m} \rd x^{\rm n} + \rd z^2~.
\end{align}
Nonlinear $\sigma$-models are then constructed from \emph{flat} 4D $\N=1$
chiral superfields parametrically dependent on $z$. The authors found
that the target space manifold must not only be hyperk\"ahler but also
be equipped with a certain Killing vector which acts as an SO(2) rotation on 
the complex structures. The main difference from our result is that
the Killing vector is \emph{manifestly} holomorphic -- that is, holomorphic
with respect to $J_3$.

The reason for this difference can be explained as follows. 
As shown in \cite{Kuzenko:2008kw}, the 5D $\cN=1$ 
AdS superspace, $\rm AdS^{5|8}$, 
is formulated
in terms of a constant isotriplet $\bm S^{ij}$, which can be interpreted as
the constant torsion tensor of AdS$^{5|8}$; the same tensor $\bm S^{ij}$
occurs in the case of $\rm AdS^{4|8}$. 
The foliation \eqref{eq_AdS5Metric},
which was recently employed in \cite{BX2011, BaggerLi}, was used
(in slightly different form) several years ago in \cite{Kuzenko:2008kw}
to realize $\rm AdS^{5|8}$ as a conformally flat 5D superspace with 
flat 4D $\cN=1$ subspaces. 
In  \cite{Kuzenko:2008kw} the above foliation 
arose naturally upon choosing an SU(2) gauge where
$\bm S^{\1\1} = \bm S^{\2\2} = 0$. 

What about 
AdS$^{4|8}$? 
Again one has
a constant isotriplet $\bm S^{ij}$. The choice $\bm S^{\1\2} = 0$ allows one to naturally
select an AdS$^{4|4}$ subspace of AdS$^{4|8}$
simply by turning off the second set of Grassmann
coordinates  \cite{Kuzenko:2008kw}.\footnote{See for example the discussion in Appendix \ref{appendixB2}.}
This is the choice we have made in this work, and it should come as
no surprise that the allowed target space geometry differs from that of \cite{BX2011, BaggerLi}
simply by a different choice of which complex structure is manifest. 
Instead of this choice, we could take $\bm S^{\1\1} = \bm S^{\2\2} = 0$ in AdS$^{4|8}$
by applying a rigid SU(2) transformation.
(The same SU(2)
which acts on $\bm S^{ij}$ should act on the two-sphere of complex structures.)
In analogy to five dimensions,
we expect that this choice be naturally respected by a foliation
involving 3D $\cN=2$ Minkowski superspace.
The allowed target space
geometries should again be hyperk\"ahler manifolds with a special SO(2) Killing
vector but with the preferred complex structure matching the manifest one as in
\cite{BX2011, BaggerLi}. It would be interesting to see this borne
out by an explicit construction.
\\

\noindent
{\bf Acknowledgements:}\\
This work is supported in part by the Australian Research Council 
and by a UWA Research Development Award. D.B. would like to thank
Jon Bagger for correspondence.

\appendix

\section{$\cN=1$ (superconformal) Killing vector fields}
\label{Appendix_A}
In this appendix we discuss superconformal and isometry transformations 
of the $\cN=1$ AdS superspace, AdS$^{4|4}$, 
which is a maximally symmetric solution of old minimal 
supergravity with a cosmological term (see \cite{BK} for more details regarding the
$\cN=1$ AdS supergeometry). 
The corresponding covariant derivatives,\footnote{We
follow the notation and conventions adopted in \cite{BK}, with the only exception that  we use lower
case {\it Roman} letters for tangent-space vector indices.} 
\bea
\cD_{\rm A}=(\cD_{\rm a}, \cD_\a,{\bar \cD}^\ad)=E_{\rm A}{}^{M}\pa_M+\hf\f_{\rm A}{}^{\rm bc}M_{\rm bc}
\equiv E_{\rm A} +\f_{\rm A}~,
\eea
obey the following (anti-)commutation relations:
\begin{subequations}\label{N=1-AdS-algebra}
\bea
\{\cD_\a,\cD_\b\}&=&
-4\bar{\mu}M_{\a\b}~, \qquad \qquad
\{ \bar \cD_\ad , \bar \cD_\bd \} = 4\m \bar M_{\ad \bd}~,
\label{N=1-AdS-algebra-1}\\
\{\cD_\a,{\bar \cD}_\bd\}&=&-2\ri(\s^{\rm c})_{\a \bd}\cD_{\rm c} \equiv -2\ri \cD_{\a\bd}
~,~~
\label{N=1-AdS-algebra-2}
\\
{[}\cD_{\rm a},\cD_\b{]}&=&
-\frac{\ri}{ 2}   {\bar \mu} ({\s}_{\rm a})_{\b\gd} {\bar \cD}^\gd~,\qquad
{[}\cD_{\rm a}, \bar \cD_\bd{]}=
\frac{\ri}{ 2}   { \mu} ({\s}_{\rm a})_{\g \bd} { \cD}^\g~,
\label{N=1-AdS-algebra-3}
\\
{[}\cD_{\rm a},\cD_{\rm b}{]}&=&-| { \mu}|^2 M_{\rm ab}~,~~~~~~
\label{N=1-AdS-algebra-4}
\eea
\end{subequations}
with $\m$ a complex non-vanishing parameter.  One can think of  $\m$ as
a square root of the curvature of the anti-de Sitter space. 
The Lorentz generators with vector indices ($M_{\rm ab}=-M_{\rm ba}$) and spinor indices
($M_{\a\b}=M_{\b\a}$ and ${\bar M}_{\ad\bd}={\bar M}_{\bd\ad}$) are related to each other 
by the rule:
$$
M_{\rm ab}=(\s_{\rm ab})^{\a\b}M_{\a\b}-(\tilde{\s}_{\rm ab})^{\ad\bd}\bar{M}_{\ad\bd}~,~~~
M_{\a\b}=\hf(\s^{\rm ab})_{\a\b}M_{\rm ab}~,~~~
\bar{M}_{\ad\bd}=-\hf(\tilde{\s}^{\rm ab})_{\ad\bd}M_{\rm ab}~.
$$ 
The Lorentz generators act on the spinor covariant derivative by the rule:
\bea
{[}M_{\a\b},\cD_{\g} {]}
&=& \hf ( \ve_{\g \a}\cD_{\b} + \ve_{\g \b}\cD_{\a}) ~,\qquad
{[}\bar{M}_{\ad\bd}, \cDB_{\gd}{]}= \hf ( \ve_{\gd\ad}\cDB_{\bd} +  \ve_{\gd\bd}\cDB_{\ad})~,
\label{generators}
\eea
while 
${[}M_{\a\b},\cDB_{\gd}{]}=
{[}\bar{M}_{\ad\bd},\cD_{\g}{]}=0.$

In accordance with \cite{BK}, 
a real vector field, $\x^{\rm A} = (\x^{\rm a} , \x^\a , \bar \x_\ad )$, on AdS$^{4|4}$
is said to be superconformal Killing if the corresponding infinitesimal   coordinate transformation 
can be accompanied by special Lorentz and super-Weyl transformations such that 
the covariant derivatives remain  unchanged. In terms of the first-order differential operator 
\bea
\x := \x^{\rm a}\cD_{\rm a} + \x^\a \cD_\a + \bar \x_\ad \bar \cD^\ad ~,
\eea
the above definition is equivalent to the condition
\bea
[ \x + \hf \l^{\rm cd} M_{\rm cd}, \cD_\a ] + (\bar \s -\hf \s) \cD_\a +(\cD^\b \s) M_{\a\b} =0~.
\label{N=1scKv1}
\eea
This same transformation acts on any \emph{primary} tensor superfield $U$ as
\begin{align}\label{eq_deltaU}
\delta U = -\xi U - \frac{1}{2} \lambda^{\rm cd} M_{\rm cd} U
	- \frac{\Delta}{2} (\sigma+\bar\sigma) U
	- \frac{3 w}{4} (\sigma - \bar\sigma) U
\end{align}
where $\Delta$ is the conformal dimension of $U$ and $w$ is its chiral
weight.
If $U = \Phi$ is chiral, then we must have
$2 \Delta = 3 w$ and the above expression can be rewritten
\begin{align}
\delta \Phi
	= -\xi \Phi - \frac{1}{2} \lambda^{cd} M_{\rm cd} \Phi
	- \sigma \Delta \Phi
\end{align}

Let us now solve for the relations that the $\N=1$ AdS parameter must obey.
Making use of the (anti-)commutation relations (\ref{N=1-AdS-algebra}) gives
\begin{subequations}\label{N=1scKv2}
\bea
\cD_\a \x^{\rm b} - 2\ri (\s^{\rm b})_{\a \bd} \bar \x^\bd &=&0~, \label{A.6a} \\
-\frac{\ri}{2}\bar \m \x^{\rm b} (\s_{\rm b})_{\a \bd} - \cD_\a \bar \x_\bd &=&0~, \label{A.6b}  \\
\cD_\a \x^\b -\l_\a{}^\b - \d_\a{}^\b (\bar \s -\hf \s)
&=&0~, \label{A.6c} \\
\cD_\a \bar \l^{\bd \gd} &=&0~, \label{A.6d} \\
2\bar \m (\d_\a{}^\b \x^\g + \d_\a{}^\g \x^\b) +\cD_\a \l^{\b \g}
-\hf (\d_\a{}^\b \cD^\g \s + \d_\a{}^\g \cD^\b \s) 
&=&0~.\label{A.6e} 
\eea
\end{subequations}
Just as in a Minkowski background \cite{BK}, this set of equations can be solved
in terms of the single parameter $\xi_{\alpha \dalpha}$ obeying the so-called
``master equation''
\begin{align}\label{eq_N1master}
\cD_{(\beta} \xi_{\alpha) \dalpha} = \bar\cD^{(\dbeta} \xi^{\dalpha) \alpha} = 0~.
\end{align}
Note that this equation implies
\begin{align}\label{eq_N0master}
\cD_{\rm a} \xi_{\rm b} + \cD_{\rm b} \xi_{\rm a} = \frac{1}{2} \eta_{\rm ab} \cD_{\rm c} \xi^{\rm c}
\end{align}
which guarantees that the lowest component of $\xi^{\rm a}$ is a conformal Killing vector.

The other parameters are given by
\begin{gather} \label{eq_N1params1}
\xi_\alpha = \frac{\ri}{8} \bar \cD^\dbeta \xi_{\alpha \dbeta}~,\qquad
\lambda_{\rm a b} = \cD_{\rm [a} \xi_{\rm b]}~, \\
\sigma = -\frac{\ri}{24} \cD_\alpha \bar \cD_\dalpha \xi^{\dalpha \alpha}
	- \frac{\ri}{12} \bar\cD_\dalpha \cD_\alpha \xi^{\dalpha \alpha}~.\label{A.15}
\end{gather}
By construction, the super-Weyl parameter $\s$ must be chiral, 
\be
\bar \cD_\ad \s =0~.
\ee 
This property follows from the relation \eqref{A.15} and the master equation \eqref{eq_N1master}.
We note that the dilatation and chiral rotation parameters are given respectively by
\begin{align}
\textrm{Re }\sigma &= -\frac{1}{8} \cD_{\alpha \dalpha} \xi^{\dalpha \alpha}
	= \frac{1}{4} \cD_{\rm a} \xi^{\rm a}~, \\
\frac{3}{2} \,\textrm{Im }\sigma &= \frac{1}{32} [\cD_\alpha, \bar \cD_\dalpha] \xi^{\dalpha \alpha}~.
\end{align}
It can be shown that the superconformal Killing vector fields generate the $\cN=1$
superconformal group $\rm SU(2,2|1)$.

Any superconformal Killing vector field $\x$ with  the additional property
\be\label{eq_N1MinKilling}
\s=0 \implies \cD_\alpha \bar\cD_\dalpha \xi^{\dalpha \alpha} = 0
\ee
is called a Killing vector field of the $\cN=1$ AdS superspace. All Killing vector fields
are characterized by the property 
\bea
[ \x + \hf \l^{\rm cd} M_{\rm cd}, \cD_{\rm A} ] =0~. \label{N=1Kv1}
\eea
This master equation implies the relations 
\begin{subequations}\label{N=1Kv2}
\bea
\cD_{(\a} \x_{\b)\bd}&=&0~, \qquad  {\bar \cD}^\bd \x_{\a\bd}   + 8\ri\x_\a=0~,\\
\cD_\a\x^\a&=&0~,
\qquad
{\bar \cD}_\ad \x_\a  + {\ri\over 2}{\mu}\x_{\a\ad}  =0~, \\
\l_{\a\b}&=&\cD_\a\x_\b~,
\eea
\end{subequations}
and these equations follow from \eqref{eq_N1params1} and \eqref{eq_N1master}
using the additional constraint \eqref{eq_N1MinKilling}.
Since \eqref{eq_N1MinKilling} implies $\cD_{\rm a} \xi^{\rm a} = 0$, it follows that
the conformal Killing equation \eqref{eq_N0master} reduces to
\bea
\cD_{\rm a} \x_{\rm b} + \cD_{\rm b} \x_{\rm a}=0~.
\eea
The AdS Killing vector fields generate the isometry group of the $\cN=1$ AdS superspace, 
$\rm OSp(1|4)$.

\section{$\cN=2$ (superconformal) Killing vector fields}
\label{appendixB}
The four-dimensional $\cN=2$ AdS superspace 
$$
{\rm AdS^{4|8} } := \frac{{\rm OSp}(2|4)}{{\rm SO}(3,1) \times {\rm SO} (2)}
$$
can be realized as a maximally symmetric geometry that originates within 
the superspace formulation of $\cN=2$ conformal supergravity developed in  \cite{KLRT-M1}.
Assuming the superspace is parametrized  by local bosonic ($x$) and fermionic ($\q, \bar \q$) 
coordinates  ${\bm z}^{\cM}=(x^{m},\q^{\mu}_{\imath},{\bar \q}_{\dot{\mu}}^{\imath})$
(where $m=0,1,\cdots,3$, $\mu=1,2$, $\dot{\mu}=1,2$ and  $\imath=\1,\2$),
the corresponding covariant derivatives 
\bea
{\bm \cD}_{\cA} =({\bm \cD}_{\rm a}, {\bm \cD}_{{\a}}^i, {\bm \cDB}^\ad_i)
= {\bm E}_{\cA}{}^\cM \pa_\cM + \hf  {\bm \f}_{\cA}{}^{\rm bc} M_{\rm bc} 
+ {\bm \f}_{\cA} {\bm S}^{ij} J_{ij}~, \qquad i,j =\1 , \2
\eea
obey the algebra \cite{KLRT-M1,KT-M-ads}:
\begin{subequations}\label{N=2acd}
\begin{gather}
\{  {\bm \cD}_\a^i, {\bm \cD}_\b^j  \} = 4 {\bm S}^{ij} M_{\alpha \beta}
     + 2 \ve_{\alpha \beta} \ve^{ij} {\bm S}^{kl} J_{kl}~, \qquad
\{ {\bm \cD}_\a^i, \bar {\bm \cD}_{\ad j }  \} = -2\ri  \delta^i_j {\bm \cD}_{\alpha \dalpha}~, \\
[  {\bm \cD}_{ \a \ad },  {\bm \cD}_\beta^i] = - \ri \ve_{\a \b} {\bm S}^{ij} \bar{\bm \cD}_{\dalpha j}~, 
\qquad [{\bm \cD}_{\rm a}, {\bm \cD}_{\rm b}]   = -{\bm S}^2 M_{\rm ab}~,
\end{gather}
\end{subequations}
where ${\bm S}^{ij} $ is a {\it covariantly constant} and {\it constant}
 real iso-triplet, ${\bm S}^{ji} = {\bm S}^{ij}$, 
 $\overline{ {\bm S}^{ij}} = {\bm S}_{ij} =\ve_{ik}\ve_{jl}{\bm S}^{kl}$, 
and  ${\bm S}^2 :=  \frac{1}{2} {\bm S}^{ij} {\bm S}_{ij} $. 
The SU(2) generators, $J_{kl}$, act on the spinor covariant derivatives 
by the rule:
\bea
[J_{kl} , {\bm \cD}_\a{}^i ] = -\hf ( \d^i_k  {\bm \cD}_{\a l} +\d^i_l  {\bm \cD}_{\a k})~.
\eea
This superspace proves to be a conformally flat solution to the equations of motion 
for $\cN=2$ supergravity with a cosmological term \cite{BK2011}.

\subsection{$\cN=2$ superconformal Killing vector fields}
We now turn to studying the algebra of superconformal Killing vector fields of  ${\rm AdS^{4|8} }$. 
By definition,  a real vector field,
${\bm \x}^{\cA} = ({\bm \x}^{\rm a} , {\bm \x}^\a_i , \bar {\bm \x}_\ad^i )$, on AdS$^{4|8}$
is said to be superconformal Killing if the corresponding infinitesimal   coordinate transformation 
can be accompanied by special local Lorentz,  local SU(2) and super-Weyl transformations such that 
the covariant derivatives remain  unchanged, 
\bea
[{\bm \x} + \hf {\bm \l}^{\rm cd} M_{\rm cd}+{\bm \l}^{kl} J_{kl}, {\bm \cD}_\a^i  ] 
+ \hf \bar {\bm \s}  {\bm \cD}_\a^i +({\bm \cD}^{\b i} {\bm \s}) M_{\a\b} 
-({\bm \cD}_{\a j} {\bm \s} ) J^{ij} =0~,
\label{N=2scKv1}
\eea
where 
\bea
{\bm \x} := {\bm \x}^{\rm a}{\bm \cD}_{\rm a} + {\bm \x}^\a_i {\bm\cD}_\a^i + \bar {\bm \x}_\ad^i \bar {\bm\cD}^\ad_i ~.
\label{B.5}
\eea
This same transformation acts on any $\N=2$ \emph{primary} tensor superfield $\bm U$ as
\begin{align}\label{eq_deltaU2}
\delta \bm U = -\bm\xi \bm U - \frac{1}{2} \bm\lambda^{\rm cd} M_{\rm cd} \bm U
	- \bm\lambda^{jk} J_{jk} \bm U
	- \frac{\Delta}{2} (\bm\sigma + \bar{\bm\sigma}) \bm U
	- \frac{w}{4} (\bm\sigma - \bar{\bm\sigma}) \bm U
\end{align}
where $\Delta$ is the conformal dimension of $\bm U$ and $w$ is its chiral
weight.
If $\bm U = \Phi$ is chiral, then we must have $2 \Delta = w$ and the
above expression can be rewritten
\begin{align}
\delta \Phi
	= -\bm\xi \Phi - \frac{1}{2} \bm\lambda^{\rm cd} M_{\rm cd} \Phi
	- \bm\lambda^{jk} J_{jk} \Phi
	- \bm\sigma \Delta \Phi
\end{align}

It follows from (\ref{N=2scKv1}) that
\begin{subequations}\label{N=2scKv2}
\bea
{\bm \cD}_\a^i {\bm \x}^{\rm b} - 2\ri ( \s^{\rm b})_{\a \bd} \bar {\bm \x}^{\bd i} &=&0~, \label{B.6a} \\
\frac{\ri}{2} {\bm S}^{ij} {\bm \x}_{\a \bd} - {\bm \cD}_\a^i \bar {\bm \x}_\bd^j &=&0~, \label{B.6b}  \\
-{\bm \cD}_\a^i {\bm \x}^\b_j  + \d^i_j {\bm \l}_\a{}^\b +\d_\a^\b{\bm \l}^i{}_j
+ \hf \d_\a^\b \d^i_j \,\bar {\bm \s} 
&=&0~, \label{B.6c} \\
{\bm \cD}_\a^i \bar {\bm \l}^{\bd \gd} &=&0~, \label{B.6d} \\
2{\bm S}^{ij} (\d_\a^\b {\bm \x}^\g + \d_\a^\g {\bm \x}^\b) - {\bm \cD}_\a^i {\bm \l}^{\b \g}
+ \hf (\d_\a^\b {\bm \cD}^{\g i} {\bm \s} + \d_\a^\g {\bm \cD}^{\b i} {\bm \s})  &=&0~,\label{B.6e} \\
2{\bm \x}_\a^i S^{kl} -{\bm \cD}_\a^i {\bm \l}^{kl} 
+\hf (\ve^{ik} {\bm \cD}_\a^l {\bm \s} +\ve^{il} {\bm \cD}_\a^k {\bm \s})&=&0~. \label{B.6f}
\eea
\end{subequations}

As in the $\N=1$ case, all the parameters above can be expressed in terms of the
single parameter $\bm \xi_{\alpha \dalpha}$, which obeys the master equation
\begin{align}\label{eq_N2master}
\bm\cD_{(\beta}^i \bm\xi_{\alpha) \dalpha} = \bar{\bm\cD}^{(\dbeta}_j \bm \xi^{\dalpha) \alpha} = 0~.
\end{align}
The other parameters are given by
\begin{gather}\label{eq_N2params1}
\bm\xi_{\alpha i} = \frac{i}{8} \bar {\bm\cD}^\dbeta_i \bm\xi_{\alpha \dbeta}~,\qquad
\bm\lambda_{\rm a b} = \bm\cD_{\rm [a} \bm \xi_{\rm b]}~, \\  \label{eq_N2params2}
\bm\lambda^{ij} = -\frac{\ri}{16} {\bm\cD}_\alpha^{(i} \bar{\bm\cD}_\dalpha^{j)} \bm\xi^{\dalpha \alpha}~, \qquad
\bm\sigma = -\frac{\ri}{16} \bar{\bm\cD}_{\dalpha j} \bm\cD_\alpha^j \bm\xi^{\dalpha \alpha}
\end{gather}
In accordance with \cite{KLRT-M1}, the super-Weyl parameter $\bm\s$ must be chiral, 
\be
\bar {\bm \cD}_{\ad i}  {\bm \s} =0~.
\ee 
This property follows from the master equation \eqref{eq_N2master}.
Note that the dilatation and U(1) parameters are given respectively by
\begin{align}
\textrm{Re }\bm \sigma &= -\frac{1}{8} \bm \cD_{\alpha \dalpha} \bm \xi^{\dalpha \alpha}
	= \frac{1}{4} \bm \cD_{\rm a} \bm \xi^{\rm a}~, \\
\frac{1}{2} \textrm{Im } \bm\sigma &=
	-\frac{1}{64} [\bm \cD_\alpha^j, \bar{\bm \cD}_{\dalpha j}] \bm\xi^{\dalpha \alpha}~.
\end{align}

The superconformal equations (\ref{N=2scKv2}) have a number of implications
of which we now list only a few, the most important for our subsequent analysis. From eq. (\ref{B.6f}) 
we deduce
\bea 
2{\bm \x}_\a^{(i}{\bm S}^{kl)} -{\bm \cD}_\a^{(i}{\bm \l}^{kl)} &=&0~.
\label{B.9}
\eea
In conjunction with the first relation in \eqref{eq_N2params2}, this leads to
\bea
{\bm \cD}^{ \a(i} {\bm \cD}_\a^{j} {\bm \l}^{kl)} +4 {\bm S}^{(kl} {\bm \l}^{ij)}=0~.
\eea
Again from (\ref{B.6f}) we derive 
\bea
4{\bm S}^2 {\bm \x}_{\a i} - \ve_{ij}{\bm \cD}_{\a }^j {\bm \l}^{kl} {\bm S}_{kl}  
+ {\bm S}_{ik } {\bm \cD}^k_\a {\bm \s}=0~.
\label{B.11}
\eea
Each of these can be checked against the explicit solutions given above
in terms of the parameter $\bm\xi_{\alpha \dalpha}$ obeying the master
equation \eqref{eq_N2master}.

The above analysis is a natural extension of that given in the $\cN=2$ super-Poincar\'e case 
in \cite{KT} (see also \cite{Park}).

The superconformal Killing vector fields of the 
 $\cN=2$ AdS superspace, ${\rm AdS^{4|8} } $, prove to generate 
the supergroup $\rm SU(2,2|2)$, which is also the superconformal group of the 
$\cN=2$ Minkowski superspace, ${\mathbb R}^{4|8}$.
The superspaces ${\rm AdS^{4|8} } $ and  ${\mathbb R}^{4|8}$ have the same superconformal group, 
since they are conformally related or, equivalently,  since ${\rm AdS^{4|8} } $ is conformally flat 
(see, e.g., \cite{KT-M-ads,BILS}).

\subsection{$\cN=1$ reduction} 
\label{appendixB2}
By applying a rigid SU(2) rotation to the covariant derivatives, 
it is always possible to choose the iso-triplet ${\bm S}^{ij}$ such that 
\be
{\bm S}^{\1 \2} =0~.
\label{S12}
\ee
We denote the non-vanishing components of ${\bm S}^{ij}$ as 
\bea
{\bm S}^{\1\1}= {\bm S}_{\2\2} = -\bar \m~, \qquad {\bm S}^{\2\2} = {\bm S}_{\1\1} = -\m~.
\label{S11S22}
\eea
With the choice (\ref{S12}) made, the above relations imply the following conditions:
\begin{subequations}
\bea
{\bm \cD}_\a^{\1} \bar{\bm \x}_\bd^{\2} &=& 0 \quad \Longrightarrow \quad 
\bar{\bm \cD}_{\ad \1} {\bm \x}^\b_{\2} = 0~,
\label{B.14a} \\
\big[ ( {\bm \cD}^{\1} )^2 - 4 \bar \m \big] {\bm \l}^{\1\1} &=&0  \quad \Longrightarrow \quad
\big[ ( \bar{\bm \cD}_{\1} )^2 - 4  \m \big] {\bm \l}^{\2\2} =0~,  \label{B.14b} \\
{\bm \cD}_\a^{\1} ( {\bm \l}^{\1\2} -\hf {\bm \s} ) &=&0  \quad \Longrightarrow \quad 
\bar{\bm \cD}_{\ad \1} ( {\bm \l}^{\1\2} +\hf \bar{\bm \s} ) =0~.
\label{B.14c} 
\eea
\end{subequations}
${}$From (\ref{B.9}) we derive 
\bea
{\bm \x}_{\a \2} = - \frac{\m}{2|\m|^2} {\bm \cD}_\a^{\1} {\bm \l}^{\1\1}
\quad \Longrightarrow \quad
\bar{\bm \x}_\ad^{\2} = - \frac{\bar \m}{2|\m|^2} \bar{\bm \cD}_{\ad \1} {\bm \l}^{\2\2}~.
\label{B-rho}
\eea
Then eq. (\ref{B.14a}) tells us that 
\bea
\bar{\bm \cD}_{\ad \1} {\bm \cD}_\a^{\1} {\bm \l}^{\1\1} = 0 \quad \Longrightarrow \quad
 {\bm \cD}_\a^{\1}   \bar{\bm \cD}_{\ad \1} {\bm \l}^{\2\2} = 0~.
\label{B.16}
\eea
Eq.  (\ref{B.11}) also implies
\bea
4 |\m|^2 {\bm \x}_{\a \2} - {\bm \cD}_{\a }^{\1} (\m {\bm \l}^{\1\1} +\bar \m {\bm \l}^{\2\2} )   
-\bar \m {\bm \cD}^{\2}_\a {\bm \s}=0~.
\label{B.11-2}
\eea

We are interested in  an $\cN=1$ AdS reduction of the $\cN=2$ 
superconformal Killing vector fields. In other words, we would like to 
derive those transformations in $\cN=1$ AdS superspace which are generated 
by an arbitrary $\cN=2$ superconformal Killing vector field.\footnote{In the case of $\cN=2$ AdS
Killing vector fields, their $\cN=1$ reduction was carried out in \cite{KT-M-ads}.}
Given a tensor superfield  $\bm U(x,\q_i,\bar \q^i)$ in $\cN=2$ 
AdS superspace, we introduce
its $\cN=1$ projection
\bea
U = \bm U | := \bm U(x,\q_i,\bar \q^i)|_{\q_\2={\bar \q}^\2=0}~
\eea
in a {\it special coordinate system} specified below. 
Given a gauge-covariant operator of the form ${\bm \cD}_{\cA_1} \dots {\bm \cD}_{\cA_n}$, 
its projection $\big({\bm \cD}_{\cA_1} \dots {\bm \cD}_{\cA_n}\big)\big|$ is defined as follows:
\bea
 \Big( \big( {\bm \cD}_{\cA_1} \dots {\bm \cD}_{\cA_n}\big)\big| {\bm U} \Big) \Big|
:=  \big({\bm \cD}_{\cA_1} \dots {\bm \cD}_{\cA_n}\bm U\big)\big|~, 
\eea
with $\bm U$ an arbitrary tensor superfield.
The conceptual possibility to have a well-defined $\cN=2 \to \cN=1$ AdS superspace reduction
was noticed in \cite{KT-M-ads}.  
Specifically, with the choice (\ref{S12}) and (\ref{S11S22}),  
it follows from  (\ref{N=2acd}) that
the operators $({\bm \cD}_{\rm a},\, {\bm \cD}_\a^\1,\, {\bm \cDB}^\ad_\1)$ form a closed  algebra
which is isomorphic to that of the covariant derivatives   for 
$\cN=1$ AdS superspace, eq. (\ref{N=1-AdS-algebra}). 

As argued in \cite{KT-M-ads}, 
the freedom to perform general coordinate, local Lorentz and  U(1) 
transformations can be used  to choose the gauge
\bea
{\bm \cD}^\1_\a|=\cD_\a~, \qquad {\bm \cDB}^\ad_\1|=\bar \cD^\ad~,
\eea
with $\cD_{\rm A} = (\cD_{\rm a} , \cD_\a ,\bar \cD^\ad)$ the covariant derivatives
for $\cN=1$ AdS superspace introduced in Appendix \ref{Appendix_A}.
In such a coordinate system,
the operators ${\bm \cD}_\a^\1|$ and ${\bm \cDB}_{\ad \1}|$ do not involve any 
partial derivatives with respect to $\q_\2$ and ${\bar \q}^\2$, 
and therefore, for any positive integer $k$,  
it holds that $\big( {\bm \cD}_{\hat{\a}_1} \cdots  {\bm \cD}_{\hat{\a}_k} \bm U \big)\big|
= {\bm \cD}_{\hat{\a}_1}| \cdots  {\bm \cD}_{\hat{\a}_k}| U$, 
where $ {\bm \cD}_{\hat{\a}} :=( {\bm \cD}_\a^\1, {\bar {\bm \cD}}^\ad_\1)$ and $U$ is a tensor superfield.
We therefore obtain
\be
{\bm \cD}_{\rm a} | = \cD_{\rm a}~.
\ee

Introduce the $\cN=1$ projection of the  $\cN=2$ superconformal Killing vector  (\ref{B.5}):
\bea
{\bm \x }| = \x + {\bm \x}^\a_{\2} | {\bm \cD}_\a^{\2} | + \bar{\bm \x}_\ad^{\2} | \bar{\bm \cD}^\ad_{\2} |
\equiv  \x + \r^\a {\bm \cD}_\a^{\2} | + \bar \r_\ad \bar{\bm \cD}^\ad_{\2} |
~,
\label{B.22}
\eea
where we have denoted
\bea
\x:= {\bm \x}^{\rm a} |\cD_{\rm a} + {\bm \x}^\a_{\1} | {\cD}_\a + \bar{\bm \x}_\ad^{\1} | \bar{\cD}^\ad
\equiv  { \x}^{\rm a} \cD_{\rm a} + { \x}^\a {\cD}_\a + \bar{\x}_\ad \bar{\cD}^\ad
~.
\label{B.23}
\eea
${}$From (\ref{B-rho}) we obtain 
\bea
\r_\a = \cD_\a \r~, \qquad \r:= - \frac{\m}{2|\m|^2} {\bm \l}^{\1\1}|~.
\label{rho-deff}
\eea
In accordance with (\ref{B.14b}) and (\ref{B.16}), the parameter $\r$ obeys the constraints 
\bea
(\cD^2 -4\bar \m) \r =0~, \qquad \bar \cD_\ad \cD_\a \r =0~.
\label{rho-constraints}
\eea
One can show that these equations also imply\footnote{The proof that 
$\cD_\a \bar \cD_\ad \r =0$ follows from analyzing the condition
$\bar\cD_\dalpha(\cD^2 - 4 \mu)\rho=0$. One finds that
$\bar\cD_\dalpha \rho = \ri \cD_{\alpha \dalpha} \cD^\alpha \rho / 2 \bar\mu$ 
from which the new constraint follows. Similarly, one can
show $(\bar\cD^2 -4 \m) \r =0$ by analyzing
$\bar\cD^2 (\cD^2 - 4 \mu)\rho=0$.}
\bea
(\bar\cD^2 -4 \m) \r =0~, \qquad \cD_\a \bar \cD_\ad \r =0~,\qquad \cD_{\a \ad} \r = 0~.
\label{rho-constraints2}
\eea
This implies that $\rho$ can be decomposed into real and imaginary parts
which each obey \eqref{rho-constraints} and \eqref{rho-constraints2}.

It turns out that $\x$ in (\ref{B.22}) and (\ref{B.23}) is an $\cN=1$ superconformal Killing vector field. 
The simplest way to see that is to rewrite eq. (\ref{N=2scKv1}) in the form 
\bea
0&=&  \big[{\bm \x} + \hf {\bm \l}^{\rm cd} M_{\rm cd} , {\bm \cD}_\a^{\1}  \big] 
+(\hf \bar{\bm \s} - {\bm \l}^{\1\2}) {\bm \cD}_\a^{\1}  
+({\bm \cD}^{\b \1} {\bm \s}) M_{\a\b} \non \\
&& + {\bm \l}^{\1\1} {\bm \cD}_\a^{\2}
+\big( {\bm \cD}_\a^{ \2} {\bm \s}  - {\bm \cD}_\a^{ \1} {\bm \l}^{\2\2} \big)J_{\2\2}
 - {\bm \cD}_\a^{ \1} {\bm \l}^{\1\1} J_{\1\1}~,
\label{B.26}
\eea
where we have used eq. (\ref{B.14c}). 
Upon $\cN=1$ projection, 
this master equation can be seen to lead to completely decoupled conditions
on the $\cN=1$ parameters  $\x$ and $\r^\a$ in (\ref{B.22}) and (\ref{B.23}).
The vector field $\x$ obeys the $\cN=1$ superconformal Killing equation
(\ref{N=1scKv1}) if we identify
\bea
\l^{\rm ab} := {\bm \l}^{\rm ab}|~, \qquad \s := {\bm \s}| ~, \qquad {\bm \l}^{\1\2} |= \hf (\s-\bar \s)~.
\label{B.27} 
\eea
The second and third relations hold in the absence of a shadow chiral rotation.

Now we are in a position to list transformations of matter superfields in $\cN=1$ AdS superspace 
which are generated by the $\cN=2$ superconformal Killing vector $\bm \x$. They are:

\begin{enumerate}
\bfseries \item \mdseries
An $\cN=1$ superconformal transformation. It corresponds to the choice 
\bea
{\bm \l}^{\1\1} |=0~, \qquad  {\bm \l}^{\1\2} |= \hf ({\bm \s}-\bar {\bm \s})|~.
\eea
One can see that all terms in  the second line of (\ref{B.26}) vanish. 

\bfseries \item \mdseries
An extended superconformal transformation. It corresponds to the choice 
\bea
\x =0~, \qquad {\bm \s}|=0~, \qquad  {\bm \l}^{\1\2} |= 0~.
\eea
It is described by the complex parameter $\r$ which is defined by (\ref{rho-deff})
and obeys the constraints
(\ref{rho-constraints}).

\bfseries \item \mdseries
A shadow chiral rotation. 
It corresponds to the choice 
\bea
\x = 0~, \qquad {\bm \l}^{\1\1} |=0~, \qquad  {\bm \l}^{\1\2} |= \hf \bar{\bm \s}| = - \hf {\bm \s}| ={\rm const}~.
\eea
This transformation does not act on the coordinates of $\cN=1$ AdS superspace.
\end{enumerate}

It is possible to give an explicit solution for the superconformal
$\N=2$ Killing vector in terms of the three independent $\N=1$ superfields
making up the $\N=2$ superfield $\bm\xi_{\alpha \dalpha}$.
In addition to the covariant derivatives, we must identify the $\N=1$
reduction of the other $\N=2$ superconformal generators.
The dilatation operator, which we denote $\mathbb D$, obeys
\begin{align}
[\mathbb D, {\bm \cD}_\alpha^i] = \frac{1}{2} {\bm \cD}_\alpha^i~, \qquad
[\mathbb D, \bar{\bm \cD}^\dalpha_i] = \frac{1}{2} \bar{\bm \cD}^\dalpha_i~, \qquad
\end{align}
for $\N=2$ covariant derivatives and
\begin{align}
[\mathbb D, {\cD}_\alpha] = \frac{1}{2} {\cD}_\alpha~, \qquad
[\mathbb D, \bar{\cD}^\dalpha] = \frac{1}{2} \bar{\cD}^\dalpha~, \qquad
\end{align}
for $\N=1$ covariant derivatives. Clearly these definitions coincide.
The Lorentz generator $M_{ab}$ should also work the same way in $\N=1$ and $\N=2$.
However, the $\N=1$ ${\rm U}(1)_R$ generator $\mathbb J$ is identified with a certain linear
combination of the ${\rm U}(1)_R$ and diagonal part of ${\rm SU}(2)_R$
from the $\N=2$ superconformal algebra,
\begin{align}\label{eq_J2toJ1}
\mathbb J = \frac{1}{3} \bm J + \frac{4}{3} J_{\1\2}~.
\end{align}
Here we recall that $\bm J$ obeys
\begin{align}
[\bm J, {\bm \cD}_\alpha^i] = - {\bm \cD}_\alpha^i~, \qquad
[\bm J, \bar{\bm \cD}^\dalpha_i] = + \bar{\bm \cD}^\dalpha_i
\end{align}
while $\mathbb J$ must be chosen to obey
\begin{align}
[\mathbb J, \cD_\alpha] = - \cD_\alpha~, \qquad
[\mathbb J, \bar\cD_\dalpha] = + \bar\cD_\dalpha~.
\end{align}
The specific relation given in \eqref{eq_J2toJ1} is selected
out by the superconformal algebra.
Another linear combination yields the so-called shadow chiral rotation
\begin{align}
\mathbb S = \frac{1}{2} \bm J - J_{\1\2}
\end{align}
which rotates the second supersymmetry generator while leaving the first fixed,
\begin{align}
[\mathbb S, \cD_\alpha^\1] = 0~,\qquad [\mathbb S, \cD_\alpha^\2] = - \cD_\alpha^\2~, \\
[\mathbb S, \bar\cD^\dalpha_\1] = 0~,\qquad [\mathbb S, \bar\cD^\dalpha_\2] = + \bar\cD^\dalpha_\2~.
\end{align}

If a tensor superfield $\bm U$ is conformally primary,
then under an $\N=2$ superconformal Killing isometry $\bm U$ transforms as
in \eqref{eq_deltaU2}. Taking the $\N=1$ projection of this gives
\begin{align}
\delta U &= -\left(\bm\xi \bm U + \frac{1}{2} \bm\lambda^{\rm cd} M_{\rm cd} \bm U
	+ \bm\lambda^{jk} J_{jk} \bm U
	+ \frac{1}{2} (\bm\sigma+\bar{\bm\sigma}) \mathbb D \bm U
	+ \frac{1}{4} (\bm\sigma-\bar{\bm\sigma}) \bm J \bm U\right)\Big| \eol
	&= -\left(\xi U + \frac{1}{2} \lambda^{\rm cd} M_{\rm cd} U
	+ \frac{1}{2} (\sigma+\bar{\sigma}) \mathbb D U
	+ \frac{3}{4} (\sigma-\bar{\sigma}) \mathbb J U\right)
	\eol & \quad
	- \rho^\alpha (\bm\cD_\alpha^\2 \bm U)\vert
	- \bar\rho_\dalpha (\bar{\bm\cD}^\dalpha_\2 \bm U)\vert
	- \bm \lambda^{\1\1} J_{\1\1} U
	- \bm \lambda^{\2\2} J_{\2\2} U
	\eol & \quad
	- \ri \alpha \mathbb S U~.
\end{align}
Identifying these two equalities implies that
a general $\N=2$ superconformal Killing isometry decomposes into
three independent transformations:
\begin{enumerate}
\bfseries \item \mdseries
An $\N=1$ superconformal Killing isometry with $\delta U$ given by \eqref{eq_deltaU},
$\xi$ given in \eqref{B.23}, and the remaining parameters obeying
\begin{align}
\lambda_{\rm ab} &= \bm \lambda_{\rm ab}\vert = \cD_{\rm [a} \xi_{\rm b]} \\
\sigma &= \frac{2}{3} \bm\sigma\vert+\frac{1}{3} \bar{\bm\sigma}\vert + \frac{2}{3} \bm \lambda^{\1\2}\vert
	= -\frac{\ri}{24} \cD_\alpha \bar \cD_\dalpha \xi^{\dalpha \alpha}
	- \frac{\ri}{12} \bar\cD_\dalpha \cD_\alpha \xi^{\dalpha \alpha}
\end{align}
\bfseries \item \mdseries
An extended superconformal transformation with
\begin{align}
\delta U &= -\rho^\alpha (\bm\cD_\alpha^\2 \bm U)\vert
	- \bar\rho_\dalpha (\bar{\bm\cD}^\dalpha_\2 \bm U)\vert
	- \bm \lambda^{\1\1} \mathbb J_{\1\1} U
	- \bm \lambda^{\2\2} \mathbb J_{\2\2} U~, \\
\rho_\alpha &= \frac{\ri}{8} (\bar{\bm\cD}^\dalpha_\2 \bm\xi_{\alpha \dalpha})\vert~,\qquad
\bm\lambda^{\1\1}\vert = -\frac{\ri}{16} \cD_\alpha (\bar{\bm\cD}_{\dalpha \2} \bm\xi^{\dalpha \alpha})\vert
	= -\frac{1}{2} \cD^\alpha \rho_\alpha
\end{align}
As discussed in eq. \eqref{rho-deff}, it is possible to define the $\N=1$ superfield $\rho$ in AdS
which simplifies these two formulae. From an $\N=1$ perspective, $\rho$ is an independent
superfield.
\bfseries \item \mdseries
A shadow chiral rotation with
\begin{align}
\delta U &= -\ri \alpha \mathbb S U~, \\
\alpha &= -\frac{\ri}{3} (\bm\sigma - \bar{\bm\sigma})\vert + \frac{2\ri}{3} \bm\lambda^{\1\2}\vert
	= \frac{1}{96} [\cD_\alpha, \bar\cD_\dalpha] \xi^{\dalpha \alpha}
	+ \frac{1}{32} ([\bm \cD_\alpha^\2, \bar{\bm\cD}_{\dalpha\2}] \bm\xi^{\dalpha \alpha})\vert~.
\end{align}
One can check that $\alpha$ is a \emph{constant}. From an $\N=1$ perspective, it is 
an independent parameter.
\end{enumerate}

\subsection{$\cN=2$ Killing vector fields and their $\cN=1$ reduction}
\label{appendix_B3}
Any superconformal Killing vector field $\x$ with  the additional property
\be
{\bm \s}=0
\ee
is called a Killing vector field of the $\cN=2$ AdS superspace. 
The Killing vector obeys the equation 
\bea
[{\bm \x} + \hf {\bm \l}^{\rm cd} M_{\rm cd}+{\bm \l}^{kl} J_{kl}, {\bm \cD}_\a^i  ] 
=0   \quad \Longrightarrow \quad {\bm \l}^{kl} = 2 {\bm \ve} \,{\bm S}^{kl} ~,
 \quad \bar{\bm \ve}={\bm \ve}~.
\label{N=2Kv1}
\eea
Making use of eq. \eqref{eq_N2params2} gives
\bea
{\bm \ve} = \frac{1}{8} {\bm S}^{ij} {\bm \cD}_{\a i} {\bm \x}^\a_j~.
\label{B.333}
\eea
As before, we fix ${\bm S}^{ij}$ as in eqs. (\ref{S12}) and (\ref{S11S22}), and thus 
\bea
{\bm \l}^{\1\1} = -2 \bar \m {\bm \ve}~.
\eea
The Killing vector fields prove to generate the isometry group of the $\cN=2$ AdS superspace, 
$\rm OSp(2|4)$. They were studied in detail in \cite{KT-M-ads}.

Consider the $\cN=1$ projection of the  $\cN=2$ Killing vector  
\bea
{\bm \x }| &=& \x + {\bm \x}^\a_{\2} | {\bm \cD}_\a^{\2} | + \bar{\bm \x}_\ad^{\2} | \bar{\bm \cD}^\ad_{\2} |
\equiv  \x + \ve^\a {\bm \cD}_\a^{\2} | + \bar \ve_\ad \bar{\bm \cD}^\ad_{\2} |~, \\
\x&:=& {\bm \x}^{\rm a} |\cD_{\rm a} + {\bm \x}^\a_{\1} | {\cD}_\a + \bar{\bm \x}_\ad^{\1} | \bar{\cD}^\ad
\equiv  { \x}^{\rm a} \cD_{\rm a} + { \x}^\a {\cD}_\a + \bar{\x}_\ad \bar{\cD}^\ad ~.
\eea
One may see that $\x$ is a Killing vector of the $\cN=1$ AdS superspace, 
see Appendix \ref{Appendix_A}.
In accordance with the relations (\ref{rho-deff}) and (\ref{rho-constraints}),
we now have
\bea
\ve_\a = \cD_\a \ve~, \qquad \ve:= {\bm \ve}| =\bar \ve ~,
\eea
where the real parameter $\ve $ obeys the constraints
\bea
(\bar \cD^2 -4 \m) \ve =(\cD^2 -4\bar \m) \ve =0~, \qquad 
 \cD_\a \bar \cD_\ad \ve =\bar \cD_\ad \cD_\a \ve =0~.
\label{B.37}
\eea

We see that there are transformations of matter superfields in $\cN=1$ AdS superspace 
which are generated by the $\cN=2$ superconformal Killing vector $\bm \x$. They are:
\begin{enumerate}
\bfseries \item \mdseries
An $\cN=1$ AdS  transformation. It is described by an $\cN=1$ AdS Killing vector
$\x$ and corresponds to the choice
\bea
{\bm \ve} |=0~.
\eea
\bfseries \item \mdseries
An extended supersymmetry transformation. It corresponds to the choice 
\bea
\x =0~.
\eea
It is described by a real parameter $\ve$ subject to the constraints (\ref{B.37}).
This parameter was introduced in \cite{GKS} where the $\cN=2$ off-shell massless  supermultiplet 
of arbitrary spin in $\rm AdS_4$ were constructed.
\end{enumerate}

\section{Direct proof of invariance for general nonlinear $\sigma$-models}\label{HK_direct}
We presented in subsection \ref{HK_indirect} an elegant proof of invariance
which was somewhat indirect and relied upon the existence of quantities such
as the superfield $B$ in \eqref{eq_B2}. We present in this appendix a more direct
(but rather technical) proof of the same result, namely that the action
\begin{align}
S = \int \rd^4x\, \rd^4\theta\, E\, \cK (\f, \bar \f)
\end{align}
is invariant under the $\N=2$ AdS transformations (\ref{N=2SUSYtr})
provided the conditions \eqref{eq_omegaCond} and \eqref{eq_KillingCond} hold.

We begin by recasting $\delta \veps S$ in the form
\begin{align}
\delta_\ve S &= \fint \left(
     \frac{1}{2} \cK_a \bar \veps_\dalpha \bar\cD^\dalpha \bar \Omega^a
     + \frac{1}{2} \veps \bar\cD_\dalpha \cK_a \bar \cD^\dalpha \bar\Omega^a
     + \HC \right)~.
\end{align}
Making use of the conditions \eqref{eq_omegaCond}, 
the second term on the right can be seen to vanish, and thus
\begin{align}
\delta_\ve S &= -\frac{1}{2} \fint 
     \bar\veps_\dalpha \,\bar\cD^\dalpha \bar\phi^{\bar a} \,\bar\Omega_{\bar a} + \HC~,
\label{eq_pfGenForm}
\end{align}
where  $\bar\O_{\bar b}:= g_{a \bar b} \bar \O^a$,
and $\bar \O^a$ is given by eq. (\ref{6.32}).
Rewriting this as a chiral integral yields
\begin{align}
\delta_\ve S &= \cint
     \Big\{
     \frac{1}{8} \bar \veps_\dalpha \bar \cD^2 (\bar\cD^\dalpha \bar\phi^{\bar a} \bar\Omega_{\bar a})
     - \frac{3 \mu}{4} \bar\veps_\dalpha \bar\cD^\dalpha \bar\phi^{\bar a} \,\bar \Omega_{\bar a}
     - \frac{\mu}{2} \veps \bar\cD_\dalpha (\bar \cD^\dalpha \bar\phi^{\bar a} \bar \Omega_{\bar a})
     \Big\} + \HC
\end{align}

To evaluate this further, it helps a great deal to make use of reparametrization-covariant
derivatives, which ensure that derivatives of superfields are packaged in a convenient manner.
(In particular, all factors of the connection $\Gamma^a{}_{bc}$ are hidden from view.)
We make use of the formalism developed in detail in \cite{Binetruy:2000zx}
at the superfield level.\footnote{The formulation is essentially a generalization of that
employed at the component level in the textbook \cite{WB}.}
On any superfield $\mathbf U^a$ which transforms as a target-space vector under 
holomorphic reparametrizations and as an arbitrary tensor under Lorentz transformations,
we may define the derivative $\nabla_A$ by
\begin{align}
\nabla_A \mathbf U^a := \cD_A \mathbf U^a + \Gamma^a{}_{b c} \cD_A \phi^b \mathbf U^c, \qquad
\nabla_A \mathbf U^{\bar a} := \cD_A \mathbf U^{\bar a}
     + \Gamma^{\bar a}{}_{\bar b \bar c} \cD_A \bar\phi^{\bar b} \mathbf U^{\bar c}
\end{align}
which is reparametrization-covariant. Similarly on a superfield $\mathbf U_a$, one has
\begin{align}
\nabla_A \mathbf U_a := \cD_A \mathbf U_a - \Gamma^c{}_{ab} \cD_A \phi^b \mathbf U_c, \qquad
\nabla_A \mathbf U_{\bar a} := \cD_A \mathbf U^{\bar a}
     - \Gamma^{\bar c}{}_{\bar a \bar b} \cD_A \bar\phi^{\bar b} \mathbf U_{\bar c}~.
\end{align}
This is essentially the pullback to superspace of the target-space covariant derivative.
(The generalization to a superfield with multiple target-space indices is straightforward.)
In particular, the metric $g_{a \bar b}$ is covariantly constant under this derivative.

Making use of these new derivatives, the variation of the action can be written
\begin{align}
\delta_\ve S &= \cint
     \Big\{
     - \frac{1}{2} \veps \mu \bar\cD_\dalpha (\bar \cD^\dalpha \bar\phi^{\bar a} \bar \Omega_{\bar a})
     - \mu \bar\veps_\dalpha \bar\cD^\dalpha \bar\phi^{\bar a} \,\bar \Omega_{\bar a}
    \eol & \quad
     + \frac{1}{8} \bar\veps_\dalpha \,\bar\cD^\dalpha \bar\phi^{\bar a}\,
          \bar\nabla_\dbeta \bar\cD^\dbeta \bar\phi^{\bar b}
          \left(\nabla_{\bar b} \bar\Omega_{\bar a} - \nabla_{\bar a} \bar\Omega_{\bar b}\right)
     + \frac{1}{8} \bar\veps_\dalpha \,\bar\cD^\dalpha \bar\phi^{\bar a}\,
          \bar\cD_\dbeta\bar\phi^{\bar b} \,\bar\cD^\dbeta\bar\phi^{\bar c}\,
          \nabla_{\bar c} \nabla_{\bar b} \bar \Omega_{\bar a}
     \Big\} + \HC
\end{align}
At this point, several simplifications occur. We  first recall that
\begin{align}\label{eq_dXi}
\nabla_{\bar b} \bar\Omega_{\bar a} = -\omega_{\bar b \bar a}
\end{align}
is both antisymmetric and covariantly constant. Several terms then
simplify to yield
\begin{align}
\delta_\ve S = \cint
     \Big\{
    & - \frac{\mu}{2} \veps \bar\cD_\dalpha (\bar\cD^\dalpha \bar\phi^{\bar a} \bar \Omega_{\bar a})
     - \mu \bar\veps_\dalpha \bar\cD^\dalpha \bar\phi^{\bar a} \,\bar \Omega_{\bar a} \non \\
&     - \frac{1}{4} \bar \veps_\dalpha \bar\cD^\dalpha \bar\phi^{\bar a} \bar\nabla_\dbeta
          \bar\cD^\dbeta \bar\phi^{\bar b} \omega_{\bar b \bar a}
     \Big\} + \HC
\end{align}

Now we must go to components using the $\cN=1$ AdS  reduction rule
(see  \cite{KS94} or standard texts on $\cN=1$ supergravity \cite{BK,WB})
\bea
\cint  \cL_{\text{chiral}} = -\frac{1}{4} \compint (\cD^2 - 12 \bar \mu) \cL_{\text{chiral}}~,
\eea
for any covariantly chiral Lagrangian $\cL_{\text{chiral}}$.
However, before doing so, there are
certain steps we may take which will drastically simplify the resulting
calculation. First we rewrite $\delta_\ve S$ as\footnote{Note that the
second and third terms involve the usual covariant derivative
since their arguments are scalars under reparametrizations.}
\begin{align}\label{eq_deltaS}
\delta_\ve S = \cint
     \Big\{
&     - \frac{\mu}{2} \bar\veps_\dalpha \bar\cD^\dalpha \bar\phi^{\bar a} \,\bar \Omega_{\bar a}
     - \frac{1}{6} \bar\cD_\dbeta \left(\bar \veps_\dalpha \bar\cD^\dalpha \bar\phi^{\bar a}
          \bar\cD^\dbeta \bar\phi^{\bar b} \omega_{\bar b \bar a}\right) \non \\
&	- \frac{\mu}{2} \bar \cD_\dalpha \left(\veps \bar\cD^\dalpha \bar\phi^{\bar a} \bar \Omega_{\bar a}\right)
     \Big\} + \HC
\end{align}
The first term gives simply
\begin{align}
\compint \bar\veps_\dalpha \Big(
     - \mu \bar \mu \bar \cD^\dalpha \bar\phi^{\bar a} \bar \Omega_{\bar a}
     + \frac{\ri}{2} \mu \cD^{\dalpha \alpha} \bar\phi^{\bar a} \nabla_\alpha \bar\Omega_{\bar a}
     + \frac{\mu}{8} \bar\cD^\dalpha \bar\phi^{\bar a} \nabla^2 \bar \Omega_{\bar a}
     \Big)~.
\end{align}
To evaluate the second and third terms of \eqref{eq_deltaS} requires the AdS identity \cite{KS94}
\begin{align}\label{eq_AdSEval2}
      \compint (\cD^2 - 12 \bar \mu) \bar\cD_\dalpha \bar\Psi^\dalpha
     =
     \compint \bar\cD_\dalpha (\cD^2 - 8 \bar \mu) \bar\Psi^\dalpha~,
\end{align}
where we have discarded a total derivative in the final equality. The second term
in \eqref{eq_deltaS} then evaluates to 
\begin{align}
&\frac{1}{3} \, \compint \bar \cD_\dalpha \left(
          \bar\veps_\dbeta \cD^{\dalpha \alpha} \bar\phi^{\bar a}
               \cD_\alpha{}^\dbeta \bar\phi^{\bar b} \omega_{\bar b \bar a}
     \right) \eol
     &= -\frac{1}{3}\, \compint \left\{
          \veps_{\dbeta} \cD^{\dalpha \alpha} (\bar\cD_\dalpha \bar\phi^{\bar a}
          \cD_\alpha{}^\dbeta \bar \phi^{\bar b} \omega_{\bar b \bar a})
          + \veps_{\dbeta} \cD^{\dbeta \alpha} (\bar\cD_\dalpha \bar\phi^{\bar a}
          \cD_\alpha{}^\dalpha \bar \phi^{\bar b} \omega_{\bar b \bar a}) \right\} \eol
     &= \ri \,\mu\compint
          \veps^\alpha \bar\cD^\dalpha \bar\phi^{\bar a}
          \cD_{\alpha \dalpha} \bar \phi^{\bar b} \omega_{\bar b \bar a}
\end{align}
after integrating by parts.
The third term in \eqref{eq_deltaS} is a bit more complicated. It evaluates to
\begin{align*}
 \frac{\mu}{8} 
    \compint  
     (\cD^2 - 12 \bar \mu)
     \bar \cD_\dalpha \left(\veps \bar\cD^\dalpha \bar\phi^{\bar a} \bar \Omega_{\bar a}\right)
     = \compint \left(\bar \veps_\dalpha \Psi^{\dalpha}_{1}
	+ \veps^\alpha \Psi_{2\, \alpha}
	+ \veps \Psi_{3}\right)~,
\end{align*}
where
\begin{align*}
\Psi^\dalpha_{1} &:=
     \frac{\mu}{8} \bar\cD^\dalpha \bar\phi^{\bar a} \nabla^2 \bar\Omega_{\bar a}
     + \frac{\ri}{2} \mu \cD^{\dalpha \alpha} \bar\phi^{\bar a} \nabla_\alpha \bar\Omega_{\bar a}~, \\
\Psi_{2\alpha} &:=
     - \frac{\ri}{2} \mu \nabla_{\alpha \dalpha} \bar\cD^\dalpha \bar\phi^{\bar a}
          \bar\Omega_{\bar a}
     - \frac{\ri}{2} \mu \cD_{\alpha \dalpha} \bar \phi^{\bar a} \bar\nabla^\dalpha \bar \Omega_{\bar a}
     - \frac{\mu}{4} \bar\cD^\dalpha \bar\phi^{\bar a} \bar\nabla_{\dalpha} \nabla_\alpha \bar \Omega_{\bar a}
     + \frac{\mu}{4} \bar\nabla_\dalpha \bar\cD^\dalpha \bar\phi^{\bar a} \nabla_\alpha \bar\Omega_{\bar a} ~,\\
\Psi_3 &:=
     \frac{\mu}{8} \bar\nabla\bar\cD \bar\phi^{\bar a} \nabla^2 \bar\Omega_{\bar a}
     + \frac{\mu}{8} \bar \cD_\dalpha \bar\phi^{\bar a} \bar\nabla^{\dalpha} \nabla^2 \bar \Omega_{\bar a}
     + \frac{\ri}{2} \mu \nabla^{\dalpha \alpha} \bar \cD_\dalpha \bar\phi^{\bar a} \nabla_\alpha \bar\Omega_{\bar a}
     + \frac{\ri}{2} \mu \cD^{\dalpha \alpha} \bar\phi^{\bar a}
          \bar \nabla_{\dalpha} \nabla_\alpha \bar \Omega_{\bar a}~.
\end{align*}

Now we put these three terms together. Performing one integration by parts
(for the first term in $\Psi_{2\alpha}$) and making use of \eqref{eq_dXi}
and $\mu \nabla_b \bar \Omega_{\bar a} = -\bar\mu \nabla_{\bar a} \Omega_{b}$,
we find
\begin{align}
\delta_\ve S &= \compint \left(\bar \veps_\dalpha \Psi'^{\dalpha}_{1}
	+ \veps^\alpha \Psi'_{2\, \alpha}
	+ \veps \Psi'_{3}\right) + \HC \eol
	&= \compint \Big\{\bar \veps_\dalpha (\Psi'^{\dalpha}_{1} + \bar\Psi'^\dalpha_2)
	+ \veps^\alpha (\Psi'_{2\, \alpha} + \bar \Psi'_{1 \,\alpha})
	+ \veps (\Psi'_{3} + \bar\Psi'_3) \Big\}~,
\end{align}
where
\begin{align*}
\Psi'^\dalpha_1 &:=
     \frac{\mu}{4} 
     \bar\cD^\dalpha \bar\phi^{\bar a} \nabla^\alpha \cD_\alpha \phi^b \nabla_b\bar\Omega_{\bar a}
     + 
     \frac{\mu}{4} 
     \bar\cD^\dalpha \bar\phi^{\bar a} \cD^\alpha \phi^b \cD_\alpha \phi^c
          \nabla_c \nabla_b \bar\Omega_{\bar a}
     +\ri 
     \mu 
     \cD^{\dalpha \alpha} \bar\phi^{\bar a} \cD_\alpha \phi^b \nabla_b \bar\Omega_{\bar a} ~,\\
\Psi'_{2\alpha} &:=
     - \frac{\bar\mu}{4}  \bar\nabla_\dalpha \bar\cD^\dalpha \bar\phi^{\bar a}
          \cD_\alpha \phi^b \nabla_{\bar a} \Omega_{b}
     - \frac{\bar\mu}{4} 
     \bar\cD_\dalpha \bar\phi^{\bar a} \bar\cD^\dalpha \bar\phi^{\bar b}
          \cD_\alpha \phi^c\, \nabla_{\bar b} \nabla_{\bar a} \Omega_{c}
     - \ri \bar \mu 
     \bar \cD^\dalpha \bar\phi^{\bar a}
          \cD_{\alpha \dalpha} \phi^b \nabla_{\bar a} \Omega_{b} ~,\\
\Psi'_3 &:=
     \frac{\mu}{8} \bar\nabla\bar\cD \bar\phi^{\bar a} \nabla^2 \bar\Omega_{\bar a}
     + \frac{\mu}{8} \bar \cD_\dalpha \bar\phi^{\bar a} \bar\nabla^{\dalpha} \nabla^2 \bar \Omega_{\bar a}
     + \frac{\ri}{2} \mu \nabla^{\dalpha \alpha} \bar \cD_\dalpha \bar\phi^{\bar a}
          \nabla_\alpha \bar\Omega_{\bar a}
     + \frac{\ri}{2} \mu \cD^{\dalpha \alpha} \bar\phi^{\bar a}
          \bar \nabla_{\dalpha} \nabla_\alpha \bar \Omega_{\bar a}~.
\end{align*}
By inspection we can see that the coefficients of $\veps_\alpha$ and $\bar\veps_\dalpha$
will cancel. The remaining terms involving $\veps$ may be rearranged into
\begin{align}
\delta_\ve S
     &= \compint \veps \Bigg\{
     \frac{1}{16} [\cD_\alpha, \bar \cD_\dalpha] \Big(\mu \cD^\alpha \phi^b 
          \bar \cD^\dalpha \bar\phi^{\bar b} \nabla_b \bar \Omega_{\bar b}
          + \bar\mu \cD^\alpha \phi^{b} \bar\cD^\dalpha \bar\phi^{\bar b} \nabla_{\bar b} \Omega_{b} \Big)
     \eol & \quad
     + \frac{\ri}{4} \mu \nabla^{\dalpha \alpha} \bar \cD_\dalpha \bar\phi^{\bar a}
          \nabla_\alpha \bar\Omega_{\bar a}
     + \frac{\ri}{4} \mu \cD^{\dalpha \alpha} \bar\phi^{\bar a}
          \bar \nabla_{\dalpha} \nabla_\alpha \bar \Omega_{\bar a}
     \eol & \quad
     + \frac{\ri}{4} \bar\mu \nabla^{\dalpha \alpha} \cD_\alpha \phi^{a} \nabla_\alpha \Omega_{a}
     + \frac{\ri}{4} \bar\mu \cD^{\dalpha \alpha} \phi^{a}
          \nabla_{\alpha} \bar\nabla_\dalpha \Omega_{a}
     \Bigg\}~,
\end{align}
where we have again discarded a total derivative.
The first line vanishes using the relation
$\mu \nabla_b \bar\Omega_{\bar b} = -\bar\mu \nabla_{\bar b} \Omega_{b}$.
The remaining terms can be rearranged using the same relation
(and after relabelling some indices) to
\begin{align}
\delta_\ve S &=  \frac{\ri}{4} \mu  \compint \veps \Big\{
     \nabla^{\dalpha \alpha} \bar \cD_\dalpha \bar\phi^{\bar b}
          \cD_\alpha \phi^b \nabla_b \bar\Omega_{\bar b}
     + 
     \bar\cD_\dalpha \bar\phi^{\bar b} \nabla^{\dalpha \alpha} \cD_\alpha \phi^{b}
          \nabla_{b} \bar \Omega_{\bar b}
     \eol & \quad
     + 
     \bar \cD_{\dalpha} \phi^{\bar b} \cD_\alpha \phi^b
          \cD^{\dalpha \alpha} \bar\phi^{\bar c}
          \nabla_{\bar b} \nabla_b \bar\Omega_{\bar c}
     + 
     \bar \cD_{\dalpha} \phi^{\bar b} \cD_\alpha \phi^b
          \cD^{\dalpha \alpha} \phi^{c}
          \nabla_b \nabla_{c} \bar\Omega_{\bar b}
     \Big\}~,
\end{align}
where all terms have been rewritten to involve only $\bar \Omega$.
This can be seen to be a total derivative by using the relations
\begin{align*}
\nabla_b \nabla_c \bar\Omega_{\bar b} &= \nabla_c \nabla_b \bar\Omega_{\bar b}~, \\
\mu \nabla_{\bar b} \nabla_b \bar \Omega_{\bar c} &= -\bar\mu \nabla_{\bar b} \nabla_{\bar c} \Omega_{b}
     = -\bar\mu \nabla_{\bar c} \nabla_{\bar b} \Omega_{b}
     = \mu \nabla_{\bar c} \nabla_{b} \bar \Omega_{\bar b}~.
\end{align*}
Thus the action is invariant and the necessary conditions \eqref{eq_omegaCond}
and \eqref{eq_KillingCond} are indeed sufficient.

\section{Improvement transformations for non-minimal supergravity}\label{appendixD}
Within non-minimal AdS supergravity \cite{BK-dual}, supercurrent conservation in an
AdS background naturally takes the form
\begin{align}\label{eq_nonminSC}
\bar \cD^\dalpha J_{\alpha \dalpha} = -\frac{1}{4} \bar\cD^2 \zeta_\alpha~,\qquad
\cD_{(\beta} \zeta_{\alpha)} = 0~.
\end{align}
This can be derived by considering the linearized coupling of a matter action to
non-minimal supergravity, which involves a real supergravity prepotential
$H_{\alpha \dalpha}$ as well as a complex linear compensator $\Gamma$ obeying
\begin{align}
(\bar \cD^2 - 4\mu)\Gamma = 0~.
\end{align}
The constraint on $\Gamma$ can be solved by taking
$\Gamma= \bar\cD_\dalpha \bar\psi^\dalpha$ for an unconstrained spinor
superfield $\bar\psi^\dalpha$ with a gauge invariance
$\delta\bar\psi^\dalpha = \bar\cD_\dbeta \bar\Omega^{\dbeta \dalpha}$
for $\bar\Omega^{\dbeta \dalpha} = \bar\Omega^{\dalpha\dbeta }$.
The linearized action can generically be written
\begin{align}
S^{(1)}= \int \rd^4x\, \rd^4\theta\, E\, \Big(
	H^{\dalpha \alpha} J_{\alpha \dalpha}
	+ \psi^\alpha \zeta_\alpha + \bar\psi_\dalpha \bar\zeta^\dalpha \Big)~.
\end{align}
The gauge invariance of $\psi$ implies that $\zeta_\alpha$ obeys the constraint
$\cD_{(\beta} \zeta_{\alpha)} = 0$. This constraint is solved by
$\zeta_\alpha = \cD_\alpha \zeta$ with $\zeta$ possessing a gauge invariance
$\zeta \rightarrow \zeta + \bar\Lambda$ for an antichiral parameter
$\bar\Lambda$.\footnote{In fact, in AdS, one can construct a global solution:
$\zeta = \cD^\alpha \zeta_\alpha / 4\bar\mu$.}
The equation \eqref{eq_nonminSC} arises when we impose invariance under the supergravity
gauge transformations
\begin{align}\label{D.4}
\delta H_{\alpha\dalpha} = \cD_\alpha \bar L_\dalpha - \bar\cD_\dalpha L_\alpha~,\qquad
\delta \psi_\alpha = -\frac{1}{4} \bar\cD^2 L_\alpha~.
\end{align}
Instead of the transformation for $\psi_\alpha$ given by \eqref{D.4},
one can consider a more general transformation law
\begin{align}
\delta \psi_\alpha = -\frac{1}{4} \bar\cD^2 L_\alpha + \frac{\kappa}{4} \bar\cD^2 L_\alpha
	+ \frac{\kappa}{4}\bar\cD^\dalpha \cD_\alpha \bar L_\dalpha
\end{align}
for some parameter $\kappa$ (for simplicity $\kappa$ is chosen real).
This may be understood as simply shifting the original field $\psi_\alpha$
by the term $\dfrac{\kappa}{4} \cD^\dalpha H_{\alpha \dalpha}$.
The new supercurrent can be shown to obey the conservation equation
\begin{align}\label{eq_newSC}
\bar\cD^\dalpha J_{\alpha \dalpha} = \frac{1}{4} \left(\kappa - 1\right)\bar \cD^2 \zeta_\alpha
	- \frac{\kappa}{4} \bar\cD_\dalpha \cD_\alpha \bar\zeta^\dalpha~.
\end{align}
This is an equally valid supercurrent conservation equation for
non-minimal supergravity.

In \cite{BK-dual}, we argued that the supercurrent \eqref{eq_AdSgenSC}
was the most general supercurrent in AdS. As a check, we should show that the
supercurrent \eqref{eq_newSC} can be rewritten in that form.
The approach is simple. Letting $\bar\zeta_\dalpha = \bar\cD_\dalpha \bar\zeta$,
which is always possible in AdS, we note that
\begin{align}
\bar\cD_\dalpha \cD_\alpha \bar\zeta^\dalpha
	&= - \frac{1}{2} \bar \cD^2 \cD_\alpha \bar\zeta + \ri [\cD^\dalpha,\cD_{\alpha \dalpha}]\bar\zeta
	- \frac{1}{2} \cD_\alpha \bar \cD^2 \bar\zeta \eol
	&= - \frac{1}{2} \bar \cD^2 \cD_\alpha \bar\zeta - \frac{1}{2} \cD_\alpha (\bar \cD^2 - 4 \mu) \bar\zeta~.
\end{align}
Therefore \eqref{eq_newSC} can indeed be written
\begin{align}
\bar\cD^\dalpha J_{\alpha \dalpha} &= \cD_\alpha X' -\frac{1}{4} \bar \cD^2 \zeta_\alpha' \eol
X' &= \frac{\kappa}{8} (\bar \cD^2 - 4 \mu) \bar\zeta ~,\qquad
\zeta_\alpha' = (1 - \kappa) \zeta_\alpha - \frac{\kappa}{2} \cD_\alpha \bar\zeta~.
\end{align}
Absorbing $X'$ into $\zeta_\alpha'$ (which is always possible in AdS), we can write this
instead as
\begin{align}
\bar\cD^\dalpha J_{\alpha \dalpha} &= -\frac{1}{4} \bar \cD^2 \zeta_\alpha' \eol
\zeta_\alpha' &= (1 - \kappa) \zeta_\alpha
	- \frac{\kappa}{8 \mu} \cD_\alpha \bar \cD_\dalpha \bar\zeta^\dalpha~.
\end{align}

\footnotesize{

}

\end{document}